\documentclass[amsmath, amssymb, floatfix, reprint, twocolumn, notitlepage, nofootinbib, 10pt]{revtex4-1}

\usepackage[utf8]{inputenc}
\usepackage{microtype,bm,bbm,graphicx,booktabs,times}
\usepackage[usenames,dvipsnames]{xcolor}
\usepackage{setspace}
\selectfont{}

\usepackage{hyperref}
\hypersetup{colorlinks=true, linkcolor=blue, linktoc=all}
\usepackage{braket}
\usepackage{subfiles} 
\usepackage{upgreek}
\usepackage{color}
\usepackage[capitalize]{cleveref}
\crefformat{section}{Sec.~#2#1#3}

\usepackage{amsthm}
\makeatletter
\newtheorem*{rep@theorem}{\rep@title}
\newcommand{\newreptheorem}[2]{%
\newenvironment{rep#1}[1]{%
 \def\rep@title{#2 \ref*{##1}, repeated}%
 \begin{rep@theorem}}%
 {\end{rep@theorem}}}
\makeatother

\newtheorem{theorem}{Theorem}
\newreptheorem{theorem}{Theorem}
\newtheorem{lemma}{Lemma}
\newreptheorem{lemma}{Lemma}
\newtheorem{proposition}{Proposition}
\newreptheorem{proposition}{Proposition}
\newtheorem{problem}{Problem}
\newreptheorem{problem}{Problem}
\newtheorem{assumption}{Assumption}
\newreptheorem{assumption}{Assumption}
\newtheorem{definition}{Definition}
\newreptheorem{definition}{Definition}

\newtheorem{example}{Example}

\newcommand{\edit}[1]{{\color{black} #1}}

\newcommand{\norm}[1]{\left\Vert{#1}\right\Vert}%
\newcommand{\abs}[1]{\left | {#1}\right |}

\begin{document}
\title{Description and complexity of Non-markovian open quantum dynamics}
\author{Rahul Trivedi$^{1, 2}$}
\email{rahul.trivedi@mpq.mpg.de, rtriv@uw.edu}
\address{Max-Planck-Institut für Quantenoptik, Hans-Kopfermann-Str.~1, 85748 Garching, Germany,\\
Munich Center for Quantum Science and Technology (MCQST), Schellingstr. 4, D-80799 Munich, Germany,\\
Department of Electrical and Computer Engineering, University of Washington, Seattle, WA -  98195, USA.}

\date{\today}
\begin{abstract}
    Understanding and simulating non-Markovian quantum dynamics remains an important challenge in open quantum system theory. A key advance in this endeavour would be to develop a unified mathematical description of non-Markovian dynamics, and classify its complexity in the many-body setting. In this paper, we identify a general class of non-Markovian memory kernels, described by complex-valued radon measures, and define their dynamics through a regularization procedure constructing the corresponding system-environment unitary groups. Building on this definition, we then consider $k-$local many-body non-Markovian systems with physically motivated assumptions on the total variation and smoothness of the memory kernels. We establish that their dynamics can be efficiently approximated on quantum computers, thus providing a rigorous verification of the Extended Church-Turing thesis for this general class of non-Markovian open quantum systems.
\end{abstract}

\maketitle
\section{Introduction}

Quantum systems invariably interact with their environment, and any model describing their behaviour should model this interaction. Often such systems are approximated as Markovian, wherein the environment does not retain any memory of the system. Markovian open quantum systems have been extensively studied in quantum information theory and quantum optics --- a mathematically rigorous description of a unitary group over the system-environment Hilbert space which generates Markovian open quantum dynamics is provided in the theory of quantum stochastic calculus \cite{hudson1984quantum, kr1992introduction} as the solution of a quantum stochastic differential equation. When evolved with this unitary group, the system's reduced state satisfies the Lindbladian master equation \cite{breuer2002theory}. Furthermore, Markovian open quantum systems have also been considered in the many-body setting, whereing the system has an exponentially large Hilbert space --- several quantum algorithms have been developed to simulate their dynamics on quantum computers \cite{zanardi2016dissipative,chenu2017quantum,cleve2016efficient, kliesch2011dissipative}.

However, a number of quantum systems arising in solid-state physics \cite{finsterholzl2020nonequilibrium, chin2011generalized, groeblacher2015observation, de2008matter}, quantum optics \cite{calajo2019exciting, andersson2019non, gonzalez2019engineering, aref2016quantum, leonforte2021vacancy} as well as quantum biology and chemistry \cite{ishizaki2005quantum, chin2012coherence, ivanov2015extension, caycedo2021exact} are not Markovian and the environment's memory needs to be explicitly taken into account. Furthermore, these systems are also many-body, and simulating their dynamics requires the development of quantum algorithms that can account for the environment's memory.  This opens up the question of formulating a mathematically rigorous description of non-Markovian dynamics, as well as studying their computational complexity.

In this paper, we provide rigorous answers to both of these questions. We first consider the problem of formulating a well-defined non-Markovian model --- while it is generically expected that non-Markovian open quantum systems satisfy a generalized Nakajima-Zwazig \cite{xu2018convergence, ivanov2015extension} or time-convolutionless master equation \cite{smirne2010nakajima, pereverzev2006time,kidon2015exact}, it is usually hard to obtain such a master equation explicitly except only for weakly interacting environments \cite{vacchini2010exact, schroder2007reduced, timm2011time, mukamel1978statistical}. Given that generalizing the Markovian master equation to the non-Markovian setting is hard, a natural question is if we can formulate a well-defined system-environment unitary group that is general enough to capture all physically relevant systems. This question is non-trivial to rigorously answer for most physical models, since the environment is often described by a quantum field theory as opposed to a finite-dimensional system.

This question has been addressed in several specific settings in open quantum systems literature --- using standard theory of second quantization, unitary groups have been constructed for non-Markovian normalizable spin boson models \cite{leggett1987dynamics} as well as for weakly interacting non-normalizable spin boson models \cite{lonigro2021generalized}. Furthermore, for time-delayed feedback systems, non-Markovian dynamics can be rigorously defined by expanding the system Hilbert space with time \cite{grimsmo2015time, whalen2017open}. In this paper, we identify a general class of non-perturbative non-Markovian models with memory kernels described by tempered radon measures, which have been studied in classical probability and distribution theory as generalization of the delta function \cite{grigis1994microlocal}. We rigorously define a unitary group, which in general generates non-Markovian system dynamics, associated with a tempered radon measure.  Our generalization ties together disparate models used for open-system dynamics within the same mathematical framework --- as special cases, we recover Markovian dynamics, quantum systems with time-delay and feedback \cite{calajo2019exciting, grimsmo2015time, pichler2016photonic} and spin-boson models described by spectral density functions with vanishing high-frequency response \cite{leggett1987dynamics}. A major difficulty in constructing this unitary group is that the Schroedinger's equation for the system-environment state cannot be naively used since it is not guaranteed to have a solution. Our key technical contribution is to construct this unitary group via a regularization procedure --- we use standard mollifiers to regularize the radon measure to obtain a non-Markovian model where the Schroedinger's equation has a guaranteed solution, show that the limit of this solution as the regularization is removed exists and hence defines the dynamics associated with the radon measure. 

We next consider the simulability of non-Markovian many-body dynamics, thus defined, on quantum computers. While the quantum simulability of Markovian many-body open quantum systems \cite{zanardi2016dissipative,chenu2017quantum,cleve2016efficient, kliesch2011dissipative}, and many-body closed quantum systems\cite{lloyd1996universal, berry2014exponential, low2019hamiltonian, berry2015hamiltonian} have been extensively studied, non-Markovian open quantum systems have remain relatively unexplored. \edit{Physical intuition suggests non-Markovian dynamics should be efficiently simulable on quantum computer since a non-Markovian open quantum system should be approximable by a Markovian dilation i.e.~a larger Markovian open quantum system which includes some environment modes. However, to the best of our knowledge, the current bounds on the approximation error incurred by a Markovian dilation grow exponentially with evolution time in the worst case \cite{trivedi2021convergence, mascherpa2017open}, and are known to grow at-most polynomially with time only for non-Markovian models with sufficiently rapidly decaying spectral density functions \cite{woods2015simulating, woods2016dynamical}. This implies that the number of environment modes needed to be retained in the Markovian dilation would need to be increased exponentially with time to control the approximation error. Consequently, any quantum algorithm based on such Markovian dilations would cease to be efficient after evolution time that scales polynomially with system size, thus leaving open the question of quantum simulability of non-Markovian dynamics. }

Here, we establish that the general class of non-Markovian many-body models with memory kernels described by tempered radon measures can be efficiently simulated on a quantum computer under physically motivated assumptions on the growth and smoothness of the memory kernel. The quantum algorithm for simulating non-Markovian dynamics relies on a Markovian dilation of non-Markovian dynamics using a Lanczos iteration, also referred to the star-to-chain transformation \cite{chin2010exact, woods2014mappings}, which has been previously analyzed for Pauli-Fierz Hamiltonians \cite{woods2015simulating, woods2016dynamical, gualdi2013renormalization}, as well as for general models with distributional memory kernels but under the assumption of a finite particle emission rate into the environment \cite{trivedi2021convergence}. The key technical contribution in our work that allows us to prove that the quantum algorithm is efficient (i.e. with run-time polynomial in system size \emph{and} the evolution time) is to establish error bounds between the non-Markovian dynamics and its Markovian dilation which grow polynomially with system size and evolution time, as opposed to the previously known bounds which in the worst case grow exponentially with evolution time \cite{trivedi2021convergence, mascherpa2017open}. 
\section{Overview of results}
\subsection{Informal Summary}
\edit{\emph{Setup}: We consider non-Markovian open quantum system models where a quantum system, described by a finite-dimensional Hilbert space $\mathcal{H}_S$, interacts with $M$ bosonic baths. Each bath is individually a symmetric Fock spaces over $L^2(\mathbb{R})$ i.e.~the Hilbert space of the $\alpha^\text{th}$ bath can be described within second quantization with an annihilation operator $a_{\alpha, \omega}$ for $\omega \in \mathbb{R}$ with the canonical commutation relations (CCR) 
\[
[a_\omega, a_\nu] = 0, [a_\omega, a_\nu^\dagger] = \delta(\omega - \nu).
\]
Physically, it is convenient to interpret each bath as a continuum of Harmonic oscillators with $a_{\alpha, \omega}$ as the annihilation operator corresponding to the mode of the $\alpha^\text{th}$ bath at frequency $\omega$. We will consider a system-environment Hamiltonian that can be formally written as
\begin{subequations}\label{eq:hamiltonian_heuristic}
\begin{align}
    H = H_S(t)  + \sum_{\alpha = 1}^M H_{\alpha, E} + \sum_{\alpha = 1}^M \bigg(L_\alpha^\dagger A_\alpha+ \text{h.c.} \bigg),
\end{align}
where
$H_S(t)$ is the Hamiltonian describing the system dynamics in the absence of its interaction with the environment and $H_{\alpha, E}$ is the Hamiltonian corresponding to the environment's bath and is given by
\begin{align}
H_{\alpha, E} = \int_{-\infty}^\infty \omega a_{\alpha, \omega}^\dagger a_{\alpha, \omega} d\omega.
\end{align}
\end{subequations}
The interaction between the $\alpha^\text{th}$ bath and the system is specified by a system operator $L_\alpha$ (which we will refer to as `jump operators' to be consistent with the terminology for Markovian systems) and $A_\alpha$ is an operator that acts on the $\alpha^\text{th}$ bath can be formally expressed as
\begin{align}\label{eq:mode_annihilation_operator}
A_\alpha = \int_{-\infty}^\infty \hat{v}_\alpha(\omega) a_{\alpha, \omega} \frac{d\omega}{\sqrt{2\pi}},
\end{align}
where $\hat{v}_\alpha$, referred to as a `coupling function' throughout this paper, describes the frequency-dependence of the interaction between the system and the $\alpha^\text{th}$ bath. It will often be convenient to analyze this Hamiltonian in the interaction picture with respect to the bath, in which case the Hamiltonian is given by
\[
H(t) = H_S(t) + \sum_{\alpha = 1}^M \bigg(L_\alpha^\dagger A_\alpha(t) + \text{h.c.}\bigg),
\]
where
\begin{align*}
A_\alpha(t) &= e^{iH_{\alpha, E} t} A_\alpha e^{-i H_{\alpha, E} t}=\int_{-\infty}^\infty \hat{v}_{\alpha}(\omega) a_{\alpha, \omega}e^{-i\omega t} \frac{d\omega}{\sqrt{2\pi}}.
\end{align*}
\begin{figure}
\centering
\includegraphics[scale=0.5]{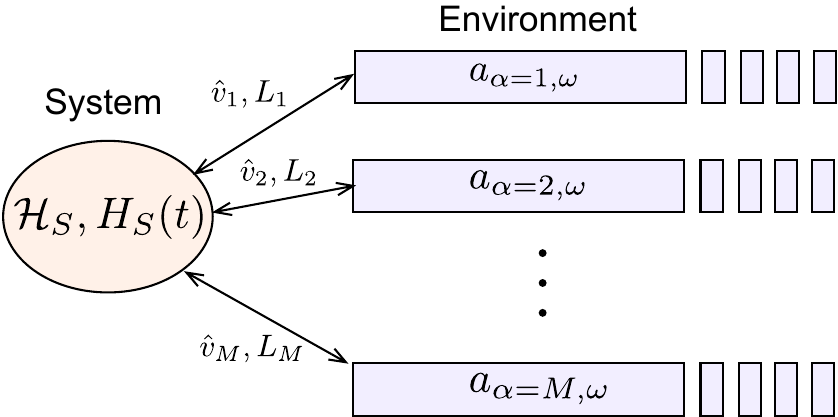}
\caption{Schematic depiction of a quantum system interacting with an environment of $M$ bosonic bath. The system dynamics is specified by, a possibly time-dependent, Hamiltonian $H_S(t)$. The interaction of the system with the individual bosonic baths is described by a jump operator $L_\alpha$ and a coupling function $\hat{v}_\alpha$.}
\end{figure}

\emph{Well definition of non-Markovian dynamics}: While one can formally write down a Hamiltonian for the system-environment interaction (Eq.~\ref{eq:hamiltonian_heuristic}), it is a time-dependent unbounded operator over an infinite-dimensional Hilbert space, and thus it is not immediate that a propagator corresponding to this Hamiltonian can be defined using the Schroedinger's equation (since its solution might not exist). Furthermore, the environment's Hilbert space is a Fock space over a continuum of modes and, in general, the operator $A_\alpha$ (Eq.~\ref{eq:mode_annihilation_operator}) does not correspond to the annihilation operator of a normalizable mode in the environment's Hilbert space, e.g.,~when the coupling function $\hat{v}_\alpha$ is the Fourier transform of a distribution as opposed to a square-integrable function. This is not just a mathematical consideration, several models of practical interest (such as that of Markovian quantum optical systems or quantum optical systems with time-delay and feedback) are modelled by distributional coupling functions. 

To make further progress on this problem, we first identify a class of coupling functions $\hat{v}_\alpha$ for which we can rigorously define a dynamical unitary group. It is useful to consider a very simple setting --- the single particle spontaneous emission problem into a single bath, i.e.~$\mathcal{H}_S \cong \mathbb{C}^2 \cong \text{span}\{ \ket{0}, \ket{1}\} $, $L = \ket{0}\bra{1}$. For an initial system-environment state $\ket{\psi(0)} = \ket{1}\otimes \ket{\textnormal{vac}}$. The amplitude of the system being in $\ket{1}$, $\varepsilon(t) = \bra{1} \otimes \bra{\text{vac}} \psi(t)\rangle$, is governed by
\begin{align}\label{eq:decay_equation}
\frac{d}{dt}\varepsilon(t) = -\int_0^t \mu(s) \varepsilon(t - s) ds
\end{align}
where $\mu$ is the non-Markovian memory kernel for the bath given by 
\[
\mu(t) = [A(t), A^\dagger(0)] = \int_{-\infty}^\infty \abs{\hat{v}(\omega)}^2 e^{-i\omega t} \frac{d\omega}{2\pi} .
\]
As we previously alluded to, $\mu$ can be a distribution. For Eq.~\ref{eq:decay_equation} to have a meaningful solution, we expect that the right hand side should be finite, else the derivative of $\varepsilon$ will be ill-defined. This expectation suggests that the memory kernel $\mu$ introduced above should be bounded i.e.~its application on a function $f$, that isnt large, should not be large. We can formalize this by demanding that for any continuous and compactly supported function $f$ with support $\Omega \subseteq \mathbb{R}$,
\begin{align}\label{eq:radon_measure_bound}
\abs{\int_{-\infty}^\infty \mu(t) f(t) dt}\leq c_\Omega \norm{f}_{L^\infty},
\end{align}
where $\norm{f}_{L^\infty} = \sup_{x\in \mathbb{R}} \abs{f(x)}$ and $c_\Omega$ is a constant that only depends on the support $\Omega$. The class of memory kernels that satisfy such a constraint have been well studied in analysis, and are called radon measures \cite{grigis1994microlocal}. Furthermore, in most physical problems of interest, the coupling function $\hat{v}$ and all of its derivatives grow at-most polynomially with $\omega$ --- since $\mu$ is a Fourier transform of $\abs{\hat{v}}^2$ for such a $\hat{v}$, it can additionally be assumed to be a tempered distribution. We will refer to such kernels as tempered radon measures throughout this paper.

Several kernels corresponding to models of common physical systems are tempered radon measures. For instance, Markovian open quantum systems are described by the delta function kernel, $\mu(t) = \gamma\delta(t)$ --- it can be immediately seen that this kernel is bounded in the sense of Eq.~\ref{eq:radon_measure_bound} with the constant $c_\Omega = \gamma$ for all $\Omega$. Non-Markovian open quantum systems with time-delays \cite{calajo2019exciting, grimsmo2015time, pichler2016photonic} and feedback are often modelled with kernels that are expressible as a sum of delta function i.e.~$\mu(t) = \sum_{j = 1}^P a_j \delta(t- \tau_j)$, and hence Eq.~\ref{eq:radon_measure_bound} is again satisfied with $c_\Omega = \sum_{j = 1}^P \abs{a_j}$. Finally, a very large class of physical models are described by memory kernels corresponding to square-integrable coupling functions ($\hat{v} \in L^2(\mathbb{R})$) --- examples of such models include lossy cavity QED systems in quantum optics \cite{tamascelli2018nonperturbative, pleasance2020generalized, mazzola2009pseudomodes, dalton2001theory, garraway2006theory, dalton2012quasimodes}, non-Markovian spin-boson models  \cite{leggett1987dynamics} and non-Markovian models with ohmic baths \cite{shi2016bound}. If $\hat{v}$ is square integrable, then it follows that the corresponding kernel $\mu(t)$ is a bounded continuous function of $t$ i.e.~for all $t$,
\[
\abs{\mu(t)} \leq {\int_{-\infty}^\infty \abs{\hat{v}(\omega)}^2 \frac{d\omega}{2\pi}} = \frac{\norm{\hat{v}}_{L^2}^2}{2\pi}.
\]
Consequently, it follows that Eq.~\ref{eq:radon_measure_bound} is again satisfied with 
\[
c_\Omega = \text{diam}(\Omega) \sup_{t \in \mathbb{R}}\abs{\mu(t)} \leq \text{diam}(\Omega)\frac{\norm{\hat{v}}_{L^2}^2}{2\pi},
\]
where $\text{diam}(\Omega) = \sup_{x, y \in \Omega} \abs{x - y}$ is the diameter of $\Omega \in \mathbb{R}$. These examples suggest that the class of tempered radon measure is general enough to encompass models for most physically relevant open-quantum systems.

Having identified a class of physically relevant coupling functions or memory kernels, we now turn to the question of the existence of the solution to the Schroedinger equation for the Hamiltonian in Eq.~\ref{eq:hamiltonian_heuristic} i.e.~we are interested in defining a unitary group $U(t, s)$ that satisfies
\[
i \frac{d}{dt} U(t, s) = H(t) U(t, s) \ \text{with} \ U(s, s) = \text{id}.
\]
If the coupling functions $\hat{v}_\alpha$ are square integrable, then the well-definition of the solution to the Schroedinger's equation follows from the standard theory of time-dependent Hamiltonians over infinite-dimensional Hilbert spaces  \cite{kato1953integration, kato1956linear}. Physically, this is an easy case since the annihilation operators $\hat{A}_\alpha$ appearing in the Hamiltonian in Eq.~\ref{eq:hamiltonian_heuristic} correspond to a well defined (normalizable) mode in the bath's Hilbert space. In order to treat the more difficult cases where the coupling functions are not square integrable, but correspond to memory kernels that are tempered distributions, we proceed via a regularization procedure. We first regularize the coupling function $\hat{v}_\alpha$ to a square integrable coupling function --- this accomplished by the transformation
\begin{align}\label{eq:reg_informal}
\hat{v}_\alpha(\omega) \to \hat{\rho}(\omega \varepsilon) \hat{v}_\alpha(\omega),
\end{align}
where $\varepsilon$ is the regularization parameter and $\hat{\rho}$ is the fourier transform of a smooth function $\rho$ that approximates the delta function. Such functions are also called mollifiers --- common examples include a gaussian or a standard mollifier \cite{grigis1994microlocal}. Physically, this regularization damps the high-frequency components of $\hat{v}_\alpha(\omega)$ --- it suppresses amplitudes of coupling functions at frequencies larger than $\sim 1 / \varepsilon$. The existence of the solution to the Schroedinger equation for the regularized coupling function is guaranteed since the coupling function is square integrable --- we then study the limit of this solution as the regularization is removed (i.e.~as $\varepsilon \to 0$). Our main result is to show that if the memory kernels corresponding to the coupling functions are tempered radon measures then this limit exists, and hence defines the solution of the Schroedinger equation. 

\begin{theorem}[Informal]
\label{theorem:non_mkv_exis}
A unitary group corresponding to the Hamiltonian in Eq.~\ref{eq:hamiltonian_heuristic} can be defined for non-Markovian model s for which the coupling functions $\hat{v}_\alpha$ correspond to memory kernels that are tempered radon measures.
\end{theorem}
\noindent We provide a more technically precise statement of this theorem in the next section. \\

\emph{Simulability of non-Markovian dynamics on quantum computers}: Now that we have a well-defined model for described non-Markovian open quantum systems, we can turn to another fundamentally important question --- is this model, in the many-body setting, simulable on a quantum computer? The quantum extended church turing thesis suggests that for physically reasonable models, this should indeed be the case else we would have a system that is potentially more powerful than a quantum computers. Apart from being a fundamental question, it could also be practically interesting to develop quantum algorithms for simulating many-body non-Markovian open quantum systems on quantum computers, since they are a model of a variety of interesting physical systems in solid-state physics \cite{finsterholzl2020nonequilibrium, chin2011generalized, groeblacher2015observation, de2008matter}, quantum optics \cite{calajo2019exciting, andersson2019non, gonzalez2019engineering, aref2016quantum, leonforte2021vacancy} and quantum biology and chemistry \cite{ishizaki2005quantum, chin2012coherence, ivanov2015extension, caycedo2021exact} .

Our next result, under some further assumptions on the growth and smoothness of the memory kernels of the non-Markovian system, provides a quantum algorithm for simulating many-body quantum dynamics whose run-time scales polynomially with the system-size i.e.~it is provably efficient. Our key contribution that enables this result is to show that the regularization of the non-Markovian environment, even when the memory kernel is a tempered distribution, results in an error that increases only polynomially (as opposed to exponentially) with system size and evolution time. While we do not focus on finding an optimal quantum algorithm for non-Markovian dynamics, the basic steps and analysis procedure that we develop in this paper could lay the foundations of future works improving the algorithm's run-time.

The general strategy behind the quantum algorithm is schematically depicted in Fig.~\ref{fig:reg_method}(a). First the bath Hilbert space, which is a continuum of modes, is discretized into a finite set of bosonic modes. Second, the infinite-dimensional Hilbert space of the resulting bosonic modes are truncated to obtain a finite-dimensional many-body problem. Then, any standard quantum algorithm for many-body Hamiltonian simulation \cite{lloyd1996universal, berry2014exponential, low2019hamiltonian, berry2015hamiltonian} to simulate the non-Markovian dynamics. To rigorously analyze the run-time of the quantum algorithm, errors in both the approximation steps need to be analyzed --- the discretization step is the more challenging one to analyze and we focus most of this paper on that step. The truncation of the Hilbert space is more straightforward and follows from a simple bound on the environment's particle number moments.
\begin{figure}
\centering
\includegraphics[scale=0.475]{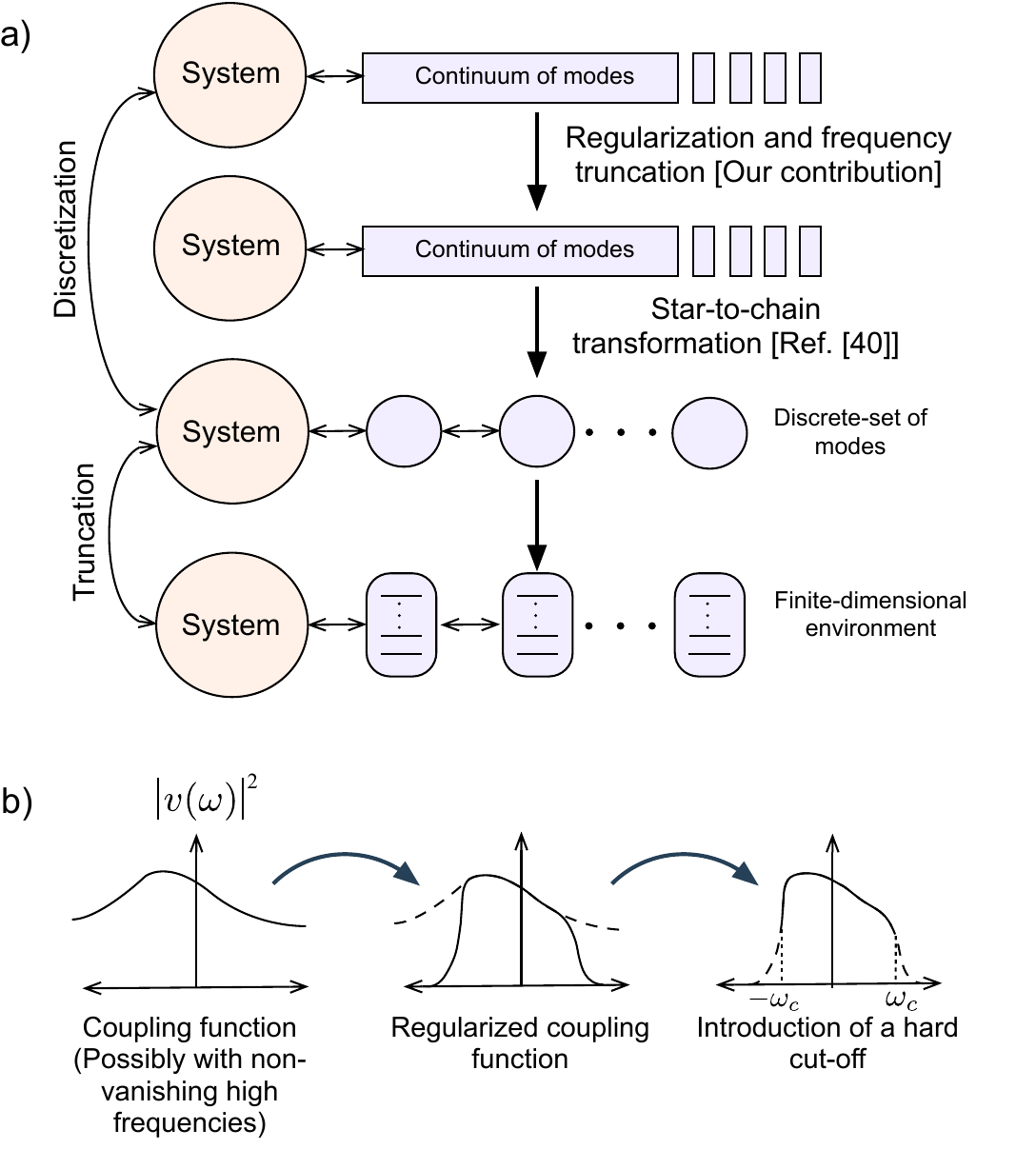}
\caption{(a) Schematic depiction of the steps involved in approximating a non-Markovian environment. First, the environment is discretized, which is done in two steps --- regularization with introduction of frequency cutoff followed an application of the star-to-chain transformation. Then, the Hilbert space of the environment, which is still infinite dimensional, is truncated to finite dimensions. (b) Transformation of a system-environment coupling function before applying the star-to-chain transformation --- the coupling function is first regularized to dampen the high frequency contribution and then a frequency cut-off is introduced.}
\label{fig:reg_method}
\end{figure}

Consider the problem of discretizing the continuum of modes --- this problem has been previously studied for square-integrable coupling functions that additionally have a hard high frequency cutoff (i.e.~$\hat{v}(\omega) = 0$ for $\abs{\omega} \geq \omega_c$ for some cut-off frequency $\omega_c$). For this case, a discrete set of modes equivalent to the continuum can be computed using a Lanczos iteration, also known as the star-to-chain transformation --- this transformation, which is described in detail in Ref.~\cite{woods2016dynamical}, approximates the bath by a semi-infinite nearest-neighbour 1D chain of bosonic modes where the first mode couples to the local system through the operator $L$. If all the infinite number of modes are retained, then this mapping is exact --- for practical simulations, only a finite number of modes is retained. An application of the Lieb-Robinson bound for 1D bosonic lattices \cite{woods2015simulating} can then be used to show that with $N_b$ modes, local system dynamics at time $t$ can be approximated within a $\text{poly}(M, g, t, N_b^{-1}, \omega_c)$ total variation error. Here $M$ is the number of baths in the environment (Eq.~\ref{eq:hamiltonian_heuristic}), and $g$ is such that $\norm{H_S(t)}, \norm{L_\alpha} \leq g$.

The more general class of tempered radon measures contains coupling functions that do not have a frequency cutoff. Furthermore, several coupling functions [e.g. memory kernels that are expressible as a sum of delta functions ($\mu(t) = \sum_{j = 1}^P a_j \delta(t - \tau_j)$)], have appreciable high frequency amplitudes. A frequency cutoff must be introduced to apply the star-to-chain transformation to these cases --- this again is done in two steps depicted schematically in Fig.~\ref{fig:reg_method}(b). First, we use the regularization defined in Eq.~\ref{eq:reg_informal} --- this leads to a coupling function that rapidly goes to 0 at frequency scales $\sim 1 / \varepsilon$, where $\varepsilon$ is the regularization parameter. We rigorously show that the error incurred in this regularization scales as $\text{poly}(M, g, t, \varepsilon)$ --- a notable contribution in our work is to prove an error bound that grows as $\text{poly}(t)$ for a very large class of memory kernels (tempered radon measures), as opposed to exponentially with $t$ as had been shown previously \cite{trivedi2021convergence, mascherpa2017open}. We numerically verify this prediction for a single-particle problem in Fig.~\ref{fig:benchmark_fig}. on a 1D XY spin chain with periodic boundary conditions. Here, the system has $n$ spins, and the system Hamiltonian is 
\[
H_S = \sum_{i = 1}^n J \big(\sigma_i^\dagger \sigma_{i + 1} + \sigma_{i + 1}^\dagger \sigma_i \big),
\]
where $\sigma_{n + 1} \cong \sigma_1$. Furthermore, the spins at sites $k = 1, \sqrt{n} + 1, 2\sqrt{n} + 1 \dots$ couple to a bath with jump operator $\sqrt{\gamma}\sigma_k$ and coupling function $v_k(\omega) = 1 + e^{i\varphi}e^{i\omega t_d}$ [which corresponds to $\mu(t) = 2\delta (t) + e^{i\varphi} \delta(t + t_d) + e^{-i\varphi} \delta(t - t_d)$]. We consider exciting the first spin, and compare the error, in the reduced density matrix of the spins, for the true and regularized models after $t = n / \gamma$. Fig.~\ref{fig:reg_method}(b) shows the dynamics of the true and regularized models and we see that the regularized model reproduces the true dynamics as $\varepsilon \to 0$ --- in Fig.~\ref{fig:reg_method}(c), we study the scalings of the regularization error with both $\varepsilon$ and system size $n$ --- as theoretically predicted, this error scales polynomially with $\varepsilon$ and $n$.

Having regularized the coupling function, the high frequency tails of such a decaying coupling function can then be ignored to introduce a hard frequency cutoff, enabling an application of the star-to-chain transformation. The analysis of the frequency truncation of the regularized model used a bound on the moments of high-frequency particle number in the environment, and can be shown to scale as $\text{poly}(t, \varepsilon^{-1}, \omega_c^{-1})$. Combining this with the estimate of the regularization step and the truncated star-to-chain transformation, we obtain that the total error in discretizing the continuum of modes in the bath to a set of $N_b$ discrete modes is given by
\[
\text{poly}\bigg(\omega_c, \frac{1}{N_b}\bigg) + \text{poly}\bigg(\frac{1}{\varepsilon}, \frac{1}{\omega_c}\bigg) + \text{poly}\big(\varepsilon\big),
\]
where we have hidden the polynomial dependence on $M, g$ and $t$ for brevity. From this expression, we note that a choice of $\varepsilon^{-1}, \omega_c$ and $N_b$ as $\text{poly}(M, g, t, 1/\epsilon)$ ensures that this error is $O(\epsilon)$. This result provides us with a bound on the number of discrete modes of the bath needed to accurately simulate the non-Markovian model, and this bound scales at-most polynomially with $t$ and with the parameters $M, g$.

\begin{figure}[t]
\centering
\includegraphics[scale=0.55]{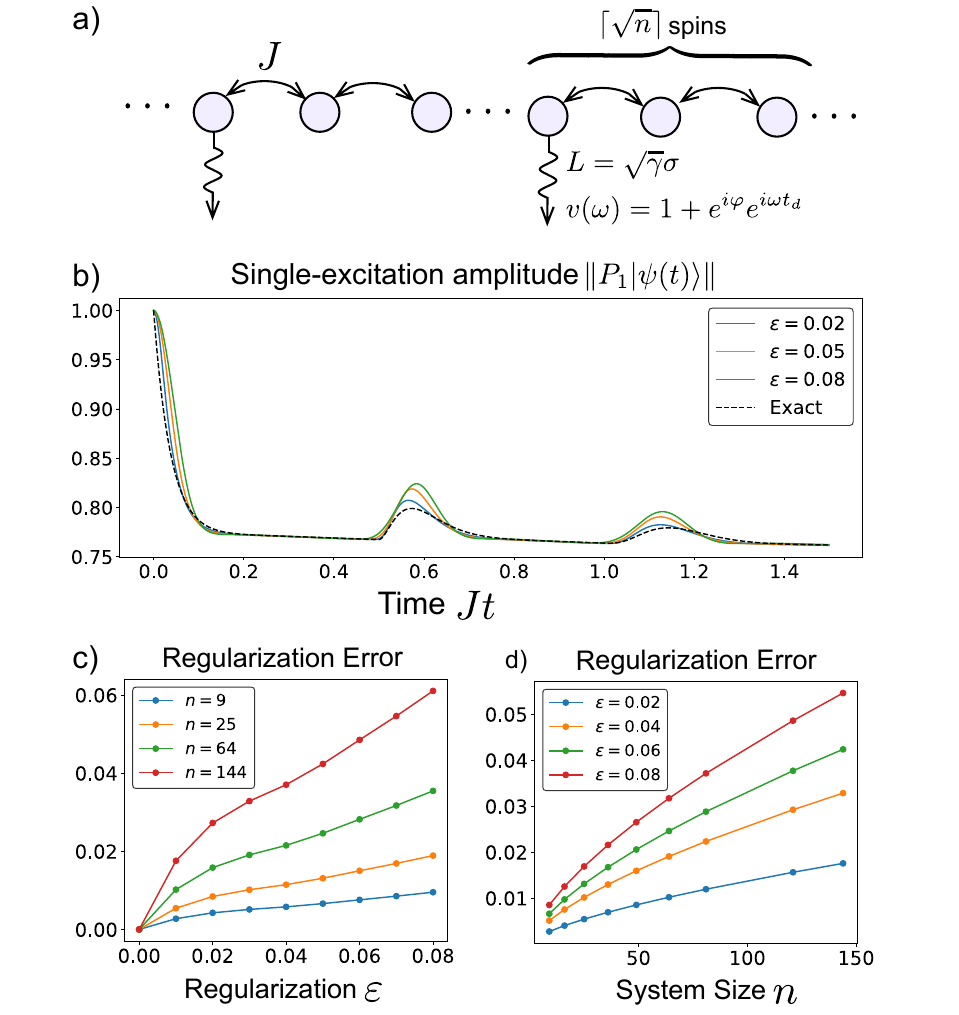}
\caption{Numerical validation study of the regularization step for a single-particle problem. (a) We consider a system which is a 1D XY model with periodic boundary conditions , $n$ spins, hopping strength $J$, and every $\sqrt{n}^\text{th}$ spin couples to a bath with time-delayed feedback. This is modelled by a jump operator $\sqrt{\gamma}\sigma$ and coupling function $1 + e^{i\varphi}e^{i\omega t_d}$. In all simulations, $\varphi = \pi$, $J\tau_d = 0.5$, $\gamma / J = 1.0$. (b) Dynamics of the exact and regularized model when starting with only the first spin in its excited state. The time-delayed feedback bath forms a bound state with the system, and thus the spins do not completely decay to 0. (c) Regularization error evaluated at time $t = n / \gamma$ as a function of the number of spins $n$ and the regularization parameter $\varepsilon$. As predicted by our analysis, the error bounds grow at-most polynomially with $n$ and decrease polynomially with $\varepsilon$.}
\label{fig:benchmark_fig}
\end{figure}

This result has an immediate consequence on the complexity classification of many-body non-Markovian quantum dynamics. For a more precise statement of this result, we specifically analyze a physically motivated many-body setting --- the problem of $k-$local non-Markovian many-body dynamics. We consider a system of $n$ spins with the local-system Hamiltonian $H_S(t)$ which is $k-$local i.e. $H_S(t) = \sum_{i} H_i(t)$, where $H_i(t)$ act on at-most $k-$spins and has a normalization $\norm{H_i(t)} \leq O(1)$. Furthermore, we assume that the jump operators individually also act on at-most $k$ spins, and that the memory kernel corresponding to each jump operator is a tempered radon measure (with some additional reasonable conditions on its growth and smoothness, that were also required earlier). This problem can be considered to be a generalization of the $k-$local Hamiltonian dynamics \cite{lloyd1996universal, berry2014exponential, low2019hamiltonian, berry2015hamiltonian} and $k-$local Markovian dynamics \cite{zanardi2016dissipative,chenu2017quantum,cleve2016efficient, kliesch2011dissipative}, simulation problems. Applying the discretization procedure described above, followed by a truncation of the Hilbert space of the resulting bosonic modes, we show that
\begin{theorem}[Informal]
\label{theorem:bqp_non_mkv} The $k-$local non-Markovian many body dynamics problem can be efficiently simulated on a quantum computer.
\end{theorem}
We note that that establishing error bounds on the discretization procedure that grow as $\text{poly}(t)$, as opposed to the previously proven bounds which were exponential in $t$, is central to establishing this result. This is so because, if the exponential in $t$ growth of errors was indeed tight, then it would indicate that using this discretization procedure would not yield an efficient algorithm for simulating the non-Markovian many-body system after $t = \text{poly}(n)$. To the best of our knowledge, our work is the first that establishes error bounds on this discretization that grows as $\text{poly}(t)$ for memory kernels that are tempered radon measures, and thus enables a proof of the theorem above.
}

%Furthermore, kernels expressed as sum of multiple delta functions ($\mu \cong \sum_{i = 1}^P \delta(t - t' - \tau_i)$ for some $\{\tau_1, \tau_2 \dots \tau_P\}$), which arise in the study of time-delay feedback systems  also fall into this class. As a final example, consider the kernels corresponding to square integrable coupling functions ($\hat{v} \in L^2(\mathbb{R})$) --- these models arise in the study of cavity QED systems in quantum optics  , or the non-Markovian spin-boson model. Such kernels are also tempered radon measures with $\textnormal{TV}_\Omega(\mu) \leq \norm{\hat{v}}_{L^2}^2 \textnormal{diam}(\Omega)$\footnote{For $\Omega \subseteq \mathbb{R}$, $\text{diam}(\Omega) = \sup_{x, y \in \Omega}\abs{x - y}$}. The supplement contains details of these specific radon measures.

\subsection{Formal Statements and proof ideas}\label{subsec:proof_ideas}
\emph{Well definition of non-Markovian dynamics}: As discussed in the previous section, a {radon measure} $\mu$ is a map from the space of continuous and compactly supported functions ($\textnormal{C}_c^0(\mathbb{R})$) to complex numbers which is bounded in the sense that $\forall f\in \textnormal{C}_c^0(\mathbb{R})$ with support $\text{supp}(f)\subseteq \Omega \subseteq \mathbb{R}$
\[
\abs{\langle \mu, f\rangle} \leq \textnormal{TV}_{\Omega}(\mu) \sup_{x \in \Omega}\abs{f(x)},
\]
with $\textnormal{TV}_\Omega(\mu)$ is defined to be the total variation of $\mu$ within the compact set $\Omega$. Furthermore, we will call a radon measures that has a fourier transform which is of at-most polynomial growth in frequency a tempered radon measure. Since most physical systems have a spectral density function that do not grow very rapidly with frequency, and since the fourier transform of $\mu_\alpha$ describes the environment's spectral density function, it is reasonable to assume it to be a tempered radon measure.

A non-Markovian model can thus be specified by the system Hamiltonian, jump operators and the coupling functions in between system and the baths. These coupling functions are provided as a tempered radon transform, which specifies the magnitude of the coupling function, and the phase of the coupling function.
\begin{definition}[Non-Markovian model]\label{def:model}
A non-Markovian open system model for a quantum system with finite-dimensional Hilbert space $\mathcal{H}_S$ is specified by
\begin{enumerate}
\item[(a)] A time-dependent system Hamiltonian $H_S(t) $ which is Hermitian, norm continuous and differentiable in $t$,
\item[(b)] A set of coupling functions $\{(\mu_\alpha, \varphi_\alpha) \}_{\alpha \in \{1, 2 \dots M\}}$, where $\mu_\alpha$ are tempered radon measures and $\varphi_\alpha : \mathbb{R}\to \mathbb{C}$ specify the phase of the coupling functions,
\item[(c)] A set of bounded jump operators \\ $\{L_\alpha\}_{\alpha \in \{1, 2 \dots M\}}$.
\end{enumerate}
\end{definition}

We now turn to the question of defining the quantum dynamics corresponding to a non-Markovian model. In general, this cannot be done simply through the Schroedinger's equation, since for general radon measure kernel it isnt clear if a meaningful solution with the Hamiltonian in Eq.~\ref{eq:hamiltonian_heuristic} exists. We approach the problem of defining the associated quantum dynamics through a regularization procedure. An elementary but important observation that enables this regularization is that if the coupling functions are square integrable ($v_\alpha \in L^2(\mathbb{R})$), then the solution to Schroedinger's equation with Hamiltonian in Eq.~\ref{eq:hamiltonian_heuristic} can be shown to exist using standard tools from the theory of non-autonomous differential equations on Banach spaces \cite{kato1953integration, kato1956linear}. Now, a coupling function $(\mu, \varphi)$ can be approximated by a square integrable function by using a mollifier\footnote{A mollifier $\rho$ is a smooth compact function ($\rho \in \textnormal{C}_c^\infty(\mathbb{R})$) which is positive and with support $\textnormal{supp}(\rho) \subseteq [-1, 1]$ with 
\[
\int_{[-1, 1]} \rho(x) dx = 1.
\]
Unless otherwise mentioned, we will assume $\rho$ to be a symmetric (even) function.
}(smoothing function) $\rho$.
\begin{definition}[Regularization]\label{def:regularization}
For $\varepsilon > 0$ and given a symmetric mollifier $\rho \in \textnormal{C}_c^\infty(\mathbb{R})$, an $\varepsilon, \rho-$regularization of a distributional coupling function $(\mu, \varphi)$ is a square integrable function $v_\varepsilon \in L^2(\mathbb{R})$ whose fourier transform\footnote{We will assume the following convention for the fourier transform $\hat{v}$ of $v \in L^2(\mathbb{R})$: \[
\hat{v}(t) = \int_{\mathbb{R}} {v(\omega)}e^{-i\omega t} \frac{d\omega}{\sqrt{2\pi}}.
\]} is given by  $\hat{v}_\varepsilon \in L^2(\mathbb{R})$ is given by
\[
\hat{v}_\varepsilon(\omega) = \sqrt{ \hat{\mu}(\omega)} \hat{\rho}({\omega}{\varepsilon}) e^{i\varphi(\omega)} \ \forall \omega \in \mathbb{R},
\]
where $\hat{\rho}$ is the fourier transform of $\rho$ and $\hat{\mu}$ is the fourier transform of $\mu$ \footnote{For a tempered distribution $\mu \in \mathcal{S}'(\mathbb{R})$, the fourier transform $\hat{\mu}$ (if it exists as a function from $\mathbb{R}$ to $\mathbb{C}$) is defined by demanding that
\[
\langle \mu, f\rangle = \int_{\mathbb{R}} \hat{\mu}(\omega) \hat{f}(\omega) \frac{d\omega}{\sqrt{2\pi}}
\]
for all smooth compact functions $f \in \textnormal{C}_c^\infty(\mathbb{R})$.
}.
\end{definition}
We note that since $\rho$ is smooth and compact, its fourier transform $\hat{\rho}$ falls off faster than any polynomial in $\omega$ as $\omega \to \infty$ and thus $v_\varepsilon$ is indeed a square integrable function. Equivalently, this regularization step can be considered as approximating the kernel $\mu$ with another radon measure $\mu_\varepsilon$ whose action on a continuous compact function $f$ is given by
\[
\langle \mu_\varepsilon, f\rangle = \langle \mu, f\star \rho_\varepsilon \star \rho_\varepsilon \rangle,
\]
where $\rho_\varepsilon(x) = \varepsilon^{-1}\rho(\varepsilon^{-1}x)$ and $\star$ denotes a convolution operation. It can be seen that as $\varepsilon \to 0$, $f \star \rho_\varepsilon \star \rho_\varepsilon $ becomes an increasingly better approximation to $f$, and thus $\mu_\varepsilon$ becomes an increasingly better approximation of $\mu$. 

Since the regularized coupling functions are square integrable, their associated dynamics can be computed by solving the Schroedinger's equation. We can now study the limit of this dynamics on removing the regularization ($\varepsilon \to 0$) --- our first result shows that this limit exists, and is independent of the choice of the mollifier. The proof of this result, provided in section \ref{sec:well_def}, relies on an upper bound bound on the rate of change of two-point correlation functions of system observables that is uniform in the regularization parameter $\varepsilon$.\\

\begin{reptheorem} {theorem:non_mkv_exis}[Non-markovian dynamics]

Given a non-Markovian open system model (definition \ref{def:model}) with $U_{\varepsilon, \rho}(t, s)$ for $t, s \in \mathbb{R}$ being the propagator corresponding to an $\varepsilon, \rho-$regularization of its coupling functions, $\lim_{\varepsilon \to 0} U_{\varepsilon, \rho}(t, s)$ exists weakly\footnote{A single parameter family of operators $\{O_x:\mathcal{D} \to \mathcal{H}\}_{x\in [0, \infty)}$ is said to converge weakly (or converge in the weak topology) to $O:\mathcal{D} \to \mathcal{H}$ as $x\to 0$ if $\forall \ket{\psi} \in \mathcal{D}$, $\lim_{x\to 0} O_x\ket{\psi} = O\ket{\psi}$. } as an isometry from a dense subspace $\mathcal{D}$ of the system-environment Hilbert space and is independent of the choice of the mollifier $\rho$. 
\end{reptheorem}
\emph{Simulability of non-Markovian dynamics on quantum computers}: Our next result considers the simulability of non-Markovian dynamics of many-body systems on quantum computers. To make further progress, we need two additional assumptions --- one on the radon measures describing the coupling functions between the many-body system and the bath, and the second on the initial state in the environment.
\begin{assumption}
\label{assump:radon_measure}
The radon measure $\mu$ corresponding to the coupling function should satisfy:
\begin{enumerate}
\item[(a)] For any compact interval $[a, b] \subseteq \mathbb{R}$, $\textnormal{TV}_{[a, b]}(\mu) \leq \textnormal{poly}(\abs{a}, \abs{b})$ and
\item[(b)] Given a compact interval $[a, b] \subseteq \mathbb{R}$, $\exists\Delta^0_{\mu; [a, b]}(\varepsilon)$, $\Delta^1_{\mu; [a, b]}(\varepsilon) = \textnormal{poly}(a, b, \varepsilon)$ which are locally integrable with respect to $a, b$, tend to $0$ as $\varepsilon \to 0$ and for any differentiable function $f \in \textnormal{C}^1(\mathbb{R})$ 
\begin{align*}
&\abs{\langle \mu, f_{[a, b]} \star \rho_\varepsilon \rangle - \lim_{\varepsilon \to 0} \langle \mu, f_{[a, b]} \star \rho_\varepsilon \rangle  } \leq \nonumber \\
&\ \Delta_{\mu; [a, b]}^0(\varepsilon) \sup_{t \in [a, b]}\abs{f(t)} + \Delta_{\mu; [a, b]}^1(\varepsilon)\sup_{t \in [a, b]}\abs{f'(t)}
\end{align*}
where $f_{[a, b]}(t) = f(t)$ if $t \in [a, b]$ and $0$ otherwise.
\end{enumerate}
\end{assumption}
Assumption \ref{assump:radon_measure}(a), which constrains the growth of the total variation of the memory kernel, can be physically interpreted as limiting the amount of ``memory" that the non-Markovian system can accumulate. Assumption \ref{assump:radon_measure}(b) is a constraint on the smoothness of the memory kernel (i.e.~the error incurred on smoothing the memory kernel with a mollifier), and it limits how rapidly the memory kernel diverges from its smooth approximations. Both of these assumptions are satisfied coupling functions encountered in most experimentally relevant physical systems.

\begin{assumption}
\label{assump:initial_state}
The initial environment state $\ket{\phi_1}\otimes \ket{\phi_2} \dots \ket{\phi_M}$ where for $\alpha \in \{1,2 \dots M\}$, $\ket{\phi_\alpha} \in \textnormal{Fock}[L^2(\mathbb{R})]$ and for its $n-$particle wavefunctions $\phi_{\alpha, n} \in L^2(\mathbb{R}^n)$, and any $j, k \geq 0$, $\exists \mathcal{N}_{j, k} > 0$ such that
\[
\sum_{n = 0}^\infty n^j \int_{\mathbb{R}^n} (1 + \omega_1^2)^k \abs{\phi_{\alpha, n}(\omega)}^2 d\omega < \mathcal{N}_{j, k}.
\]
\end{assumption}
Assumption \ref{assump:initial_state} demands that both high particle number or high frequency amplitude of the initial wavefunction vanishes superpolynomially. This assumption is reminiscent of assumption on particle number and energies of initial states made in studying the simulatability of quantum field theories \cite{jordan2012quantum, jordan2018bqp}. A number of commonly used initial environment states (such as thermal states) in physically relevant open systems have exponentially vanishing high energy amplitudes, and satisfy this assumption.

The computational problem of simulating $k-$local many-body non-Markovian dynamics can now be stated as
\begin{problem}[$k-$local non-Markovian dynamics]
\label{prob:k_local_non_mkv}
Consider a system of $n$ qudits $(\mathcal{H}_S = \big(\mathbb{C}^d\big)^{\otimes n})$ interacting with $M = \textnormal{poly}(n)$ baths with
\begin{enumerate}
\item[(a)] System Hamiltonian $H_S(t)$ is $k-$local i.e. $H_S(t) = \sum_{i = 1}^{N} H_i(t)$, where $N = \textnormal{poly}(n)$, and $H_i(t)$ is an operator acting on atmost $k$ qudits and $\norm{H_i(t)} \leq 1$.
\item[(b)] Jump operators $\{L_\alpha \}_{\alpha \in \{1, 2 \dots M\}}$ such that $L_\alpha$ acts on at-most $k$ qudits and $\norm{L_\alpha} \leq 1$.
\item[(c)] Coupling functions $\{(\mu_\alpha, \varphi_\alpha) \}_{\alpha \in \{1, 2 \dots M\}}$ such that $\mu_\alpha$ satisfies assumption \ref{assump:radon_measure}.
\item[(d)] An initial state $\ket{\Psi} = \ket{0}^{\otimes n} \otimes \ket{\Phi}$, where $\ket{\Phi} = \ket{\phi_1} \otimes \ket{\phi_2}\otimes \dots \ket{\phi_M}$ is an initial state which satisfies assumption \ref{assump:initial_state}. Furthermore, the initial state is computable in the sense that for $v_1, v_2 \dots v_m \in L^2(\mathbb{R})$ and $P \in\mathbb{Z}_{>0}$, all the amplitudes
\begin{align*}
 \bra{\textnormal{vac}}\prod_{i = 1}^m \bigg(\int_{\mathbb{R}}v_i(\omega)a_\omega d\omega\bigg)^{n_i} \ket{\phi_\alpha} 
\end{align*}
with $n_1 + n_2 \dots n_m \leq P$ can be computed in $\textnormal{poly}(m, P)$ time on a classical or quantum computer.
.
\end{enumerate}
Denoting by $\rho_S(t)$ the reduced state of the system at time $t$ for this non-Markovian model, then for $t = \textnormal{poly}(n)$, prepare a quantum state $\hat{\rho}$ such that $\norm{\hat{\rho} - \rho_S(t)}_\textnormal{tr} \leq 1 / \textnormal{poly}(n)$.
\end{problem}
The key ingredient to analyzing this problem is a Markovian dilation of the non-Markovian model, which identifies a finite number of modes in the environment's Hilbert space and then approximates the non-Markovian model by a Hamiltonian simulation of the system only interacting with these modes. We make this precise in the definition below.
\begin{definition}[Chain Dilation] Consider a non-Markovian model specified by a system Hamiltonian $H_S(t)$, coupling functions $\{(\mu_\alpha, \varphi_\alpha)\}_{\alpha \in \{1, 2 \dots M\}}$ and jump operators $\{L_\alpha\}_{\alpha \in \{1, 2 \dots M\}}$. A chain dilation, with $N_m$ modes and bandwidth $B$, of this model is described by the following Hamiltonian over the system-environment Hilbert space $(\mathcal{H}_S \otimes \textnormal{Fock}[L^2(\mathbb{R})]):$
\begin{align*}
&H(t) = \nonumber\\
& H_S(t) + \sum_{\alpha = 1}^M\bigg(g_\alpha a_{\alpha, 1} L_\alpha^\dagger + \sum_{i = 1}^{N_m - 1} t_{\alpha, i} a_{\alpha, i}^\dagger a_{\alpha, i + 1} + \textnormal{h.c.}\bigg),
\end{align*}
where for $\alpha \in \{1, 2 \dots M\}, i \in \{1, 2 \dots N_m\}$,
\begin{enumerate}
\item[(a)] The operator $a_{\alpha, i} = \int_{\mathbb{R}} \varphi_{\alpha, i}(\omega) a_{\alpha, \omega} d\omega$ is the annihilation operator corresponding to the $i^\textnormal{th}$ mode of the $\alpha^\textnormal{th}$ bath described by the orthonormal mode functions $\varphi_{\alpha, i} \in \textnormal{L}^2(\mathbb{R})$ $($i.e. $\langle \varphi_{\alpha, i}, \varphi_{\alpha, i'}\rangle = \delta_{i, i'})$.
\item[(b)] The coupling constants $g_\alpha, t_{\alpha, i}$ are upper bounded by the bandwidth $B$ i.e.~$\abs{g_\alpha}, \abs{t_{\alpha, i}} \leq B$.
\end{enumerate}
\end{definition}
Our next lemma, which is used in the analysis of problem \ref{prob:k_local_non_mkv}, uses the well-known star-to-chain transformation \cite{chin2010exact, woods2014mappings} to systematically construct a Markovian dilation to the non-Markovian system. We analyze the error between the dynamics of the non-Markovian system and its Markovian dilation and estimate the number of modes and bandwidth of a Markovian dilation needed to approximate the non-Markovian model.
\begin{proposition}[Chain dilation]
\label{theorem:final_dilation}
Consider a non-Markovian model specified by a system Hamiltonian $H_S(t)$, jump operators $\{L_\alpha\}_{\alpha \in \{1, 2 \dots M\}}$ and coupling functions $\{(\mu_\alpha, \varphi_\alpha)\}_{\alpha \in \{1, 2 \dots M\}}$ where $\mu_\alpha$ satisfy assumption \ref{assump:radon_measure} and $\hat{\mu}_\alpha(\omega) < O(\omega^{2k}$) for some $k > 0$. For $\ket{\Psi_0} := \ket{\sigma} \otimes \ket{\Phi_0} \in \mathcal{H}_S\otimes \textnormal{Fock}[L^2(\mathbb{R})]^{\otimes M}$, where $\ket{\sigma} \in \mathcal{H}_S$ and $\ket{\Phi_0}$ is an initial environment state that satisfies assumption \ref{assump:initial_state}, then $\exists$ a Markovian dilation of the  non-Markovian model with 
\begin{align*}
&N_m, B \leq O\bigg(\textnormal{poly}\bigg(\frac{1}{\epsilon}, t, M, \sup_{\alpha}\norm{L_\alpha},\nonumber \\
&\qquad \qquad \qquad \qquad   \sup_{\alpha, s \in [0,t]}\norm{[H_S(s), L_\alpha]},\nonumber\\
 &\qquad \qquad \qquad \qquad \mathcal{N}_{1, k + 1}, \mathcal{N}_{1, k + 2}, \mathcal{N}_{1, 0}\bigg)\bigg)
\end{align*}
whose system-environment state at time $t$ is within $\epsilon$ norm distance of the exact state.
\end{proposition}
Our analysis of the Markovian dilation, detailed in the supplement, is performed in three steps. First, we analyze the error incurred in regularizing the non-Markovian model with a mollifier $\rho$, then we introduce a sharp frequency cutoff on the resulting regularized square-integrable functions. A key technical contribution in our analysis of the regularization and frequency cutoff is to prove bounds on error that grow only polynomially with time, which improves previous bounds that, in the worst case, grow exponentially with time \cite{trivedi2021convergence, mascherpa2017open}. After introduction of this cut-off, we perform a star-to-chain transformation --- the analysis of this step closely follows previous works that have studied the convergence of the star-to-chain transformation for the spin-boson models with a hard frequency cutoff \cite{woods2015simulating, woods2016dynamical, gualdi2013renormalization}. 

Using this lemma, we can map problem \ref{prob:k_local_non_mkv} into a Hamiltonian simulation problem with a finite number of modes. This problem is still infinite-dimensional --- however, we can easily show that the moments of the particle number operator for the environment can grow at-most polynomially with the problem size $n$. Therefore, we can truncate the Hilbert space of this model and obtain a finite-dimensional Hilbert space --- an application of the sparse Hamiltonian lemma \cite{aharonov2003adiabatic} then yields the the second main result of our paper. A detailed proof is provided in section \ref{sec:complexity}.

\begin{reptheorem}{theorem:bqp_non_mkv}[$k-$local Non-Markovian dynamics $\in$ BQP] 
Problem \ref{prob:k_local_non_mkv} can be solved in $\textnormal{poly}(n)$ time on a quantum computer.
\end{reptheorem}

\section{Notation and preliminaries}\label{sec:notation_prelim}
This section describes the notation used throughout this paper. For the convenience of the reader, and for the sake of completeness, we also collect some basic definitions and facts from the theory of function spaces and analysis that we use in this paper --- the interested reader can refer to Refs.~\cite{reed1972methods, reed1975ii} for more detailed discussion. The reader can choose to skip this section, and refer back to it as and when it is referenced in the following sections.\\

\noindent \emph{General}: For $x := (x_1, x_2 \dots x_n) \in \mathbb{R}^n, y:= (y_1, y_2 \dots y_m) \in \mathbb{R}^m$ we will denote by $(x, y) \in \mathbb{R}^{n + m}$ defined by $(x, y) = (x_1, x_2 \dots x_n, y_1, y_2 \dots y_m)$. For an ordered subset $B$ of $\{1, 2 \dots n\}$ and $x \in \mathbb{R}^n$, $Bx = (x_{B(1)}, x_{B(2)} \dots )$. We will denote by $\alpha^n$ the $n-$element constant vector $(\alpha, \alpha \dots \alpha)$, and by $\alpha^\infty$ the constant sequence $(\alpha, \alpha, \alpha \dots)$. \\

\noindent\emph{Function spaces and analysis}: Throughout this paper, all integrals over $\mathbb{R}^n$ will be Lesbesgue integrals with respect to the Lesbesgue measure over $\mathbb{R}^n$. Two measurable functions $f, g:\mathbb{R}^n \to \mathbb{C}$ are equal almost everywhere, denoted by $f =_{a.e.} g$, if the set $\{x | f(x) \neq g(x)\}$ is a zero measure set. For a measurable function $f:\mathbb{R}^n \to \mathbb{R}$ and a measurable set $\Omega \subseteq \mathbb{R}^n$, $\text{ess sup}_{x \in \Omega} f(x) = c$ if the set $\{x \in \Omega | f(x) > c\}$ is a zero measure set. We will use the standard $L^p$ function spaces throughout this paper --- for $p \geq 1$, we will denote by
\[
L^p(\mathbb{R}^n) =\bigg\{ f:\mathbb{R} \to \mathbb{C} \bigg | \norm{f}_{L^p} < \infty\bigg\} \bigg/ =_{a.e.}
\]
{where }
\[
 \norm{f}_{L^p} := \bigg(\int_{\mathbb{R}^n}|f(x)|^p dx\bigg)^{1/p}.
\]
and 
\[
L^\infty(\mathbb{R}^n) = \bigg\{ f:\mathbb{R} \to \mathbb{C} \bigg | \norm{f}_{L^\infty} < \infty\bigg\}\bigg/ =_{a.e.},
\]
\text{where }
\[
 \norm{f}_{L^\infty} = \textnormal{ess\ sup}_{x \in \mathbb{R}^n} |f(x)|.
\]
Recall that for $g:\mathbb{R}^n \to \mathbb{R}$, $\textnormal{ess\ sup}_{x\in \mathbb{R}^n} g(x)$ is the smallest $M \in \mathbb{R}$ such that the set $\{x \in \mathbb{R}^n | g(x) \geq M\}$ has measure 0.
A map $f:\mathbb{R} \to \mathbb{C}$ is said to be a compactly supported function with support $\textnormal{supp}(f) \subseteq \mathbb{R}$ if $\overline{\textnormal{supp}(f)}$ is compact, and the set $\{ x \in \mathbb{R}\setminus \textnormal{supp}(f) | f(x) \neq 0\}$ is a zero measure set. For $k \in \mathbb{Z}_{\geq 0}$, we will denote by $\textnormal{C}^k(\mathbb{R})$ the set of $k-$differentiable functions (with $k = 0$ being continuous functions, and $k = \infty$ being smooth) from $\mathbb{R}$ to $\mathbb{C}$, and by $\textnormal{C}^k_c(\mathbb{R})$ the set of such functions with compact support. 

A function $\rho \in \textnormal{C}_c^\infty(\mathbb{R})$ is said to be a mollifier if $\rho(x)\geq 0 \ \forall x \in \mathbb{R}$ and $\int_{\mathbb{R}} \rho(x) dx = 1$ --- unless otherwise mentioned, we will assume that $\textnormal{supp}(\rho) = [-1, 1]$. The mollifier is symmetric if $\rho(x) = \rho(-x) \ \forall x \in \mathbb{R}$. Given a mollifier $\rho$ and $\varepsilon > 0$, we will denote by $\rho_\varepsilon \in \textnormal{C}_c^\infty(\mathbb{R})$ the map
\[
\rho_\varepsilon(x) = \frac{1}{\varepsilon} \rho\bigg(\frac{x}{\varepsilon}\bigg) \ \forall \ x \in \mathbb{R}.
\]
Note that $\rho_\varepsilon$ is also a mollifier with $\textnormal{supp}(\rho_\varepsilon) \subseteq [-\varepsilon, \varepsilon]$. Given a subset $\Omega \subseteq \mathbb{R}$, its indicator function $\mathcal{I}_\Omega : \mathbb{R} \to \mathbb{C}$ is defined by
\[
\mathcal{I}_\Omega(x) = \begin{cases}
1 & \text{ if } x \in \Omega, \\
0 & \text{ otherwise}.
\end{cases}
\]
A linear map $\mu: \textnormal{C}^0_c(\mathbb{R}) \to \mathbb{C}$ is a radon measure if $\forall \Omega \subset \mathbb{R}$ which are compact, $\exists u_\Omega > 0$ such that
\[
\langle \mu,  f\rangle \leq u_\Omega \sup_{x \in \mathbb{R}} |f(x)| \ \forall f \in \textnormal{C}^0_c(\mathbb{R}) \textnormal{ with }\textnormal{supp}(f) \subseteq \Omega.
\]
The smallest such $u_\Omega$ is defined to be the total variation of $\mu$ in $\Omega$ and will be denoted by $\textnormal{TV}_\Omega(\mu)$. The set of all radon measures will be denoted by $\mathcal{M}(\mathbb{R})$. By the Lesbesgue decomposition theorem \cite{ambrosio2000functions}, any $\mu \in \mathcal{M}(\mathbb{R})$ can be uniquely expressed as
\[
\mu = \mu_c + \mu_d,
\]
where $\mu_c \in \mathcal{M}(\mathbb{R})$ is called the continuous part of $\mu$ and $\mu_a$ is called its atomic part. The continuous part can be characterized by a function $\phi_c \in \textnormal{C}^0(\mathbb{R})$ where $\forall f \in \textnormal{C}^1_c(\mathbb{R})$,
\[
\langle \mu_c , f\rangle = -\int_\mathbb{R} f'(x) \phi_c(x) dx.
\]
The function $\phi_c$ is also often denoted by $\phi_c(x) = \mu_c((-\infty, x])$ to be consistent with the `cumulative function' of $\mu_c$ in the measure-theoretic definition of the radon measure. $\mu_c$ can further be decomposed into an absolutely continuous part, which can be described by a density function, and a Cantor part --- we will not require this decomposition in this paper. The atomic part, $\mu_a$, which can be expressed as
\[
\mu_a \cong \sum_{i \in I} a_i \delta(x - x_i),
\]
for some finite or countably infinite sequence $\{a_i \in \mathbb{C}\}_{i\in I}$ and $\{x_i \in \mathbb{R}\}_{i \in I}$ such that for any compact $\Omega \subseteq \mathbb{R}$,
\[
\sum_{i \in I | x_i \in \Omega} \abs{a_i} < \infty.
\]
For any compact $\Omega \subseteq \mathbb{R}$, it can be shown that
\[
\textnormal{TV}_\Omega(\mu) = \textnormal{TV}_\Omega(\mu_c) +  \textnormal{TV}_\Omega(\mu_a) 
\]
\text{where }
\begin{align*}
 &\textnormal{TV}_\Omega(\mu_c) = \sup_{\substack{f\in \textnormal{C}_c^1(\mathbb{R})  \norm{f}_{L^\infty} = 1}} \abs{\int_\mathbb{R} \phi_c(x) f'(x) dx}\text{ and} \\ &\textnormal{TV}_\Omega(\mu_d) = \sum_{i \in I | x_i \in \Omega} |a_i| .
\end{align*}
We will use standard notation for Schwartz space, $\mathcal{S}(\mathbb{R})$ and tempered distributions by $\mathcal{S}'(\mathbb{R})$. Note that every radon measure is a distribution (i.e.~a continuous map from compact smooth function to complex numbers) --- a radon measure $\mu \in \mathcal{M}(\mathbb{R})$ which is additionally a tempered distribution will be called a tempered radon measure. From the Schwartz representation theorem \cite{adams2003sobolev}, it follows that any tempered distribution can be expressed as
\[
\langle \mu , f\rangle = \sum_{\alpha = 0}^m   \int_{\mathbb{R}} u_{\alpha}(\omega) \frac{\partial^\alpha}{\partial \omega^\alpha} \hat{f}(\omega) \frac{d\omega}{\sqrt{2\pi}} \ \forall \ f \in \mathcal{S}(\mathbb{R}),
\]
where $\hat{f}$ is the fourier transform of $f$, and $u_{0}, u_1 \dots u_m$ are continuous functions of at-most polynomial growth. Of particular interest will be distributions which contain only the term corresponding to $\alpha = 0$ i.e.
\[
\langle \mu , f\rangle = \int_{\mathbb{R}} \hat{\mu}(\omega) \hat{f}(\omega) \frac{d\omega}{\sqrt{2\pi}} \ \forall \ f\in \mathcal{S}(\mathbb{R}),
\]
where $\hat{\mu}$ is a continuous function of at-most polynomial growth. Such a distribution will be called a tempered distribution with a fourier transform being a function of at-most polynomial growth, and $\hat{\mu}$ will be referred to as the fourier transform of $\mu$.

Given a Banach space $X$ and an operator $O: X \to X$, the operator norm will be denoted by $\norm{O} = \sup_{x \in X} \norm{Ox} / \norm{x}$. The space of bounded linear operators on a Banach space $X$ will be denoted by $\mathfrak{L}(X)$ i.e. $\mathfrak{L}(X) = \{O: X\to X | \norm{O} < \infty\}$. A map $F: \mathbb{R} \to \mathfrak{L}(X)$ is norm continuous at $t$ if
\[
 \lim_{s\to t} \norm{F(t) - F(s)}=0
\]
\text{ and strongly continuous at} $t$ if
\[
\forall x \in X, \lim_{s \to t}\norm{F(t) x - F(s)x} = 0.
\]
Similarly, it is \text{norm differentiable at }$t$ if
\begin{align*}
\exists F'(t): \lim_{s \to t} \norm{F'(t) - \frac{F(t) - F(s)}{t - s}} = 0,
\end{align*}
and strongly differentiable at $t$ if
\begin{align*}
\exists F'(t): \forall x \in X \lim_{s\to t} \norm{F'(t) x - \frac{F(t) x - F(s) x}{t - s}} = 0.
\end{align*}
Note that if $X$ is finite dimensional, then the notion of norm continuity/differentiability and strong continuity/differentiability are equivalent.

A sequence $\{\mu_n : X \to \mathbb{C}\}_{n \in \mathbb{N}}$ weakly converges to $\mu^* : X \to \mathbb{C}$, denoted by $\mu^* = \textnormal{wlim}_{n \to \infty} \mu_n$, if $\forall x \in X, \mu^* x = \lim_{n \to \infty} \mu_n x$. 

Given a Hilbert space $\mathcal{H}$, a densely defined operator $O:\textnormal{dom}[O] \to \mathcal{H}$ is said to be closed if $\forall \psi \in \textnormal{dom}[O]$, such that $\forall$ sequences $\{\psi_n \in \textnormal{dom}[O]\}_{n\in \mathbb{N}}$ which converge to $0$ such that the sequence $\{O\psi_n \in \mathcal{H}\}_{n \in \mathbb{N}}$ also converges, $\lim_{n \to \infty} O\psi_n = 0$. A densely defined operator $O$ is said to be closable if it has a closed extension, called the closure of the operator and denoted by $\overline{O}$. We will use the following property of the domain of the closure, $\textnormal{dom}[\overline{O}]$: $\psi \in \textnormal{dom}[\overline{O}]$ if and only if $\exists$ a sequence $\{\psi_n \in \textnormal{dom}[O]\}_{n \in \mathbb{N}}$ such that $\lim_{n \to \infty} \psi_n = \psi$, and the sequence $\{O\psi_n \in \mathcal{H}\}_{n \in \mathbb{N}}$ also converges. Furthermore if $O$ is closable, the limit of the sequence $\{O\psi_n \in \mathcal{H}\}_{n \in \mathbb{N}}$ is independent of the sequence $\{\psi_n \in \mathcal{H}\}_{n \in \mathbb{N}}$, and is equal to $\overline{O}\psi$. The adjoint of a densely defined operator $O: \textnormal{dom}[O] \to \mathcal{H}$ is an operator $O^\dagger : \textnormal{dom}[O^\dagger] \to \mathcal{H}$ where $\textnormal{dom}[O^\dagger] = \{\psi \in \mathcal{H} | \langle \psi, O\cdot\rangle : \textnormal{dom}[O] \to \mathbb{C} \text{ is bounded}\}$ and by the Riesz' representation theorem, $\forall \psi \in \mathcal{H}, O^\dagger \psi$ is identified as the unique vector which satisfies $\langle O^\dagger \psi, \phi\rangle = \langle \psi, O\phi\rangle \ \forall \phi \in \textnormal{dom}[O].$ An operator is self adjoint if $\textnormal{dom}[O] = \textnormal{dom}[O^\dagger]$. A closable operator is essentially self adjoint if it has a self adjoint extension, which then coincides with its closure.

\noindent\emph{Fock Spaces}: For a separable Hilbert space $\mathcal{H}$, and $n\in \mathbb{Z}_{\geq 1}$, we will denote by $\textnormal{Sym}_n(\mathcal{H}) \subseteq \mathcal{H}^{\otimes n}$ the set of symmetric (permutationally invariant) states in $\mathcal{H}^{\otimes n}$. We will denote by $\textnormal{Fock}[\mathcal{H}] := \mathbb{C}\oplus \bigoplus_{n \in \mathbb{Z}_{\geq 1}}\textnormal{Sym}_n(\mathcal{H}) $ the symmetric (bosonic) Fock space generated by $\mathcal{H}$.  We will denote by $\Pi_{n}:\textnormal{Fock}[\mathcal{H}] \to \textnormal{Fock}[\mathcal{H}]$ the projector onto $\textnormal{Sym}_{n}(\mathcal{H})$ (i.e.~the $n$ particle sector), by $\Pi_{\leq n} := \sum_{i = 0}^{n}\Pi_i$ and by $\Pi_{>n} = \textnormal{id} - \Pi_{\leq n}$.

We will denote by $\textnormal{F}_\infty[\mathcal{H}] \subseteq \textnormal{Fock}[\mathcal{H}]$ the space of all states with a finite number of particles i.e.~
\[
\textnormal{F}_\infty[\mathcal{H}] = \big\{\ket{\Psi} \in \textnormal{Fock}[\mathcal{H}] \big| \exists N_0 \in \mathbb{N}: \Pi_n\ket{\Psi} = 0 \ \forall n > N_0 \big\}.
\]
The space of states with finite $k^\text{th}$ particle number moment will be denoted by $\textnormal{F}_k(\mathcal{H}) \subseteq \textnormal{Fock}[\mathcal{H}]$ i.e.~for $k \in \mathbb{Z}_{\geq 1}$ 
\[
\textnormal{F}_k[\mathcal{H}] = \bigg\{\ket{\Psi} \in \textnormal{Fock}[\mathcal{H}]  \bigg| \sum_{n = 0}^\infty n^k \bra{\Psi}\Pi_n \ket{\Psi} < \infty \bigg\},
\]
and by $\textnormal{F}_\mathcal{S}[\mathcal{H}]$ we will denote the space of states where all the particle number moments are finite i.e.
\begin{align*}
&\textnormal{F}_{\mathcal{S}}[\mathcal{H}] := \bigcap_{k = 1}^\infty \textnormal{F}_k[\mathcal{H}] \nonumber\\
&\qquad=  \bigg\{\ket{\Psi} \in \textnormal{Fock}[\mathcal{H}]  \bigg| \sum_{n = 0}^\infty n^k \bra{\Psi}\Pi_n \ket{\Psi} < \infty \ \forall k \in \mathbb{Z}_{\geq 1}\bigg\}.
\end{align*}
We remark that $\textnormal{F}_\infty[\mathcal{H}]$, $\textnormal{F}_\mathcal{S}[\mathcal{H}]$ and $\textnormal{F}_k[\mathcal{H}]$ (for any $k \in \mathbb{Z}_{\geq 1}$) are dense in $\textnormal{Fock}[\mathcal{H}]$. Note also the inclusions $\textnormal{F}_{\infty}[\mathcal{H}] \subseteq \textnormal{F}_\mathcal{S}[\mathcal{H}] \subseteq \textnormal{F}_k[\mathcal{H}]$. For $\ket{\Psi} \in \textnormal{F}_k[\mathcal{H}]$, we will denote by $\mu^{(k)}_{\ket{\Psi}}$ the $k^\textnormal{th}$ moment of the photon number operator i.e.
\[
\mu^{(k)}_{\ket{\Psi}} = \sum_{n = 0}^\infty n^k \bra{\Psi} \Pi_n \ket{\Psi}. 
\]

For any $v \in \mathcal{H}$, we will denote by $a^{-}_{v}$ and $a^{+}_v$ the corresponding annhilation and creation operator. These operators can be explicitly defined over the domain $\textnormal{F}_\infty[\mathcal{H}]$ --- $a^{-}_v:\textnormal{F}_\infty[\mathcal{H}] \to \textnormal{Fock}[\mathcal{H}]$ is an operator defined by
\begin{align*}
    &a_v^{-}(\alpha, 0^\infty) = 0 \ \forall \ \alpha \in \mathbb{C}, 
\end{align*}
and for all $\ u \in \mathcal{H}, n \in \mathbb{Z}_{\geq 1}$,
\begin{align*}
    &a_v^{-}(0^n, u^{\otimes n}, 0^\infty) =  \langle v, u \rangle(0^{n - 1}, \sqrt{n} u^{\otimes {n - 1}}, 0^\infty) .
\end{align*}
Since for every $n \in \mathbb{Z}_{\geq 1}$, the set $\textnormal{span}\{u^{\otimes n} | u \in \mathcal{H} \}$ is dense in $\textnormal{Sym}_n(\mathcal{H})$, and when domain-restricted to $\textnormal{span}\{u^{\otimes n} | u \in \mathcal{H} \}$, $a_v^{-}$ as defined above is a bounded operator, it can be uniquely extended to $\textnormal{Sym}_n(\mathcal{H})$ as a consequence of the bounded linear transformation theorem, and then extended to $\textnormal{F}_\infty[\mathcal{H}]$ by linearity. Similarly, $a_v^+: \textnormal{F}_\infty[\mathcal{H}] \to \textnormal{Fock}[\mathcal{H}]$ is defined via
\begin{align*}
    &a_v^+(\alpha, 0^\infty ) = (0, \alpha v, 0^\infty) \ \forall \ \alpha \in \mathbb{C},
\end{align*}
and for all $u \in \mathcal{H}, n \in \mathbb{Z}_{\geq 1}$,
\begin{align*}
    &a_v^+(0^n, u^{\otimes n}, 0^\infty) =\nonumber\\
    &\qquad \bigg(0^{n + 1}, \frac{1}{\sqrt{n + 1}} \sum_{i = 0}^{n} u^{\otimes i}\otimes v \otimes u^{\otimes (n - i)}, 0^\infty\bigg).
\end{align*}
As with $a_v^{-}$, this definition of $a_v^{+}$ can be extended uniquely to $\textnormal{F}_0$.

In this paper, we will encounter finite tensor products of Fock spaces. Given a Hilbert space $\mathcal{H}$, $\textnormal{Fock}[\mathcal{H}]^{\otimes M} \simeq \textnormal{Fock}\big[\mathcal{H}^{\oplus M}\big]$, where the tensor product is taken as a tensor product over Hilbert spaces. We will use the notation $\textnormal{F}_\infty^M[\mathcal{H}] = \textnormal{F}_\infty[\mathcal{H}^{\oplus M}] $ and $\textnormal{F}_k^M[\mathcal{H}] = \textnormal{F}_k[\mathcal{H}^{\oplus M}]$ for $k \in \mathbb{Z}_{\geq 1}$. For $\alpha \in \{1, 2 \dots M\}$ and $v\in \mathcal{H}$, we define $a_{\alpha, v}^{-}: \textnormal{F}^M_\infty[\mathcal{H}] \to \textnormal{Fock}[\mathcal{H}]^{\otimes M}$ via
\begin{align*}
&a_{\alpha, v}^{-}(\alpha, 0^\infty) = 0 \ \forall \ \alpha \in \mathbb{C},
\end{align*}
and for all $u \in \mathcal{H}^{\oplus M}, n \in \mathbb{Z}_{\geq 1}$,
\begin{align*}
&a_{\alpha, v}^{-}(0^n, u^{\otimes n}, 0^\infty) = (0^{n - 1}, \sqrt{n}\langle v_\alpha, u \rangle u^{\otimes (n - 1)}, 0^\infty),
\end{align*}
where $v_\alpha = 0^{\oplus (\alpha - 1)} \oplus v \oplus 0^{\oplus(M - \alpha)}$. Similarly, we define $a_{\alpha, v}^+:F_\infty^M[\mathcal{H}] \to \textnormal{Fock}[\mathcal{H}]^{\otimes M}$ via
\begin{align*}
    &a_{\alpha, v}^+(c, 0^\infty ) = (0, c v_\alpha, 0^\infty) \ \forall \ c \in \mathbb{C},
\end{align*}
and for all $u \in \mathcal{H}^{\oplus M}, n \in \mathbb{Z}_{\geq 1}$,
\begin{align*}
    &a_{\alpha, v}^+(0^n, u^{\otimes n}, 0^\infty) =\\
    &\qquad \bigg(0^{n + 1}, \frac{1}{\sqrt{n + 1}} \sum_{i = 0}^{n} u^{\otimes i}\otimes v_\alpha \otimes u^{\otimes (n - i)}, 0^\infty\bigg).
\end{align*}
We will denote by $\Pi_n : \textnormal{Fock}[\mathcal{H}]^{\otimes M} \to \textnormal{Fock}[\mathcal{H}]^{\otimes M}$ the projector onto $\textnormal{Sym}_n(\mathcal{H}^{\oplus M})$, by $\Pi_{\leq n} := \sum_{i = 0}^n \Pi_i$ and by $\Pi_{> n} := \textnormal{id} - \Pi_{\leq n}$. 

\section{Well-definition of non-Markovian models}
\label{sec:well_def}

\subsection{Square-integrable coupling functions}

We first state some simple results about the non-Markovian model for the case where the coupling functions $v_\alpha$ are square integrable. As described in the previous section, we will assume the system to be finite-dimensional with Hilbert space $\mathcal{H}_S$, and the environment described by $M$ fock spaces $\text{Fock}[L^2(\mathbb{R})]^{\otimes M}$. The system-environment Hamiltonian is assumed to be of the form:
\begin{align}\label{eq:sq_int_hamil}
H = H_S(t) + \sum_{\alpha = 1}^M H_{\alpha, E} + \sum_{\alpha = 1}^M  \big(a_{\alpha, v_\alpha} L_\alpha^\dagger  +a_{\alpha, v_\alpha}^\dagger L_\alpha \big).
\end{align}
where $H_{\alpha, E}$ is the Hamiltonian describing the dynamics of the $\alpha^\text{th}$ bath and $a_{\alpha, f}$ for $f \in L^2(\mathbb{R})$ is the annihilation operator corresponding to $f$ (see section \ref{sec:notation_prelim}). Assuming the environment to be non-interacting and particle number conserving, we can specify $H_{\alpha, E}$ by a strongly-continuous single-parameter unitary group $\uptau_{\alpha, t} : L^2(\mathbb{R}) \to L^2(\mathbb{R})$ such that $\forall f \in L^2(\mathbb{R})$,
\[
e^{iH_{\alpha, E}t } a_{\alpha, f}  e^{-iH_{\alpha, E}t } =  a_{\alpha, \uptau_{\alpha, t} f}.
\]
The unitary group $\uptau_{\alpha, t}$ can also be physically interpreted as specifying the dynamics within the single-particle sector of the non-interacting bath. In this paper, we will mostly consider $\uptau_{\alpha, t}$ to be the time-translation unitary group [i.e. $(\uptau_{\alpha, t} f)(\omega) = f(\omega)e^{-i\omega t}$ for all $f \in L^2(\mathbb{R})$)], which is equivalent to $H_{\alpha, E}$ as given by Eq.~\ref{eq:hamiltonian_heuristic}. However, the results stated below for square-integrable $v_\alpha$ hold for more general unitary groups.

The basic data needed to specify a non-Markovian model with square-integrable coupling functions is provided in the definition below.
\begin{definition}
\label{def:non_markovian_sq_int}
A non-Markovian open system model for a quantum system with Hilbert space $\mathcal{H}_S$ with square integrable system-environment coupling functions is specified by
\begin{enumerate}
\item[(a)] A time-dependent system Hamiltonian $H_S(t) \in \mathfrak{L}(\mathcal{H}_S)$ which is Hermitian, norm continuous and differentiable in $t$.
\item[(b)] $M$ square integrable functions $\{v_\alpha \in L^2(\mathbb{R}) \}_{\alpha \in \{1,2 \dots M\}}$,
\item[(c)] $M$ bounded operators on the system Hilbert space $\{L_\alpha \in \mathfrak{L}(\mathcal{H}_S)\}_{\alpha \in \{1, 2 \dots M\}}$.
\item[(d)] $M$ strongly continuous single-parameter unitary groups on $L^2(\mathbb{R})$, $\{\uptau_{\alpha, t} : L^2(\mathbb{R}) \to L^2(\mathbb{R})\}_{\alpha \in \{1, 2 \dots M\}}$.
\end{enumerate}
\end{definition}
\noindent It is convenient to consider the Hamiltonian in Eq.~\ref{eq:sq_int_hamil} in the interaction picture with respect to the enivornment Hamiltonian, as provided in the definition below.
\begin{definition}
\label{def:hamil}
For a non-Markovian model with square integrable function as specified by definition \ref{def:non_markovian_sq_int} and for $t \in \mathbb{R}$,  $H(t): \mathcal{H}_S\otimes \textnormal{F}_\infty^M[L^2(\mathbb{R})] \to \mathcal{H}$ is defined via
\[
H(t) = H_S(t) +  \sum_{\alpha = 1}^M \big(L^\dagger_\alpha a_{\alpha, \uptau_{\alpha, t} v_\alpha}^-  + L_\alpha a^{+}_{\alpha, \uptau_{\alpha, t} v_\alpha}\big).
\]
\end{definition}
\noindent We note that we have been careful in specifying the domain of $H(t)$, since it is an unbounded operator and cannot be defined over the entire system-environment Hilbert space. Following the standard approach to treat quadratic field theories \cite{reed1972methods}, we restrict it to only states which have a finite-number of particles in the environment (i.e.~$\text{F}^M_\infty[L^2(\mathbb{R})]$). Then, using standard results from the theory of unbounded operators, we can ``close" the operator $H(t)$ and extend this domain. While precisely charaterizing the domain of the closure of $H(t)$ is in general a difficult problem, we only need the characterization of its domain provided in the following lemma (proved in appendix \ref{app:sq_int}).
\begin{lemma}\label{lemma:domain}
For all $t \in \mathbb{R}$,
\begin{enumerate}
    \item[(a)] $H(t):  \mathcal{H}_S\otimes \textnormal{F}_\infty^M[L^2(\mathbb{R})] \to \mathcal{H}$ is essentially self adjoint.
    \item[(b)] $H(t)$ is closable, and if $\overline{H(t)}:\textnormal{dom}[\overline{H(t)}] \to \mathcal{H}$ is its closure then $\mathcal{H}_S\otimes \textnormal{F}_1^M[L^2(\mathbb{R})] \subseteq \textnormal{dom}[\overline{H(t)}]$ and $\forall \ket{\Psi} \in \textnormal{dom}[\overline{H(t)}]$,
    $\overline{H(t)} \ket{\Psi} = \sum_{n = 0}^\infty H(t) \big(\Pi_n \ket{\Psi}\big).$
\end{enumerate}
\end{lemma}
\noindent For square integrable coupling function, we can then show the well definition of non-Markovian dynamics, as given in the following proposition (proved in appendix \ref{app:sq_int}).
\noindent\begin{proposition}\label{thm:schr_sq_int}
Given a non-Markovian model with square integrable coupling functions and for $t, s \in \mathbb{R}$, there exists a unique isometry $U(t, s) : \mathcal{H}_S \otimes \textnormal{F}_\infty^M[L^2(\mathbb{R})] \to \mathcal{H}_S \otimes \textnormal{F}_\infty^M[L^2(\mathbb{R})] \subseteq \mathcal{H}$ which is strongly continuous and differentiable in both $t, s$ and satisfies
\begin{align}\label{eq:prop_schr_eq}
\frac{d}{dt} U(t, s) = -i \overline{H(t)} U(t, s),\ \frac{d}{ds} U(t, s) = i U(t, s) \overline{H(s)},
\end{align}
with $U(s, s) = \textnormal{id} \ \forall s \in \mathbb{R}$.
\end{proposition}
\noindent In the following lemmas, we provide some useful properties of $U(t, s)$ corresponding to non-Markovian models with square integrable functions which will be useful for the analysis of distributional coupling functions. The next lemma can be considered to be an input-output equation for the bath modes \cite{gardiner1985input}.
\begin{lemma}\label{lemma:inp_out_eq}
Given $u \in L^2(\mathbb{R})$ and a non-Markovian model with square integrable coupling functions, $\forall \alpha \in \{1, 2 \dots M\},  s, t \in [0, \infty)$, 
\begin{align*}
[a^{-}_{\alpha, u}, U(t, s)] =  -i \int_s^t \langle u,  \uptau_{\alpha, \tau} v_\alpha\rangle U(t, \tau) L_\alpha U(\tau, s)d\tau
\end{align*}
over the domain $\mathcal{H}_S \otimes \textnormal{F}_\mathcal{S}^M[L^2(\mathbb{R})]$, where $U(t, s)$ is the propagator corresponding to the non-Markovian model as defined in lemma \ref{thm:schr_sq_int}. 
\end{lemma}
\noindent \emph{Proof}: Throughout this proof, all operators are considered to be over the domain $\mathcal{H}_S \otimes \textnormal{F}_\infty^M[L^2(\mathbb{R})]$ --- in particular, we extend $a_{\alpha, u}^\pm$ from the domain $\mathcal{H}_S \otimes \textnormal{F}_\infty^M[L^2(\mathbb{R})]$ to $\mathcal{H}_S \otimes \textnormal{F}_{\infty}^M[L^2(\mathbb{R})]$ via
\[
a_{\alpha, u}^\pm \ket{\Psi} = \sum_{n = 0}^\infty a_{\alpha, u}^\pm \Pi_n \ket{\Psi} \ \forall \ket{\Psi} \in \mathcal{H}_S \otimes \textnormal{F}_\mathcal{S}^M[L^2(\mathbb{R})].
\]
Note that since $U(t, s)$ is strongly differentiable with respect to $t$, and it maps $\mathcal{H}_S \otimes \textnormal{F}_{\infty}^M[L^2(\mathbb{R})]$ to $\mathcal{H}_S \otimes \textnormal{F}_{\infty}^M[L^2(\mathbb{R})]$, it follows that the operator $a_{\alpha, u}^{-}(t, s) = U(s, t) a_{\alpha, u}^{-} U(t, s)$ is strongly differentiable in both $t$ and $s$. Differentiating it with respect to $t$, and using the characterization of $\overline{H(t)}$ when acting on  $\mathcal{H}_S \otimes \textnormal{F}_{\infty}^M[L^2(\mathbb{R})]$ as provided in lemma \ref{lemma:domain}, we obtain
\[
\frac{d}{dt}a_{\alpha, u}^{-}(t, s) = -i \langle u, \uptau_{\alpha, t} v_\alpha\rangle U(s, t) L_{\alpha} U(t, s).
\]
Noting that since $U(t, s)$ is strongly continuous in both of its arguments and since $L_\alpha$ is a bounded operator, the right hand side in the above equation is strongly continuous in $t$ and thus the equation can be integrated to obtain
\begin{align*}
&a_{\alpha, u}^{-} U(t, s) \noindent\\
&\qquad= U(t, s) a_{\alpha, u}^{-} - i\int_s^t \langle u, \uptau_{\alpha, \tau} v_\alpha\rangle U(t, \tau) L_\alpha U(\tau, s) d\tau,
\end{align*}
which proves the lemma. \hfill \(\square\)

\noindent We now define an object that we will use repeatedly in the following sections --- the system Green's function. For an open quantum system, it is traditionally defined as the expectation value of a time-ordered product of system operators  \cite{xu2015input, trivedi2018few}. We generalize its definition slightly to allow for the propagators appearing in its definition to be from different non-Markovian models. The motivation behind this generalization will be clear from the following lemma \ref{lemma:pert_theory_bound}  --- it arises from the fact that while deriving error bounds in between dynamics of two non-Markovian models (e.g. a non-Markovian model and its regularization), we often need to evolve an operator as per the propagator of one model followed by evolving it backwards in time with the propagator of another model.
\begin{definition}[System Green's functions]\label{def:gfunc}
Consider $k + 1$ non-Markovian models specified by coupling functions $v_i = \{v_{i, \alpha} \in L^2(\mathbb{R})\}_{\alpha \in \{1, 2 \dots M\}}$ for $i \in \{1, 2 \dots k+ 1\}$, but with the same time-dependent system Hamiltonian $H_S(t)$ and jump operators $\{L_\alpha \in \mathfrak{L}(\mathcal{H}_S)\}_{\alpha \in \{1, 2 \dots M\}}$. For $O_1, O_2 \dots O_k \in \mathfrak{L}(\mathcal{H}_S)$, $\ket{\Psi_1}, \ket{\Psi_2} \in \mathcal{H}$ and $t_1, t_2 \dots t_k \in [0, \infty)$, then the Green's function $G^{v_1, v_2 \dots v_{k + 1}}_{O_1, O_2 \dots O_k}(t_1, t_2 \dots t_k)$ is defined by
\begin{align*}
&G^{v_1, v_2 \dots v_{k + 1}}_{O_1, O_2 \dots O_k; \ket{\Psi_1}, \ket{\Psi_2}}(t_1, t_2 \dots t_k) =\nonumber\\
&\qquad \bra{\Psi_1} U_{v_{k + 1}}(0, t_{k + 1}) \prod_{j = k}^1 O_j U_{v_j}(t_j, t_{j - 1}) \ket{\Psi_2},
\end{align*}
where $t_0 = 0$ and $U_{v_i}(\cdot, \cdot)$ is the propagator, as defined in lemma \ref{thm:schr_sq_int}, for the $i^\textnormal{th}$ model for $i \in \{1, 2 \dots k + 1\}$.
\end{definition}
\begin{lemma}
\label{lemma:pert_theory_bound}
Consider two non-Markovian models with coupling functions $v = \{v_\alpha \in L^2(\mathbb{R}) \}_{\alpha \in \{1, 2\dots M\}}$ and $u = \{u_\alpha \in L^2(\mathbb{R})\}_{\alpha \in \{1, 2 \dots M\}}$ respectively but with the same system Hamiltonian $H_S(t)$, jump operators $\{L_\alpha\}_{\alpha \in \{1, 2 \dots M\}}$ and single-particle environment dynamics described by the time-translation unitary group. Let $\ket{\Psi_v(t)} = U_v(t, 0)\ket{\Psi_0}, \ket{\Psi_u(t)} = U_u(t, 0) \ket{\Psi_0}$, where $U_v(t, s), U_u(t, s)$ are the propagators corresponding to the two non-Markovian models, and $\ket{\Psi_0} \in \mathcal{H}_S \otimes \textnormal{F}_\infty^M[L^2(\mathbb{R})]$ , then
\begin{align*}
&\norm{\ket{\Psi_u(t)} - \ket{\Psi_v(t)}}^2 \leq \nonumber\\
&\qquad\sum_{\alpha = 1}^M \bigg(\int_0^t \mathcal{D}^{u, v}_\alpha(\tau)d\tau + \int_0^t \mathcal{E}^{u, v}_\alpha(\tau) d\tau\bigg),
\end{align*}
where for $\tau \in [0, t]$ and $\alpha \in \{1, 2 \dots M\}$,
\begin{align*}
&\mathcal{D}^{u, v}_\alpha(\tau) = 4 \norm{L}_\alpha \norm{a^-_{\alpha, \uptau_\tau(u_\alpha - v_\alpha)} \ket{\Psi_0}}, \text{ and }\\
&\mathcal{E}^{u, v}_\alpha(\tau) = \nonumber\\
&\qquad 2 \bigg| \int_0^\tau \langle \uptau_\tau (u_\alpha - v_\alpha), \uptau_s v_\alpha\rangle G_{L_\alpha, L_\alpha^\dagger; \ket{\Psi_0}, \ket{\Psi_0}}^{v, v, u}(s, \tau)ds\bigg| +\nonumber \\
&\qquad 2\bigg | \int_0^\tau \langle\uptau_s u_\alpha, \uptau_\tau(u_\alpha - v_\alpha)\rangle G^{u, u, v}_{L_\alpha, L_\alpha^\dagger; \ket{\Psi_0}, \ket{\Psi_0}}(\tau, s) ds \bigg|,
\end{align*}
where the Green's functions $G_{L_\alpha, L_\alpha^\dagger; \ket{\Psi_0}, \ket{\Psi_0}}^{v, v, u}(s, \tau)$ and $G^{u, u, v}_{L_\alpha, L_\alpha^\dagger; \ket{\Psi_0}, \ket{\Psi_0}}(\tau, s)$ are defined in definition \ref{def:gfunc}.
\end{lemma}
\noindent\emph{Proof}: Note that
\[
\norm{\ket{\Psi_u(t)} - \ket{\Psi_v(t)}}^2 = 2\bigg(1 - \ \text{Re}[\bra{\Psi_u(t)}\Psi_v(t)\rangle\bigg).
\]
Consider now the inner product $\bra{\Psi_u(t)}\Psi_v(t)\rangle$ --- differentiating this with respect to $t$, we obtain that
\begin{align*}
&\frac{d}{dt} \bra{\Psi_u(t)}\Psi_v(t)\rangle = i\sum_{\alpha = 1}^M \bigg(\bra{\Psi_u(t)} L_\alpha^\dagger  a^-_{\alpha, \uptau_t(u_\alpha - v_\alpha)}\ket{\Psi_v(t)} + \nonumber\\
&\qquad \qquad \bra{\Psi_u(t)} L_\alpha  a^+_{\alpha, \uptau_t(u_\alpha - v_\alpha)}\ket{\Psi_v(t)}\bigg).
\end{align*}
We note that from lemma \ref{lemma:inp_out_eq}, it follows that $\forall \alpha $,
\begin{align*}
&\bra{\Psi_u(t)} L_\alpha^\dagger  a^-_{\alpha, \uptau_t(u_\alpha - v_\alpha)}\ket{\Psi_v(t)} \nonumber\\
& =\bra{\Psi_0} U_u(0, t) L_\alpha^\dagger  a^-_{\alpha, \uptau_t (u_\alpha - v_\alpha)}U_v(t, 0)\ket{\Psi_0}, \nonumber\\
& =G^{v, u}_{L_\alpha^\dagger; a^-_{\alpha, \uptau_t (u_\alpha - v_\alpha)}\ket{\Psi_0}, \ket{\Psi_0}} -\nonumber\\
&\qquad \quad i \int_0^t \langle \tau_t (u_\alpha - v_\alpha), \tau_s v_\alpha\rangle G_{L_\alpha, L_\alpha^\dagger; \ket{\Psi_0}, \ket{\Psi_0}}^{v, v, u}(s, t)ds.
\end{align*}
Similarly,
\begin{align*}
&\bra{\Psi_u(t)} L_\alpha a^+_{\alpha, \uptau_t(u_\alpha - v_\alpha)}\ket{\Psi_v(t)}  \nonumber \\
&=  \bra{\Psi_0} U_u(0, t)L_\alpha a^+_{\alpha, \uptau_t(u_\alpha - v_\alpha)} U_v(t, 0)\ket{\Psi_0} ,\nonumber\\
&=G^{v, u}_{L_\alpha; \ket{\Psi_0}, a_{\alpha, \uptau_t(u_\alpha - v_\alpha)}^{-}\ket{\Psi_0}} + \nonumber\\
&\qquad i\int_0^t \langle\tau_s u_\alpha, \tau_t(u_\alpha - v_\alpha)\rangle G^{u, u, v}_{L_\alpha, L_\alpha^\dagger; \ket{\Psi_0}, \ket{\Psi_0}}(t, s) ds.
\end{align*}
Here, we have used the notation for Green's functions defined in definition \ref{def:gfunc}. Furthermore, we can note that $\forall \alpha \in \{1,2  \dots M\}$,
\begin{align*}
&\bigg| G^{v, u}_{L_\alpha^\dagger; a^-_{\alpha, \uptau_t (u_\alpha - v_\alpha)}\ket{\Psi_0}, \ket{\Psi_0}}\bigg |, \bigg |G^{v, u}_{L_\alpha; \ket{\Psi_0}, a_{\alpha, \uptau_t(u_\alpha - v_\alpha)}^{-}\ket{\Psi_0}}  \bigg |\nonumber\\
&\qquad\leq \norm{L_\alpha} \norm{a^-_{\alpha, \uptau_t (u_\alpha - v_\alpha)}\ket{\Psi_0}},
\end{align*}
and thus we obtain
\begin{align}\label{eq:bound_olap_deriv}
&\bigg|\frac{d}{dt} \bra{\Psi_u(t)}\Psi_v(t)\rangle \bigg| \leq 2\sum_{\alpha = 1}^M \norm{L_\alpha} \norm{a^-_{\alpha, \uptau_t (u_\alpha - v_\alpha)}\ket{\Psi_0}} +\nonumber\\
 &\quad\sum_{\alpha = 1}^M \bigg( \bigg|  \int_0^t \langle \tau_t (u_\alpha - v_\alpha), \tau_s v_\alpha\rangle G_{L_\alpha, L_\alpha^\dagger; \ket{\Psi_0}, \ket{\Psi_0}}^{v, v, u}(s, t)ds\bigg | \nonumber\\
 &\quad + \bigg | \int_0^t \langle\tau_s v_\alpha, \tau_t(u_\alpha - v_\alpha)\rangle G^{u, u, v}_{L_\alpha, L_\alpha^\dagger; \ket{\Psi_0}, \ket{\Psi_0}}(t, s) ds \bigg |\bigg).
\end{align}
Finally, note that $\norm{\ket{\Psi_u(t)}  - \ket{\Psi_v(t)}}^2 = 2(1 - \bra{\Psi_u(t)} \Psi_v(t)\rangle)$ and therefore,
\begin{align}\label{eq:bound_norm_olap}
& \norm{\ket{\Psi_u(t)} - \ket{\Psi_v(t)}}^2 \leq 2 \int_0^t \bigg| \frac{d}{d\tau} \bra{\Psi_u(\tau)}\Psi_v(\tau)\rangle\bigg| d\tau.
\end{align}
Combining the estimates in Eq.~\ref{eq:bound_olap_deriv} and \ref{eq:bound_norm_olap}, we obtain the lemma statement. \hfill \(\square\).

\noindent In preparation for the next subsection, we now provide a lemma that follows from a straightforward application of lemma \ref{lemma:inp_out_eq} and characterizes the rate of change of this Green's function with respect to its time arguments.
\begin{lemma}
\label{lemma:gfunc_deriv}
Consider two non-Markovian models described by coupling functions $v = \{v_{ \alpha} \in L^2(\mathbb{R})\}_{\alpha \in \{1, 2 \dots M\}}, u = \{u_{\alpha} \in L^2(\mathbb{R})\}_{\alpha \in \{1, 2 \dots M\}}$, but with the same system Hamiltonian $H_S(t)$, jump operators $\{L_\alpha \in \mathfrak{L}(\mathcal{H}_S)\}_{\alpha \in \{1, 2 \dots M\}}$ and environment single-particle unitary group $\{\uptau_{\alpha, t}: L^2(\mathbb{R}) \to L^2(\mathbb{R})\}$. For $A, B \in \mathfrak{L}(\mathcal{H}_S)$, $\ket{\Psi}, \ket{\Phi} \in \mathcal{H}_S \otimes \textnormal{F}_{\mathcal{S}}^M[L^2(\mathbb{R})]$ and $t, s \in [0, \infty)$,
\begin{widetext}
\begin{align*}
&\frac{d}{ds} G^{v, v, u}_{A, B; \ket{\Psi}, \ket{\Phi}}(s, t) = iG_{[H_S(s), A], B; \ket{\Psi}, \ket{\Phi}}^{v, v, u}(s, t) + i\sum_{\alpha = 1}^M \bigg(G^{v, v, u}_{[L_\alpha^\dagger, A], B; \ket{\Psi}, a^-_{\alpha, \tau_{s}v_{\alpha}} \ket{\Phi}}(s, t) +z G^{v, v, u}_{[L_\alpha, A], B;  a^-_{\alpha, \tau_{s}v_{\alpha}}\ket{\Psi}, \ket{\Phi}}(s, t)  \bigg) - \nonumber\\
&\quad \quad i\sum_{\alpha = 1}^M \bigg(\int_0^{s} \langle \uptau_{s}v_{\alpha}, \uptau_\tau v_{\alpha}\rangle G^{v, v, v, u}_{L_\alpha, [L_\alpha^\dagger, A], B; \ket{\Psi}, \ket{\Phi}}(s, \tau, t) d\tau + \int_0^{s} \langle \uptau_\tau u_{\alpha}, \uptau_{s}v_{\alpha}\rangle G^{v, v, u, u}_{[L_\alpha, A], B, L_\alpha^\dagger; \ket{\Psi}, \ket{\Phi}}(s, t, \tau) d\tau -\nonumber\\
&\qquad \quad \qquad \int_{s}^{t}
\langle \uptau_\tau u_{\alpha}, \uptau_{s}v_{\alpha} \rangle G^{v, v, u, u}_{[L_\alpha, A], L_\alpha^\dagger, B; \ket{\Psi}, \ket{\Phi}}(s, \tau, t) d\tau\bigg)
\end{align*}
\end{widetext}
\end{lemma}
\noindent\emph{Proof}: Differentiating $G^{v, v, u}_{A, B; \ket{\Psi}, \ket{\Phi}}(s, t)$ with respect to $s$ and using lemma \ref{lemma:domain}(b), we obtain that
\begin{align*}
&\frac{d}{ds} G^{v, v, u}_{A, B;  \ket{\Psi}, \ket{\Phi}}(s, t) = \nonumber\\
&\quad i\bra{\Psi} U_u(0, t) B U_v(t, s) [H_S(s), A] U_v(s, 0) \ket{\Phi} + \nonumber\\
&\quad  i\sum_{\alpha = 1}^M \bigg( \bra{\Psi} U_u(0, t) B U_v(t, s) a^+_{\alpha,\uptau_{s} v_{\alpha}}[L_\alpha, A] U_v(s, 0)\ket{\Phi} +\nonumber\\
&\quad \bra{\Psi} U_u(0, t) B U_v(t, s) [L_\alpha^\dagger, A]a^-_{\alpha,\uptau_{s} v_{\alpha}} U_v(s, 0)\ket{\Phi} \bigg).
\end{align*}
We note that
\begin{align*}
&\bra{\Psi} U_u(0, t) B U_v(t, s) [H_S(s), A] U_v(s, 0) \ket{\Phi} = \\
&\qquad \qquad \qquad \qquad \qquad \quad G_{[H_S(s), A], B; \ket{\Psi}, \ket{\Phi}}^{v, v, u}(s, t).
\end{align*}
Using lemma \ref{lemma:inp_out_eq}, we obtain that
\begin{align*}
&\bra{\Psi} U_u(0, t) B U_v(t, s) [L_\alpha^\dagger, A]a^-_{\alpha,\uptau_{s} v_{\alpha}} U_v(s, 0)\ket{\Phi} =\\
 &\qquad \quad G^{v, v, u}_{[L_\alpha^\dagger, A], B; \ket{\Psi}, a^-_{\alpha, \tau_{s}v_{\alpha}} \ket{\Phi}}(s, t) -\nonumber \\
 &\qquad \quad i\int_0^{s} \langle \uptau_{s}v_{\alpha}, \uptau_\tau v_{\alpha}\rangle G^{v, v, v, u}_{L_\alpha, [L_\alpha^\dagger, A], B; \ket{\Psi}, \ket{\Phi}}(s, \tau, t) d\tau,
\end{align*}
and
\begin{align*}
 &\bra{\Psi} U_u(0, t) B U_v(t, s) a^+_{\alpha,\uptau_{s} v_{\alpha}}[L_\alpha, A] U_v(s, 0)\ket{\Phi} =\\
 &\qquad \quad G^{v, v, u}_{[L_\alpha, A], B; a^-_{\alpha, \uptau_{s }v_{\alpha}} \ket{\Psi}, \ket{\Phi}}(s, t) + \nonumber\\
 & \qquad \quad  i\int_{s}^{t}
\langle \uptau_\tau u_{\alpha}, \uptau_{s}v_{\alpha} \rangle G^{v, v, u, u}_{[L_\alpha, A], L_\alpha^\dagger, B; \ket{\Psi}, \ket{\Phi}}(s, \tau, t) d\tau - \nonumber\\
&\qquad \quad i\int_0^{s} \langle \uptau_\tau u_{\alpha}, \uptau_{s}v_{\alpha}\rangle G^{v, v, u, u}_{[L_\alpha, A], B, L_\alpha^\dagger; \ket{\Psi}, \ket{\Phi}}(s, t, \tau) d\tau,
\end{align*} 
which completes the proof of the lemma. \hfill \(\square\)

\subsection{Extension to general radon measures}
\label{sec:ext_gen_rad}
%\begin{definition}[Distributional coupling function]
%\label{def:distr_coup_fun}
%A distributional coupling function is specified by a tuple $(\mu, \varphi)$ where $\mu \in \mathcal{M}(\mathbb{R}) \cap \mathcal{S}'(\mathbb{R})$ is a radon measure and a tempered distribution whose Fourier transform is a positive continuous function of at-most polynomial growth and $\varphi \in \textnormal{C}^\infty(\mathbb{R})$.
%\end{definition}
n this section, we analyze the regularization procedure used to define the dynamics for non-Markovian models that have coupling functions that are not necessarily square integrable. We first begin by analyzing the regularization of a single radon measure and establishing some of its general properties. While it is possible that some of the lemmas relating radon measures that we prove below can be found in analysis literature, we could not find a proof of the precise statements that we needed and so we include our own proof of these statements. After that, we analyze the full unitary group corresponding to the regularized radon measure and study its limiting behaviour as the regularization is removed.

\subsubsection{Some generalities about radon measures}
Recall that a radon measure $\mu$ is a map from the space of continuous compact functions to complex numbers i.e.~$\mu: \textnormal{C}_c^0(\mathbb{R}) \to \mathbb{C}$. However, we need to extend its domain to apply it to certain discontinuous functions --- for e.g.~for the single-particle case described previously (Eq.~\ref{eq:decay_equation}), at time $t$, the radon measure is applied to the excited state amplitude restricted to the interval $[0, t]$ which is discontinuous as a function over $\mathbb{R}$. To extend $\mu$ to a space of discontinuous functions we use a mollifier to first smoothen the discontinuous function to a continuous function, and then apply $\mu$.

The space of discontinuous functions that we will focus on is one which is formed by windowing a differentiable (and thus continuous) function within a given interval. More specifically, we will consider the space $\textnormal{PWC}^1(\mathbb{R})$ is space of all functions which are expressible as $g \cdot \mathcal{I}_{[a, b]}$, where $\mathcal{I}_{[a,b]}(x) = 1$ if $x \in [a, b]$ and $0$ otherwise is the indicator function for the interval $[a, b]$, for some $g \in \textnormal{C}^1(\mathbb{R})$ and $[a, b]  \subseteq \mathbb{R}$.  The functions belonging to this space thus have discontinuities at two points --- the end points of the windowing intervals. Given a symmetric mollifier $\rho$ and $\varepsilon >0$, we can then define a map $\mu_\varepsilon: \text{PWC}^1(\mathbb{R}) \to \mathbb{C}$ via
\[
\langle \mu_\varepsilon, f\rangle = \langle \mu, \rho_\varepsilon \star f\rangle \ \forall \ f\in \textnormal{PWC}^1(\mathbb{R}),
\]
i.e.~given a discontinuous function from $\text{PWC}^1(\mathbb{R})$, we first smooth the discontinuities by convolving it with a mollifier $\rho_\varepsilon$ and then apply $\mu$ to the resulting continuous function. We can now take the limit of $\varepsilon \to 0$ and attempt to define $\mu^*:\text{PWC}^1(\mathbb{R}) \to \mathbb{C}$ via
\begin{align}\label{eq:mu_star}
\langle \mu^*, f\rangle = \lim_{\varepsilon \to 0} \langle \mu_\varepsilon, f\rangle \forall \ f\in \textnormal{PWC}^1(\mathbb{R}).
\end{align}

In the following lemmas, we establish that $\mu^*$ is well defined (i.e.~the limit defining $\mu^*$ exists), is independent of the precise choice of the mollifier and that when restricted to the function space $ \textnormal{C}_c^1(\mathbb{R})$ (i.e.~the space of continuously differentiable compactly supported functions), its action coincides with that of the radon measure $\mu$. We also derive certain properties of the map $\mu^*$ which will be useful in the following subsection. We first present a technical lemma, whose proof is in appendix \ref{app:distributional_mem_ker}.
\begin{lemma}\label{lemma:main_error_bound_mu}
Consider $\mu \in \mathcal{M}(\mathbb{R})$ with the Lesbesgue decomposition $\mu = \mu_c + \mu_d$ with $\phi_c \in \textnormal{C}^0(\mathbb{R})$ given by $\phi_c(x) = \mu_c((-\infty, x]) \ \forall \ x \in \mathbb{R}$, and $\mu_d \cong \sum_{i \in I} \alpha_i \delta(x - y_i)$ for some $\{\alpha_i \in \mathbb{C}\}_{i \in I}, \{y_i \in \mathbb{R}\}_{i \in I}$ and finite and countably infinite index set $I$. Given a compact interval $[a, b] \subseteq \mathbb{R}$ and $f \in \textnormal{C}^1(\mathbb{R})$, define $\langle\mu_{[a, b]}^*, f\rangle $ by
\begin{align*}
&\langle \mu^*_{[a, b]}, f\rangle =\langle \mu_{c, [a, b]}^*, f\rangle +  \langle \mu_{d, [a, b]}^*, f\rangle \ \text{where} \\ 
&\langle\mu_{c, [a, b]}^*, f\rangle =\nonumber\\
&\qquad  f(b) \phi_c(b) - f(a) \phi_c(a) - \int_a^b \phi_c(x) f'(x) dx \ \text{ and}\\
&\langle\mu_{d, [a, b]}^*, f\rangle = \frac{1}{2} \sum_{i \in I | y_i \in \{a, b\}} \alpha_i f(y_i) + \sum_{i \in I | y_i \in (a, b)} \alpha_i f(y_i).
\end{align*}
Then, for every compact intervals $[a, b] \subseteq \mathbb{R}$, $\exists \Delta^0_{\mu; [a, b]}(\varepsilon), \Delta^1_{\mu; [a,b]}(\varepsilon) > 0 $ where $ \Delta^0_{\mu; [a, b]}(\varepsilon), \Delta^1_{\mu; [a,b]}(\varepsilon) \to 0$ as $\varepsilon \to 0$ such that $\forall \varepsilon \in (0, (b- a) / 2)$ and for any even (symmetric about $0$) positive function $\alpha \in \textnormal{C}_c^\infty(\mathbb{R})$ with $\textnormal{supp}(\alpha) \subseteq [-\varepsilon,\varepsilon]$ and $\int_{[-\varepsilon, \varepsilon]}\alpha(x) dx=1$,
\begin{align*}
&\abs{ \langle \mu^*_{ [a, b]}, f\rangle  - \langle \mu, \alpha \star (f\cdot \mathcal{I}_{[a, b]}) \rangle} \leq \nonumber\\
&\qquad \Delta^0_{\mu; [a, b]}(\varepsilon) \sup_{x \in [a, b]} \abs{f(x)} + \Delta^1_{\mu; [a, b]}(\varepsilon)\sup_{x \in [a, b]} \abs{f'(x)}.
\end{align*}
The functions $\Delta^0_{\mu; [a, b]}, \Delta^1_{\mu; [a, b]}$ will be called the error functions corresponding to $\mu$.
\end{lemma}
This lemma characterizes the regularization of a radon measure $\mu$. Furthermore, it shows that the rate of convergence of the error between the radon measure and its regularization result as the regularization is removed is dependent only on the maximum value of the function within its support, and the maximum value of its derivative. This lemma now allows us to prove the existence of the limit defining $\mu^*$ (Eq.~\ref{eq:mu_star}).
\begin{lemma}[Existence of $\mu^*$]
\label{lemma:well_def_mu_star}
Given a $\mu \in \mathcal{M}(\mathbb{R})$, consider the map $\mu_\varepsilon : \textnormal{PWC}^1(\mathbb{R}) \to \mathbb{C}$ given by $\langle \mu_\varepsilon, \cdot\rangle = \langle \mu , \rho_\varepsilon \star (\cdot)\rangle$ with $\rho$ being a symmetric mollifier, then
\begin{enumerate}
\item[(a)] $\forall f\in \textnormal{PWC}^1(\mathbb{R})$, $\langle \mu^*, f\rangle := {\lim_{\varepsilon \to 0}} \langle\mu_\varepsilon, f\rangle$ exists and is independent of the choice of the mollifier,
\item[(b)] $\forall f \in \textnormal{C}_c^1(\mathbb{R})$, $\langle \mu^*, f \rangle = \langle \mu, f \rangle$.
\end{enumerate}
\end{lemma}
\noindent\emph{Proof}:  Suppose that $f \in \textnormal{PWC}^1(\mathbb{R})$ has the representation $f = g\cdot \mathcal{I}_{[a, b]}$ for some $g \in \textnormal{C}^1(\mathbb{R})$ and compact $[a, b] \subseteq \mathbb{R}$. Part \emph{(a)} of the lemma follows directly from lemma \ref{lemma:main_error_bound_mu} from which it follows that $\lim_{\varepsilon \to 0} \langle \mu, \rho_\varepsilon \star f\rangle = \langle \mu^*_{[a, b]}, g\rangle$, which can be identified as $\langle \mu^*, f\rangle$. Furthermore, we note that by construction, $\mu^*$ is independent of the choice of the mollifier.

For part \emph{(b)}, we note that for $f = g\cdot \textnormal{I}_{[a, b]} \in \textnormal{C}^1_c(\mathbb{R}) \subseteq  \textnormal{PWC}^1(\mathbb{R})$, $g(a) = g(b) = 0$, and therefore from the definition of $\langle \mu^*_{[a, b]}, g\rangle$ in lemma \ref{lemma:main_error_bound_mu}
\begin{align*}
&\langle \mu^*, f\rangle = \langle \mu^*_{[a, b]}, g\rangle \nonumber\\
&\qquad =-\int_a^b \phi_c(x) g'(x) dx + \sum_{i \in I}\alpha_i g(y_i) \nonumber\\
&\qquad= -\int_a^b \phi_c(x) f'(x) dx + \sum_{i \in I}\alpha_i f(y_i) = \langle \mu, f\rangle,
\end{align*}
which proves the lemma statement. \hfill \(\square\)
%Finally, lemma \ref{lemma:main_error_bound_mu} also straightforwardly yields a convergence estimate for the limit that defines $\mu^*$ --- the functions appearing in the convergence estimate ($\Delta^0_{\mu; [a, b]}, \Delta^1_{\mu; [a, b]}$ in lemma \ref{lemma:main_error_bound_mu}) will be key to our analysis in the following sections, and we collect them into the following definition.

%\begin{definition}[Error functions for $\mu$]\label{def:error_funcs_mu}  Given $\mu \in \mathcal{M}(\mathbb{R})$ and its associated $\mu^*:\textnormal{PWC}^1(\mathbb{R}) \to \mathbb{C}$ (Definition \ref{def:mu_star}), the error functions of $\mu$ for a compact interval $[a, b] \subseteq \mathbb{R}$, $\Delta^0_{\mu; [a, b]}, \Delta^1_{\mu; [a, b]} : (0, (b - a)/2) \to \mathbb{R}^+$ are functions such that $\Delta^0_{\mu; [a, b]}(\varepsilon), \Delta^1_{\mu; [a, b]}(\varepsilon) \to 0$ as $\varepsilon\to 0^+$, and $\forall f \in \textnormal{PWC}^1(\mathbb{R})$ expressible as $g\cdot \mathcal{I}_{[a, b]}$ for some $g \in \textnormal{C}^1(\mathbb{R})$ and $\forall$ symmetric mollifiers $\rho \in \textnormal{C}_c^\infty(\mathbb{R})$ with $\textnormal{supp}(\rho) \subseteq [-\varepsilon, \varepsilon]$ with $\varepsilon < (b - a)/2$, 
%\[
%\abs{\langle \mu^*, f\rangle - \langle \mu, \rho \star f\rangle} \leq \Delta^0_{\mu; [a, b]}(\varepsilon) \sup_{x \in [a, b]}\abs{f(x)} + \Delta^1_{\mu; [a,b]}(\varepsilon) \sup_{x \in [a, b]}\abs{f'(x)}.
%\]
%\end{definition}
We next provide some examples of distributional coupling functions that appear in a number of physical problems. All of these coupling functions are radon measures, and we also provide their extensions $\mu^*$ (Eq.~\ref{eq:mu_star}) to the space of discontinuous functions as well as the error functions defined in lemma \ref{lemma:main_error_bound_mu}.
\begin{example}[Square integrable coupling functions] Although we analyzed square-integrable coupling functions separately in the previous section, they can also be represented as, and thus are a special case of, distributional coupling functions. In particular, for $v \in L^2(\mathbb{R}) $, with Fourier transform $\hat{v} \in L^2(\mathbb{R}) $, note that
\[
\kappa(t) = \int_{-\infty}^\infty |\hat{v}(\omega)|^2 e^{-i\omega t} d\omega,
\]
is a continuous function with $\norm{\kappa}_{L^\infty} \leq \norm{v}_{L^2}^2$, and thus can be described by the radon measure $\mu$ defined by
\begin{align}\label{eq:eq_mu_st}
\langle \mu, f\rangle = \int_{-\infty}^\infty f(t) \kappa(t) dt \ \forall \ f \in \textnormal{C}^0_c(\mathbb{R})
\end{align}
is a Radon measure. Furthermore, its extension to $\textnormal{PWC}^1(\mathbb{R})$, $\mu^*$ (Eq.~\ref{eq:mu_star}), is given by
\[
\langle\mu^*, f\rangle= \int_{-\infty}^\infty  f(t) \kappa(t) dt \ \forall \ f \in \textnormal{PWC}^1(\mathbb{R}),
\]
Consider now $f = g\cdot \mathcal{I}_{[a, b]} \in \textnormal{PWC}^1(\mathbb{R})$ for some compact interval $[a, b] \subseteq \mathbb{R}$ and $g \in \textnormal{C}^1(\mathbb{R})$. Since $\kappa \in \textnormal{C}^1(\mathbb{R})$, $\exists \delta_{\kappa, [a, b]}(\varepsilon) > 0,$ where $\delta_{\kappa; [a, b]}(\varepsilon)\to 0$ as $\varepsilon \to 0$, such that $\forall x, x' \in [(3a - b) / 2, (3b - a)/2]$, with $\abs{x - x'} < \varepsilon$, $\abs{f(x) - f(x')} \leq  \delta_{\kappa; [a,b]}(\varepsilon)$. Consequently,
\[
\abs{\langle \mu^*, f \rangle - \langle \mu, f\star \rho\rangle} \leq  \delta_{\kappa; [a,b]}(\varepsilon)\sup_{x \in [a, b]}\abs{f(x)},
\]
and thus the error functions (lemma \ref{lemma:main_error_bound_mu}) for $\mu$ defined in Eq.~\ref{eq:eq_mu_st} are $\Delta_{\mu; [a, b]}^0 = \delta_{\kappa; [a, b]}, \Delta_{\mu; [a, b]}^1= 0$. As a specific example, we consider coupling functions which in frequency-domain are expressible as sum of lorentzians i.e.~
\begin{align*}
&\abs{\hat{v}(\omega)}^2 = \sum_{i = 1}^M \frac{\alpha_i}{(\omega - \omega_i)^2 + \gamma_i^2} \ \forall \omega\in \mathbb{R}, \text{ or  } \noindent\\
&\kappa(t) = \sum_{j = 1}^M \frac{\alpha_j}{2\gamma_j} e^{-\gamma_j \abs{t}} e^{-i\omega_j t}
\end{align*}
for some $\{\alpha_i \in \mathbb{R}_{> 0}\}_{i \in \{1, 2 \dots M\}}, \{\omega_i \in \mathbb{R}\}_{i \in \{1, 2 \dots M\}}$ and $\{\gamma_i \in \mathbb{R}_{> 0}\}_{i\in \{1, 2 \dots M\}}$. Such coupling functions arise commonly in modelling resonant light-matter interactions in quantum optics \cite{}. For this model, since $\kappa$ is differentiable almost everywhere and $\norm{\kappa'}_{L^\infty} \leq \sum_{j = 1}^M \alpha_j \sqrt{\gamma_j^2 + \omega_j^2} / 2\gamma_j$ and hence $\delta_{\kappa; [a, b]}(\varepsilon) = \varepsilon\sum_{j = 1}^M \alpha_j \sqrt{\gamma_j^2 + \omega_j^2} / 2\gamma_j$.
\end{example}
\begin{example}[Delta trains] This class of coupling functions arise frequently in models studying quantum systems with time-delay and feedback. Consider the coupling function specified by radon measure $\mu$
\[
 \mu \cong \sum_{i =1}^M \alpha_i \delta(x - x_i)
\]
for some $\{x_i \in \mathbb{R}\}_{i \in \{1, 2 \dots M\}}, \{\alpha_i \in \mathbb{C}\}_{i \in \{1, 2 \dots M\}}$ with $x_1 < x_2 \dots <x_M$. Furthermore, its extension to $\textnormal{PWC}^1(\mathbb{R})$, $\mu^*$ (Eq.~\ref{eq:mu_star}), is given by
\[
\langle \mu^*, f\rangle = \sum_{i = 1}^M  \frac{\alpha_i}{2}\bigg(\lim_{x \to x_i^+} f(x) + \lim_{x\to x_i^-} f(x)\bigg).
\]
The error functions (lemma \ref{lemma:main_error_bound_mu}) for $\mu$ can be chosen to be
\begin{align*}
&\Delta^0_{\mu, [a, b]}(\varepsilon) = \sum_{\substack{i \in \{1, 2 \dots M\} | \\  y_i \in [a - \varepsilon, a)\cup (b, b+ \varepsilon]}} \abs{\alpha_i} + \sum_{\substack{i \in \{1, 2 \dots M\} | \\  y_i \in (a , a + \varepsilon]\cup (b - \varepsilon, b]}} 2 \abs{\alpha_i}, \nonumber\\
&\Delta^1_{\mu, [a, b]}(\varepsilon) = \varepsilon\bigg( \sum_{\substack{i \in \{1, 2 \dots M\} | \\ y_i \in (a + \varepsilon, b - \varepsilon)}}  \abs{\alpha_i}+ \frac{1}{2} \sum_{\substack{i \in \{1, 2 \dots M\} | \\ y_i \in \{a, b\}} }\abs{\alpha_i} \bigg).
\end{align*}
We refer the reader to the proof of lemma \ref{lemma:main_error_bound_mu} (appendix \ref{app:distributional_mem_ker}) for a derivation of these error functions in a more general setting of a delta train with a countably finite number of delta functions.
\end{example}

\begin{example} [Complex gaussian] Consider the coupling function specified by the radon measure $\mu \in \mathcal{M}(\mathbb{R})$
\[
\langle \mu,  f\rangle = \sum_{j =1}^M c_j \int_{-\infty}^\infty e^{ik_j x^2} f(x) dx \ \forall \ f \in \textnormal{C}_c^0(\mathbb{R}),
\]
where $k_j \in \mathbb{R}, c_j \in \mathbb{C}$ for $j \in \{1, 2 \dots M\}$. Such a coupling function arises frequently in the study of quantum optical systems where the bath is a wire or channel with group velocity dispersion. Furthermore, its extension to $\textnormal{PWC}^1(\mathbb{R})$, $\mu^*$ defined in Eq.~\ref{eq:mu_star}, is given by
\[
\langle \mu^*,  f\rangle = \sum_{j =1}^M c_j \int_{-\infty}^\infty e^{ik_j x^2} f(x) dx \ \forall \ f \in \textnormal{C}_c^0(\mathbb{R}),
\]
independent of the mollifier $\rho$. The error functions (lemma \ref{lemma:main_error_bound_mu}) for $\mu$ can be chosen to be
\begin{align*}
&\Delta^0_{\mu, [a, b]}(\varepsilon) =\varepsilon \sum_{j = 1}^M \abs{c_j k_j} \max\bigg(\abs{\frac{3b - a}{2}}, \abs{\frac{3a - b}{2}}\bigg) \ \text{ and } \nonumber\\
&\Delta^1_{\mu, [a, b]}(\varepsilon) = 0.
\end{align*}
\end{example}

\subsubsection{Analyzing the quantum dynamics}

We now turn our attention to analyzing non-Markovian quantum dynamics --- the basic data needed to specify a non-Markovian model is repeated in the definition below.

\begin{repdefinition}{def:model} [Non-Markovian model] A non-Markovian open system model for a quantum system with Hilbert space $\mathcal{H}_S$ is specified by
\begin{enumerate}
\item[(a)] A time-dependent system Hamiltonian $H_S(t) \in \mathfrak{L}(\mathcal{H}_S)$ which is Hermitian, norm continuous and differentiable in $t$,
\item[(b)] A set of distributional coupling functions $\{(\mu_i, \varphi_i) \}_{i \in \{1, 2 \dots M\}}$, where each coupling function is specified by a tempered radon measure (for its magnitude) and a phase.
\item[(c)] A set of bounded jump operators $\{L_i \in \mathfrak{L}(\mathcal{H}_S)\}_{i \in \{1, 2 \dots M\}}$.
\end{enumerate}
\end{repdefinition}
As we mentioned previously, the regularization of the non-Markovian model amounts to approximating the coupling functions, specified by tempered radon measures, by square integrable functions. We repeat this definition below.
\begin{repdefinition}{def:regularization}[Regularization]
For $\varepsilon > 0$ and given a symmetric mollifier $\rho \in \textnormal{C}_c^\infty(\mathbb{R})$, an $\varepsilon, \rho-$regularization of a distributional coupling function $(\mu, \varphi)$ is a square integrable function $v_\varepsilon \in L^2(\mathbb{R})$ whose fourier transform $\hat{v}_\varepsilon \in L^2(\mathbb{R})$ is given by
\[
\hat{v}_\varepsilon(\omega) = \sqrt{\hat{\mu}(\omega)} \hat{\rho}({\omega}{\varepsilon}) e^{i\varphi(\omega)} \ \forall \omega \in \mathbb{R}.
\]
\end{repdefinition}

It is easily seen that $\hat{v} \in L^2(\mathbb{R})$, since $\hat{\mu}(\omega)$ has atmost polynomial growth in $\omega$ due to $\mu$ being a tempered distribution, and $\hat{\rho} \in \mathcal{S}(\mathbb{R})$ decays faster than any polynomial since it is the fourier transform of a smooth compact function. For square integrable coupling functions, proposition \ref{thm:schr_sq_int} guarantees the existence of the solution to the Schroedinger's equation --- we can then study whether the solution to the Schroedinger's equation converges as $\varepsilon \to 0$ and define the limit as the dynamics associated with the non-Markovian model.

In order to proceed further with this convergence study, we need to impose some suitable restrictions on the initial environment state since in general it can have amplitudes at arbitrarily high energies and would thus incur a large error upon regularization. We restrict ourselves to a dense subspace of the environment Hilbert space in which every state has particles only in modes whose high frequency amplitudes decrease faster than any polynomial in $\omega$. This is formalized by constructing this dense subspace using modal functions from the Schwartz spaces, $\mathcal{S}(\mathbb{R})$, which are functions that decay superpolynomially at large values of its argument --- the subspace is provided in the definition below.
\begin{definition}
For $M \in \mathbb{Z}_{\geq 1}$, define $\textnormal{F}^M_{\infty, \mathcal{S}} \subset \textnormal{F}^M_{\infty}[L^2(\mathbb{R})]$ as the set of vectors $\ket{\Phi}$ such that
\[
\forall n \in \mathbb{Z}_{\geq 1}: \Pi_n \ket{\Phi} \in \textnormal{span}\bigg(\bigg\{u^{\otimes n} \bigg|  u = \bigoplus_{\alpha = 1}^M u_\alpha, u_\alpha \in \mathcal{S}(\mathbb{R})   \bigg\} \bigg).
\]
\end{definition}
The following lemma establishes error bounds on the action of the annihilation operator corresponding to a regularized coupling function, when applied on the states from $\textnormal{F}^M_{\infty, \mathcal{S}}$. These simply follow from the rapid decay properties of the mode functions corresponding to these states --- we provide a proof of this lemma in appendix \ref{app:lemma_inc_state_trunc}.
\begin{lemma}
\label{lemma:inc_state_trunc}
Let $(\mu,\varphi)$ be a distributional coupling function. Given two mollifiers $\rho, \sigma \in \textnormal{C}_c^\infty(\mathbb{R})$ and $\varepsilon, \delta > 0$, let $v_\varepsilon$ and $v_\delta \in L^2(\mathbb{R})$ be the $\varepsilon, \rho-$ and $\delta, \sigma-$regularization of $(\mu, \varphi)$ respectively. Let $\mathcal{H}_S$ be Hilbert space, then, 
\begin{enumerate}
\item[(a)]$\forall \ket{\Phi} \in \mathcal{H}_S \otimes \textnormal{F}_{\infty, \mathcal{S}}^M$, $\exists c_{\mu, \ket{\Phi}} > 0$, $\forall \tau \geq 0$, $\forall \alpha \in \{1, 2 \dots M\}, \varepsilon > 0$ such that $\norm{a_{\alpha, \uptau_\tau v_\varepsilon}^- \ket{\Phi}} \leq c_{\mu, \ket{\Phi}}$.
\item[(b)] $\forall \ket{\Phi} \in \mathcal{H}_S \otimes \textnormal{F}_{\infty, \mathcal{S}}^M$, $\exists c_{\mu,\rho, \ket{\Phi}}, d_{\mu, \sigma, \ket{\Phi}} > 0$, $\forall \tau \geq 0, \forall \alpha \in \{1, 2 \dots M\}$, $\varepsilon, \delta > 0$ such that $\norm{a_{\alpha, \uptau_\tau(v_\varepsilon - v_\delta)}^- \ket{\Phi}} \leq c_{\mu,\rho, \ket{\Phi}}\varepsilon + c_{\mu, \sigma, \ket{\Phi}} \delta$.
\end{enumerate}
\end{lemma}
Our next lemma, which builds on lemma \ref{lemma:pert_theory_bound}, is key to showing convergence of the non-Markovian dynamics on removing regularization. 
\begin{lemma}
\label{lemma:gfunc_error_bound}
Consider a non-Markovian model specified by a system Hamiltonian $H_S(t)$, $M$ jump operators $\{L_\alpha \}_{\alpha \in \{1, 2 \dots M\}}$ and coupling functions, $\{(\mu_\alpha, \varphi_\alpha)\}_{\alpha \in \{1, 2 \dots M\}}$. Given two mollifiers $\rho, \sigma \in \textnormal{C}_c^\infty(\mathbb{R})$ and $\varepsilon, \delta \in (0, 1/2)$, consider two non-Markovian models with the same system Hamiltonian and jump operators, but with square integrable coupling functions given by $v_\varepsilon = \{v_{\alpha, \varepsilon} \in L^2(\mathbb{R})\}_{\alpha \in \{1, 2 \dots M\}}$ and $v_\delta = \{v_{\alpha, \delta} \in L^2(\mathbb{R})\}_{\alpha \in \{1, 2 \dots M\}}$, where $v_{\alpha, \varepsilon}$ and $v_{\alpha, \delta}$ are $\varepsilon, \rho-$ and $\delta, \sigma-$regularizations of $(\mu_\alpha, \varphi)$ respectively. Given an initial state $\ket{\Psi_0} \in \mathcal{H}_S\otimes \textnormal{F}_{\infty, \mathcal{S}}^M[L^2(\mathbb{R})]$, then the errors $\mathcal{E}^{v_\delta, v_\varepsilon}_{\alpha, \ket{\Psi_0}}(t)$ and $\mathcal{D}^{v_\delta, v_\varepsilon}_{\alpha, \ket{\Psi_0}}(t)$ defined in lemma \ref{lemma:pert_theory_bound} satisfy the estimates
\begin{enumerate}
\item[(a)]For all $\alpha \in \{1, 2 \dots M\}$ and $t > 0$,
\begin{align*}
\mathcal{E}_{\alpha, \ket{\Psi_0}}^{v_\delta, v_\varepsilon}(t) \leq 4\norm{L_\alpha}^2\textnormal{TV}_{[-1, t + 1]}(\mu_\alpha).
\end{align*}
\item[(b)]For all $\alpha \in \{1, 2 \dots M\}$ and $t > 0$,
\begin{widetext}
\begin{align*}
&\mathcal{E}_{\alpha, \ket{\Psi_0}}^{v_\delta, v_\varepsilon}(t)  \leq 2\bigg(2\Delta^1_{\mu_\alpha; [0, t]}(\varepsilon + \delta) + \Delta^1_{\mu_\alpha; [0, t]}(2\varepsilon) + \Delta^1_{\mu_\alpha; [0, t]}(2\delta)\bigg)\times \nonumber\\
&\qquad   \bigg( \norm{L_\alpha}\sup_{s\in[0, t]}\norm{[H_S(s), L_\alpha]}  + 4\norm{L_\alpha}^2 \sum_{\alpha' = 1}^M \norm{L_\alpha'} c_{\mu_{\alpha'}, \ket{\Psi_0}} + 6\norm{L_\alpha}^2 \sum_{\alpha' = 1}^M \norm{L_{\alpha'}}^2 \textnormal{TV}_{[-1, t + 1]}(\mu_{\alpha'}).\bigg) + \nonumber\\
&\qquad 2\bigg(2\Delta^0_{\mu_\alpha; [0, t]}(\varepsilon + \delta) + \Delta^0_{\mu_\alpha; [0, t]}(2\varepsilon) + \Delta^0_{\mu_\alpha; [0, t]}(2\delta)\bigg)\norm{L_\alpha}^2,
\end{align*}
\end{widetext}
where $\Delta^0_{\mu_\alpha; [-t, 0]}, \Delta^1_{\mu_\alpha; [-t, 0]}$ are the error functions corresponding to $\mu_\alpha$ $($defined in lemma \ref{lemma:main_error_bound_mu}$)$ and $c_{\mu_\alpha, \ket{\Psi_0}}$ is the constant introduced in lemma \ref{lemma:inc_state_trunc}(a).
\item[(c)] For all $\alpha \in \{1, 2 \dots M\}$ and $t > 0$,
\begin{align*}
\mathcal{D}_{\alpha, \ket{\Psi_0}}^{v_\delta, v_\varepsilon}(t) \leq 4\norm{L_\alpha} \big(c_{\mu_\alpha, \rho, \ket{\Psi_0}}\varepsilon + c_{\mu_\alpha, \sigma, \ket{\Psi_0}}\delta\big),
\end{align*}
where $c_{\mu_\alpha, \rho, \ket{\Psi_0}}, c_{\mu_\alpha, \sigma, \ket{\Psi_0}}$ are constants introduced in lemma \ref{lemma:inc_state_trunc}(b).
\end{enumerate}
\end{lemma}
\noindent\emph{Proof}: For this proof, it is convenient to note that for $\alpha \in \{1, 2 \dots M\}$ and any $s, \in [0, t]$,
\[
\langle \uptau_{\tau} v_{\alpha, \delta}, \uptau_{s} v_{\alpha, \varepsilon}\rangle =2\pi \int_{\mathbb{R}}\hat{\mu}_{\alpha}(\omega) \hat{\rho}(\varepsilon \omega)\hat{\sigma}^*(\delta \omega) e^{-i\omega(s - t)} d\omega.
\]
Since if $\varepsilon, \delta \in (0, 1/2)$, $\text{supp}(\rho_\varepsilon \star \rho_\varepsilon), \text{supp}(\sigma_\delta \star \sigma_\delta), \text{supp}(\sigma_\delta \star \rho_\varepsilon) \subseteq [-1, 1]$. Consequently, $\forall f \in \textnormal{C}^0(\mathbb{R})$ and $t_1, t_2 \in (0, t]$, 
\begin{subequations}\label{eq:tv_norm_estimate}
\begin{align}
&\abs{\int_{t_1}^{t_2} \langle \uptau_t v_{\alpha, \delta}, \uptau_s v_{\alpha, \varepsilon} \rangle f(s) ds} \nonumber\\
&\qquad= \abs{\langle \mu_\alpha, \sigma_\delta \star \rho_\varepsilon \star \uptau_t\big(f \cdot \mathcal{I}_{(t_1, t_2]}\big)\rangle}\nonumber\\
& \qquad \leq \textnormal{TV}_{[-1, t + 1]}(\mu_\alpha) \sup_{s \in [0, t]} \abs{f(s)}.
\end{align}
Similarly,
\begin{align}
&\abs{\int_{t_1}^{t_2} \langle \uptau_t v_{\alpha, \varepsilon},\uptau_s v_{\alpha, \delta} \rangle f(s) ds}, \abs{\int_{t_1}^{t_2} \langle \uptau_t v_{\alpha, \varepsilon}, \uptau_s v_{\alpha, \varepsilon} \rangle f(s) ds},\nonumber\\
& \abs{\int_{t_1}^{t_2} \langle \uptau_t v_{\alpha, \delta}, \uptau_s v_{\alpha, \delta} \rangle f(s) ds} \leq \textnormal{TV}_{[-1, t + 1]}(\mu_\alpha) \sup_{s \in [0, t]} \abs{f(s)}.
\end{align}
\end{subequations}

\noindent\emph{(a)} We note that
\begin{align*}
&\abs{\int_0^t \langle \uptau_t (v_{\alpha, \delta} - v_{\alpha, \varepsilon}), \uptau_s v_{\alpha, \varepsilon}\rangle G^{v_\varepsilon, v_\varepsilon, v_\delta}_{L_\alpha, L_\alpha^\dagger}(s, t) ds}  \nonumber\\
&\qquad  \leq \abs{\int_0^t \langle \uptau_t v_{\alpha, \delta}, \uptau_s v_{\alpha, \varepsilon}\rangle G^{v_\varepsilon, v_\varepsilon, v_\delta}(s, \tau) ds} + \nonumber\\
&\qquad \quad \abs{\int_0^t \langle \uptau_t v_{\alpha, \delta}, \uptau_s v_{\alpha, \varepsilon}\rangle G^{v_\varepsilon, v_\varepsilon, v_\delta}(s, \tau) ds}.
\end{align*}
Since $\forall s \in [0, t] : \abs{G^{v_\varepsilon, v_\varepsilon, v_\delta}_{L_\alpha, L_\alpha^\dagger; \ket{\Psi_0}, \ket{\Psi_0}}(s, \tau)},  \leq \norm{L_\alpha}^2$, and using Eq.~\ref{eq:tv_norm_estimate}, we obtain that
\begin{align*}
&\abs{\int_0^t \langle \uptau_t (v_{\alpha, \delta} - v_{\alpha, \varepsilon}), \uptau_s v_{\alpha, \varepsilon}\rangle G^{v_\varepsilon, v_\varepsilon, v_\delta}_{L_\alpha, L_\alpha^\dagger}(s, t) ds} \leq \nonumber\\
&\qquad \qquad \qquad \qquad \qquad \qquad 2 \norm{L_\alpha}^2\textnormal{TV}_{[-1, t + 1]}(\mu_\alpha).
\end{align*}
Similarly, we can obtain that
\begin{align*}
&\abs{\int_0^t \langle \uptau_s v_{\alpha, \delta}, \uptau_t(v_{\alpha, \delta} - v_{\alpha, \varepsilon}) \rangle G_{L_\alpha, L_\alpha^\dagger; \ket{\Psi_0}, \ket{\Psi_0}}^{v_\varepsilon, v_\delta, v_\delta}} \leq \nonumber\\
&\qquad \qquad \qquad \qquad \qquad \qquad  2 \norm{L_\alpha}^2\textnormal{TV}_{[-1, t + 1]}(\mu_\alpha).
\end{align*}
Combining these two estimates, we obtain the part \emph{(a)} of the lemma statement.

\noindent\emph{(b)} We begin by noting that 
\begin{align*}
&\langle \uptau_t (v_{\alpha, \delta} - v_{\alpha, \varepsilon}), \uptau_s v_{\alpha, \varepsilon}\rangle = \nonumber\\
&\qquad 2\pi\int_{-\infty}^\infty \hat{\mu}_\alpha(\omega) \hat{\rho}(\varepsilon \omega) \big(\hat{\rho}^*(\varepsilon \omega) - \hat{\sigma}^*(\delta \omega) \big) e^{-i\omega(s - t)}d\omega,
\end{align*}
and therefore,
\begin{align*}
&\int_0^t \langle \uptau_t (v_{\alpha, \delta} - v_{\alpha, \varepsilon}), \uptau_s v_{\alpha, \varepsilon }\rangle G_{L_\alpha, L_\alpha^\dagger; \ket{\Psi_0}, \ket{\Psi_0}}^{v_\varepsilon, v_\varepsilon, u_\delta}(s, t) ds =\nonumber\\
&\qquad\big\langle\mu_\alpha, \rho_\varepsilon \star \rho_\varepsilon \star g_{\alpha, t}^{\varepsilon, \delta}  - \rho_\varepsilon \star \sigma_\delta \star g^{\alpha, t}_{\varepsilon, \delta} \big\rangle,
\end{align*}
where
\[
g^{\varepsilon, \delta}_{\alpha, t} =  \uptau_t \bigg(G_{L_\alpha, L_\alpha^\dagger; \ket{\Psi_0}, \ket{\Psi_0}}^{v_\varepsilon, v_\varepsilon, u_\delta}(\cdot, t)\cdot \mathcal{I}_{[0, t]}\bigg).
\]
We note that $g^{\varepsilon, \delta}_{\alpha, t}$ is a continuous and differentiable function when restricted to $[-t, 0]$ and hence $ \in \textnormal{PWC}^1(\mathbb{R})$. Consequently,
\begin{align}\label{eq:error_g}
&\abs{\big\langle\mu_\alpha, \rho_\varepsilon \star \rho_\varepsilon \star g_{\alpha, t}^{\varepsilon, \delta}  - \rho_\varepsilon \star \sigma_\delta \star g^{\alpha, t}_{\varepsilon, \delta} \big\rangle}\nonumber\\
& \leq \abs{\big\langle\mu_\alpha^*, g_{\alpha, t}^{\varepsilon, \delta} \big\rangle - \big\langle \mu_\alpha, \rho_\varepsilon \star \rho_\varepsilon \star g_{\alpha, t}^{\varepsilon, \delta}\big\rangle} +\nonumber\\
&\qquad \qquad \abs{\big\langle \mu_\alpha^*, g_{\alpha, t}^{\varepsilon, \delta} \big\rangle-\big\langle\mu_\alpha,  \rho_\varepsilon \star \sigma_\delta \star g^{\alpha, t}_{\varepsilon, \delta} \big\rangle} \nonumber \\
& \leq \bigg(\Delta_{\mu_\alpha; [-t, 0]}^0(2\varepsilon) + \Delta_{\mu_\alpha; [-t, 0]}^0(\delta + \varepsilon)\bigg) \sup_{s \in [0, t]} \abs{g_{\alpha, t}^{\varepsilon, \delta}(s)} +  \nonumber\\
&\qquad \bigg(\Delta_{\mu_\alpha; [-t, 0]}^1(2\varepsilon) + \Delta_{\mu_\alpha; [-t, 0]}^1(\delta + \varepsilon)\bigg) \sup_{s \in [0, t]} \abs{\partial_s g_{\alpha, t}^{\varepsilon, \delta}(s)}
\end{align}
Furthermore, note that 
\begin{align}\label{eq:sup_norm_g}
\sup_{s \in [-t, 0]} \abs{g_{\alpha, t}^{\varepsilon, \delta}(s)} \leq \norm{L_\alpha}^2,
\end{align}
We next provide a bound on the derivative ($\sup_{s \in [-t, 0]} |\partial_s g^{\varepsilon, \delta}_{\alpha, t}(s)|$) which is uniform in $\varepsilon, \delta$. An application of lemma \ref{lemma:gfunc_deriv}, yields
\begin{align*}
&\abs{\partial_s g^{\varepsilon, \delta}_{\alpha, t}(s)}\leq \nonumber\\
&\abs{G^{v_\varepsilon, v_\varepsilon, v_\delta}_{[H_S(s + t), L_\alpha], L_\alpha^\dagger; \ket{\Psi_0}, \ket{\Psi_0}}(s+t, t)} + \nonumber\\
&  \sum_{\alpha' = 1}^M \bigg(\abs{G^{v_\varepsilon, v_\varepsilon, v_\delta}_{[L_{\alpha'}^\dagger, L_\alpha], L_\alpha^\dagger; \ket{\Psi_0}, a^{-}_{\alpha', \uptau_{(s + t)}v_{\alpha}}\ket{\Psi_0}}(s + t, t)} + \nonumber\\
&\qquad \quad \abs{G^{v_\varepsilon, v_\delta, v_\delta}_{[L_{\alpha'}, L_\alpha], L_\alpha^\dagger; \ket{\Psi_0}, a^{-}_{\alpha', \uptau_{(s + t)}v_{\alpha}}\ket{\Psi_0}}(s + t, t)}\bigg) +   \nonumber\\
&\sum_{\alpha' = 1}^M \bigg(\bigg|\int_0^{s + \tau}\langle \uptau_{s + \tau} v_{\alpha', \varepsilon}, \uptau_{s'} v_{\alpha', \varepsilon}\rangle \times \nonumber\\
& \qquad \qquad \qquad G^{v_\varepsilon, v_\varepsilon, v_\varepsilon, v_\delta}_{L_{\alpha'}, [L_{\alpha'}^\dagger, L_\alpha], L_{\alpha}^\dagger; \ket{\Psi_0}, \ket{\Psi_0}}(s', s + \tau, \tau)ds'\bigg| + \nonumber\\
&\qquad \quad  \bigg|\int_0^\tau \langle \uptau_{s'} v_{\alpha', \delta}, \uptau_{s + \tau} v_{\alpha', \varepsilon}\rangle \times \nonumber\\
&\qquad \qquad \qquad G_{[L_{\alpha'}, L_\alpha], L_{\alpha}^\dagger, L_{\alpha'}^\dagger; \ket{\Psi_0}, \ket{\Psi_0}}^{v_\varepsilon, v_\varepsilon, v_\delta, v_\delta}(s + \tau, \tau, s')ds' \bigg|+\nonumber \\
&\qquad \quad \bigg|\int_{s + \tau}^\tau \langle \uptau_{s'} v_{\alpha', \delta}, \uptau_{s + \tau} v_{\alpha', \varepsilon}\rangle \times \nonumber\\
&\qquad \qquad \qquad G_{[L_{\alpha'}, L_\alpha], L_{\alpha'}^\dagger, L_{\alpha}^\dagger;\ket{\Psi_0}, \ket{\Psi_0}}^{v_\varepsilon, v_\varepsilon, v_\delta, v_\delta}(s + \tau, s', \tau)ds'\bigg|\bigg).
\end{align*}
Using Eq.~\ref{eq:tv_norm_estimate}, we thus obtain that
\begin{align*}
&\sup_{s\in(-t, 0)} \abs{\partial_s g_{\alpha, t}^{\varepsilon, \delta}(s)} \leq \norm{L_\alpha} \sup_{s \in [0, t]}\norm{[H_S(s), L_\alpha]} + \nonumber\\
&\qquad 4\norm{L_\alpha}^2 \sum_{\alpha' = 1} ^M \norm{L_{\alpha'}}\sup_{s \in [0, t]} \norm{a^-_{\alpha', \uptau_s v_{\alpha'}} \ket{\Psi_0}} + \nonumber\\
&\qquad 6\norm{L_\alpha}^2 \sum_{\alpha' = 1}^M \norm{L_\alpha'}^2 \textnormal{TV}_{[-1, t + 1]}(\mu_\alpha)
\end{align*}
From lemma \ref{lemma:inc_state_trunc}(a), it follows that $\forall \alpha' \in \{1, 2 \dots M\}, s \in [0, t]$, $\norm{a^-_{\alpha', \uptau_s v_{\alpha'}}\ket{\Psi_0}}\leq c_{\mu_{\alpha'}, \ket{\Psi_0}}$ and therefore, we obtain that
\begin{align}\label{eq:sup_norm_g_prime}
&\sup_{s\in(-t, 0)} \abs{\partial_s g_{\alpha, t}^{\varepsilon, \delta}(s) } \leq \sup_{s\in[0, \tau]} \norm{[H_S(s), L_\alpha]} \norm{L_\alpha} + \nonumber\\
&\qquad 4\norm{L_\alpha}^2 \sum_{\alpha' = 1}^M \norm{L_\alpha'} c_{\mu_{\alpha'}, \ket{\Psi_0}} +\nonumber\\
&\qquad 6\norm{L_\alpha}^2 \sum_{\alpha' = 1}^M \norm{L_{\alpha'}}^2 \textnormal{TV}_{[-1, t + 1]}(\mu_{\alpha'}).
\end{align}
From Eqs.~\ref{eq:error_g}, \ref{eq:sup_norm_g} and \ref{eq:sup_norm_g_prime}, we obtain that
\begin{align*}
&\abs{\int_0^t \langle \uptau_t (v_{\alpha, \delta} - v_{\alpha, \varepsilon}), \uptau_s v_{\alpha, \varepsilon }\rangle G_{L_\alpha, L_\alpha^\dagger; \ket{\Psi_0}, \ket{\Psi_0}}^{v_\varepsilon, v_\varepsilon, u_\delta}(s, t) ds } \leq \nonumber\\
&\qquad \bigg(\Delta_{\mu_\alpha; [-t, 0]}^0(2\varepsilon) + \Delta_{\mu_\alpha; [-t, 0]}^0(\delta + \varepsilon)\bigg) \norm{L_\alpha}^2 +\nonumber\\
&\qquad \bigg(\Delta_{\mu_\alpha; [-t, 0]}^1(2\varepsilon) + \Delta_{\mu_\alpha; [-t, 0]}^1(\delta + \varepsilon)\bigg)\times \\
&\qquad \bigg(\sup_{s\in[0, \tau]} \norm{[H_S(s), L_\alpha]} \norm{L_\alpha} + \nonumber\\
&\qquad \qquad 4\norm{L_\alpha}^2 \sum_{\alpha' = 1}^M \norm{L_{\alpha'}} c_{\mu_{\alpha'}, \ket{\Psi_0}} + \nonumber\\
&\qquad \qquad 6\norm{L_\alpha}^2 \sum_{\alpha' = 1}^M \norm{L_{\alpha'}}^2 \textnormal{TV}_{[-1, t + 1]}(\mu_{\alpha'})\bigg).
\end{align*}
A similar bound can be obtained for the term $\int_0^t \langle  \uptau_s v_{\alpha, \delta }, \uptau_t (v_{\alpha, \delta} - v_{\alpha, \varepsilon})\rangle G_{L_\alpha, L_\alpha^\dagger; \ket{\Psi_0}, \ket{\Psi_0}}^{v_\delta, v_\delta, v_\varepsilon}(t, s) ds $  in the expression for $\mathcal{E}^{v_\delta, v_\varepsilon}_{\alpha, \ket{\Psi_0}}(t)$ provided in lemma \ref{lemma:pert_theory_bound}. Combining these estimates, we obtain part (d) of the lemma statement.
%\begin{align*}
%&\abs{\int_0^t \langle  \uptau_s v_{\alpha, \delta }, \uptau_t (v_{\alpha, \delta} - v_{\alpha, \varepsilon})\rangle G_{L_\alpha, L_\alpha^\dagger; \ket{\Psi_0}, \ket{\Psi_0}}^{v_\delta, v_\delta, v_\varepsilon}(t, s) ds } \leq \nonumber\\
%& \bigg(\Delta^0_{\mu_\alpha; [-t, 0]}(2\delta) + \Delta^0_{\mu_\alpha; [-t, 0]}(\delta + \varepsilon)\bigg) \norm{L_\alpha}^2  +\bigg(\Delta_{\mu_\alpha; [-t, 0]}^1(2\delta) + \nonumber\\
%&\qquad  \qquad 
%\Delta_{\mu_\alpha; [-t, 0]}^1(\delta + \varepsilon)\bigg) \bigg(\sup_{s\in[0, \tau]} \norm{[H_S(s), L_\alpha]} \norm{L_\alpha} + 4\norm{L_\alpha}^2 \sum_{\alpha' = 1}^M \norm{L_{\alpha'}} c_{\mu_{\alpha'}, \ket{\Psi_0}} + 6\norm{L_\alpha}^2 \sum_{\alpha' = 1}^M \norm{L_{\alpha'}}^2 \textnormal{TV}_{[-1, t + 1]}(\mu_{\alpha'})\bigg).
%\end{align*}

\noindent (c) This follows from a direct application of lemma \ref{lemma:inc_state_trunc}.
\begin{reptheorem}{theorem:non_mkv_exis}[Formal, Non-markovian dynamics] Consider a non-Markovian model specified by a system Hamiltonian $H_S(t)$, $M$ jump operators $\{L_\alpha \}_{\alpha \in \{1, 2 \dots M\}}$ and distributional coupling functions, $\{(\mu_\alpha, \varphi_\alpha)\}_{\alpha \in \{1, 2 \dots M\}}$. Construct a square-integral non-Markovian model with the same system Hamiltonian and jump operators, but with coupling functions $v_\varepsilon := \{v_{\alpha, \varepsilon} \in L^2(\mathbb{R})\}_{\alpha \in \{1, 2 \dots M\}}$, where for $\alpha \in \{1, 2 \dots M\}$, $v_{\alpha, \varepsilon}$ is an $\varepsilon, \rho-$regularization of $(\mu_\alpha, \varphi_\alpha)$ for a symmetric mollifier $\rho \in \textnormal{C}_c^\infty(\mathbb{R})$, $\varepsilon > 0$ and let $U_{v_\varepsilon}(\cdot, \cdot)$ be its propagator. Then, for $t > 0$, $U(t): \textnormal{F}^M_{\infty, \mathcal{S}}\otimes \mathcal{H}_S \to \mathcal{H}$ defined via $U(t)\ket{\Psi_0} = \lim_{\varepsilon \to 0} U_{v_\varepsilon}(t, 0) \ket{\Psi_0}$ exists, is an isometry and is independent of the choice of mollifier $\rho$.
\end{reptheorem}
\noindent\emph{Proof}: For simplicity, we will assume that $\varepsilon, \delta \in (0, 1)$. Consider two symmetric mollifiers $\rho, \sigma \in \textnormal{C}_c^\infty(\mathbb{R})$ --- let $v_\varepsilon:= \{v_{\alpha, \varepsilon} \in L^2(\mathbb{R})\}_{\alpha \in \{1, 2 \dots M\}}$ and $v_\delta:= \{v_{\alpha, \delta} \in L^2(\mathbb{R})\}_{\alpha \in \{1, 2 \dots M\}}$ be the $\varepsilon, \rho-$ and $\delta, \sigma-$regularizations of the distributional coupling functions. For $\ket{\Psi_0} \in \textnormal{F}^M_{\infty, \mathcal{S}}\otimes \mathcal{H}_S $, let $\ket{\Psi_{v_\varepsilon}(t)} = U_{v_\varepsilon}(t, 0) \ket{\Psi_0}$ and $\ket{\Psi_{v_\delta}(t)} = U_{v_\delta}(t, 0) \ket{\Psi_0}$, where $U_{v_\varepsilon}(\cdot, \cdot), U_{v_\delta}(\cdot, \cdot)$ are the propagators corresponding to the two models. We note that from lemma \ref{lemma:gfunc_error_bound}(c) that $\forall \alpha \in \{1, 2 \dots M\}$ and $\tau \in [0, t]$
\begin{align*}
&\lim_{\varepsilon, \delta \to 0}  \mathcal{D}^{v_\delta, v_\varepsilon}_{\alpha, \ket{\Psi_0}}(\tau) = 0 \ \text{ and } \\
&\quad \mathcal{D}^{v_\delta, v_\varepsilon}_{\alpha, \ket{\Psi_0}}(\tau) \leq 4\norm{L_\alpha} \big(c_{\mu_\alpha, \rho, \ket{\Psi_0}} + c_{\mu_\alpha, \sigma, \ket{\Psi_0}}\big).
\end{align*}
From the dominated convergence theorem, we then obtain that
\[
\lim_{\varepsilon, \delta \to 0}  \int_{0}^t \mathcal{D}^{v_\delta, v_\varepsilon}_{\alpha, \ket{\Psi_0}}(\tau) d\tau = \int_0^t \lim_{\varepsilon, \delta \to 0} \mathcal{D}^{v_\delta, v_\varepsilon}_{\alpha, \ket{\Psi_0}}(\tau) d\tau = 0.
\]
Similarly, we note from lemma \ref{lemma:gfunc_error_bound}(b) that $\forall \alpha \in \{1, 2 \dots M\}$ and $\tau \in [0, t]$
\[
\lim_{\varepsilon, \delta \to 0}  \mathcal{E}^{v_\delta, v_\varepsilon}_{\alpha, \ket{\Psi_0}}(\tau) = 0.
\]
From lemma \ref{lemma:gfunc_error_bound}(a), we obtain that $\forall \alpha \in \{1, 2 \dots M\}$ and $\tau \in [0, t]$
\begin{align}
&\mathcal{E}^{v_\delta, v_\varepsilon}_{\alpha, \ket{\Psi_0}}(\tau) \leq 4\norm{L_\alpha}^2 \textnormal{TV}_{[-1, \tau + 1]}(\mu_\alpha)\nonumber\\
 &\qquad \qquad \leq 4\norm{L_\alpha}^2 \textnormal{TV}_{[-1, t + 1]}(\mu_\alpha).
\end{align}
Hence, again by dominated convergence theorem, we obtain that
\[
\lim_{\varepsilon, \delta \to 0}   \int_0^t \mathcal{E}^{v_\delta, v_\varepsilon}_{\alpha, \ket{\Psi_0}}(\tau) d\tau =  \int_0^t \lim_{\varepsilon, \delta \to 0}  \mathcal{E}^{v_\delta, v_\varepsilon}_{\alpha, \ket{\Psi_0}}(\tau) d\tau = 0.
\]
We thus obtain from lemma \ref{lemma:pert_theory_bound} that
\begin{align}\label{eq:error_between_two_diff_seq}
\lim_{\varepsilon, \delta\to 0} \norm{\ket{\Psi_{v_\varepsilon}(t)} - \ket{\Psi_{v_\delta}(t)}} = 0,
\end{align}
for all symmetric mollifiers $\rho, \sigma$. From this condition, using $\rho = \sigma$, we obtain that $\lim_{\varepsilon \to 0} \ket{\Psi_{v_\varepsilon}(t)}$ exists. Furthermore, $\norm{\lim_{\varepsilon \to 0} \ket{\Psi_{v_\varepsilon}(t)}} = \lim_{\varepsilon \to 0}\norm{ \ket{\Psi_{v_\varepsilon}(t)}} = \norm{\ket{\Psi_0}}$, and hence the operator mapping $\ket{\Psi_0}$ to $\lim_{\varepsilon \to 0} \ket{\Psi_{v_\varepsilon}(t)}$ is an isometry. Furthermore, since the limit exists, Eq.~\ref{eq:error_between_two_diff_seq} additionally implies that the limit is independent of the choice of the mollifier. \hfill \(\square\)

\section{Complexity of non-Markovian dynamics}
\label{sec:complexity}
\subsection{Certifiable Markovian dilations}
\label{subsec:dilation}
In this section, we develop a certifiable dilation of the non-Markovian problem i.e.~we provide a systematic method to approximate the continuum of modes in the environment by a finite number of modes. More specifically, we will construct a `chain dilation' of the non-Markovian model as defined in section \ref{subsec:proof_ideas} and repeated below.
\begin{definition}[Chain dilation] Consider a non-Markovian model specified by a system Hamiltonian $H_S(t)$, coupling functions $\{(\mu_\alpha, \varphi_\alpha)\}_{\alpha \in \{1, 2 \dots M\}}$ and jump operators $\{L_\alpha\}_{\alpha \in \{1, 2 \dots M\}}$. A chain dilation, with $N_m$ modes and bandwidth $B$, of this model is described by the following Hamiltonian over the system-environment Hilbert space $(\mathcal{H}_S \otimes \textnormal{Fock}[L^2(\mathbb{R})]):$
\begin{align*}
&H(t) = \nonumber\\
&\qquad H_S(t) + \sum_{\alpha = 1}^M\bigg(g_\alpha a_{\alpha, 1} L_\alpha^\dagger + \sum_{i = 1}^{N_m - 1} t_{\alpha, i} a_{\alpha, i}^\dagger a_{\alpha, i + 1} + \text{h.c.}\bigg),
\end{align*}
where for $\alpha \in \{1, 2 \dots M\}, i \in \{1, 2 \dots N_m\}$,
\begin{enumerate}
\item[(a)] The operator $a_{\alpha, i} = \int_{\mathbb{R}} \varphi_{\alpha, i}(\omega) a_{\alpha, \omega} d\omega$ is the annihilation operator corresponding to the $i^\textnormal{th}$ mode of the $\alpha^\textnormal{th}$ bath described by the orthonormal mode functions $\varphi_{\alpha, i} \in \textnormal{L}^2(\mathbb{R})$ $($i.e. $\langle \varphi_{\alpha, i}, \varphi_{\alpha, i'}\rangle = \delta_{i, i'})$.
\item[(b)] The coupling constants $g_\alpha, t_{\alpha, i}$ are upper bounded by the bandwidth $B$ i.e.~$\abs{g_\alpha}, \abs{t_{\alpha, i}} \leq B$.
\end{enumerate}
\end{definition}
Thus, in a chain dilation of the non-Markovian model, each bath is approximated by a 1D chain of bosonic modes with nearest neighbour interactions, with the first mode coupling to the system. In this subsection, we will provide both the construction of such a dilation, as well as bounds on the error incurred in this dilation in terms of the number of modes in the environment. We remark that the error bounds that we provide grow only polynomially with the time for which the non-Markovian system is evolved as opposed to previously obtained bounds which grow exponentially with time {\cite{trivedi2021convergence, mascherpa2017open}. } Apart from being of independent interest to simulating non-Markovian dynamics, the result of this section is central to the proof of non-Markovian many-body dynamics being simulable on a quantum computer that is provided in the following subsection.

One of the key ingredients in our proof is the previously studied star-to-chain mapping of non-Markovian models --- this can be used to systematically construct a chain dilation when the coupling functions are square integrable and have finite-frequency support (i.e. $v_\alpha(\omega) = 0$ for $\omega$ outside $[-\omega_c, \omega_c]$). The lemma below provides a bound on the error incurred by the star-to-chain transformation --- while this result is known in the literature \cite{woods2015simulating, trivedi2021convergence}, we include a proof of this in appendix \ref{app:chain_appx}. Our proof closely follows Ref.~\cite{trivedi2021convergence}.
\begin{lemma}[Star-to-chain transformation]
\label{lemma:star_to_chain}
Consider a non-Markovian model with system Hamiltonian $H_S(t)$, square-integrable coupling functions $\{v_\alpha \in \text{L}^2(\mathbb{R})\}_{\alpha \in \{1, 2 \dots M\}}$ and jump operators $\{L_\alpha\}_{\alpha \in \{1, 2 \dots M\}}$. Furthermore, assume that $\exists\ \omega_c > 0$ such that for $\abs{\omega} \geq \omega_c, v_\alpha(\omega) = 0$. Then there exists a chain dilation of this non-Markovian model with $N_m$ modes and bandwidth $\leq \omega_c$ such that
\begin{align*}
&\norm{\ket{\Psi(t)} - \ket{\hat{\Psi}(t)}}^2 \leq \nonumber\\
&\qquad 4\ell t \bigg(1 + \ell^2 t^2  + \mu^{(1)}_{\ket{\Psi_0}}\bigg)^{1/2} N_m \bigg(\frac{2e\omega_c t}{N_m}\bigg)^{N_m/2},
\end{align*}
where $\ket{\Psi(t)}$ and $\ket{\hat{\Psi}(t)}$ are the system-environment states obtained from the model and its chain dilation, $\ket{\Psi(0)} = \ket{\hat{\Psi(0)}} = \ket{\Psi_0}$, $\ell = \sum_{\alpha = 1}^M \norm{v_\alpha}_{L^2} \norm{L_\alpha}$ and $\mu^{(1)}_{\ket{\Psi_0}}$ is the initial expectation value of the particle number operator of the environment.
\end{lemma}
\noindent From this lemma, it follows that a choice of $N_m = \textnormal{poly}(\ell, t, \omega_c, 1/\epsilon)$ ensures that the approximation error incurred using the star-to-chain transformation can be made smaller than $\epsilon$.

To use this lemma for more general non-Markovian models, we need to approximate them with coupling functions that have finite frequency support. As we noted previously, several coupling functions of interest are tempered distributions and do not necessarily fall off to zero at high frequencies --- this complicates the introduction of the frequency cutoff. We perform this in two steps --- first we regularize the coupling function, which damps that the high frequency components, and then we introduce a frequency cutoff. The analysis of the regularization error heavily uses lemma \ref{lemma:gfunc_error_bound} --- however, to predict the scalings of the regularization error with problem parameters, we need two additional assumptions. The first assumption is on the radon measures describing the memory kernels, and restricts their growth and smoothness and the second assumption ensures that the initial environment state has particle number moments that decay sufficiently rapidly with frequency. We state them mathematically below.
\begin{repassumption}{assump:radon_measure}
The radon measure $\mu$ corresponding to the coupling function should satisfy:
\begin{enumerate}
\item[(a)] For any interval $[a, b] \subseteq \mathbb{R}$, $\textnormal{TV}_{[a, b]}(\mu) \leq \textnormal{poly}(\abs{a}, \abs{b})$ and
\item[(b)] The error functions corresponding to $\mu$, $\Delta^0_{\mu; [a, b]}(\varepsilon)$ and $\Delta^1_{\mu; [a, b]}(\varepsilon)$ as specified in lemma \ref{lemma:main_error_bound_mu} are individually locally integrable with respect to $a, b$ and grow at most polynomially with $\abs{a}, \abs{b}$, and fall off polynomially with $\varepsilon$ i.e.~
\[
\Delta^0_{\mu; [a, b]}(\varepsilon), \Delta^1_{\mu; [a, b]}(\varepsilon)  \leq \textnormal{poly}(|a|, |b|, \varepsilon).
\]
\end{enumerate}
\end{repassumption}
\begin{repassumption}{assump:initial_state}
The initial environment state $\ket{\phi_1}\otimes \ket{\phi_2} \dots \ket{\phi_M}$ where for $\alpha \in \{1,2 \dots M\}$, $\ket{\phi_\alpha} \in \textnormal{Fock}[L^2(\mathbb{R})]$ and for its $n-$particle wavefunctions $\phi_{\alpha, n} \in L^2(\mathbb{R}^n)$, and any $j, k \geq 0$, $\exists \mathcal{N}_{j, k} > 0$ such that
\[
\sum_{n = 0}^\infty n^j \int_{\mathbb{R}^n} (1 + \omega_1^2)^k \abs{\phi_{\alpha, n}(\omega)}^2 d\omega < \mathcal{N}_{j, k}.
\]
\end{repassumption}
We next provide a lemma that, subject to these assumptions, bounds the regularization error.
\begin{lemma}[Regularization error]\label{lemma:reg_error}
Consider a non-Markovian model specified by a system Hamiltonian $H_S(t)$, jump operators $\{L_\alpha\}_{\alpha \in \{1, 2 \dots M\}}$ and coupling functions $\{(\mu_\alpha, \varphi_\alpha)\}_{\alpha \in \{1,2 \dots M\}}$. Furthermore, assume that $\mu_\alpha$ satisfy assumption \ref{assump:radon_measure}, and its fourier transform satisfies $\hat{\mu}_\alpha(\omega) \leq O(\omega^{2k})$ for some $k > 0$ and the initial environment state $\ket{\Psi_0}$ satisfies assumption \ref{assump:initial_state}. Then, if $\ket{\Psi(t)}$ is the system-environment state at time $t$, and $\ket{\Psi_\varepsilon(t)}$ is the system environment state of the regularized non-Markovian model, then
\begin{align*}
&\norm{\ket{\Psi(t)} - \ket{\Psi_\varepsilon(t)}} \leq \nonumber\\
&\qquad O\bigg(\varepsilon^q \textnormal{poly}\bigg(t, M, \norm{L_\alpha}, \sup_{\alpha, s\in[0, t]} \norm{[H_S(s), L_\alpha]}, \nonumber\\
&\qquad \qquad \qquad \ \ \mathcal{N}_{1, k + 1}, \mathcal{N}_{1, k + 2}\bigg)\bigg),
\end{align*}
for some $q > 0$.
\end{lemma}
\noindent\emph{Proof}: Suppose $\rho \in \textnormal{C}_c^\infty(\mathbb{R})$ is a symmetric mollifier, and $\hat{\rho}$ be its fourier transform. Consider a regularized non-Markovian model which has the same system Hamiltonian and jump operator but square integrable coupling functions $v_\varepsilon := \{v_{\alpha, \varepsilon} \in L^2(\mathbb{R})\}_{\alpha \in \{1, 2\dots M\}}$, where $v_{\alpha, \varepsilon}$ is the $\varepsilon, \rho-$regularization of $(\mu_\alpha, \varphi_\alpha)$ (definition \ref{def:regularization}). Denote the system-environment state at time $t$ corresponding to the actual model by $\ket{\Psi(t)}$, and the regularized model by $\ket{\Psi_\varepsilon(t)}$. We note from theorem \ref{thm:} that $\ket{\Psi(t)} = \lim_{\varepsilon'\to 0}\ket{\Psi_\varepsilon(t)}$. Thus, we obtain from lemma \ref{lemma:pert_theory_bound} that
\begin{align*}
&\norm{\ket{\Psi(t)} - \ket{\Psi_\varepsilon(t)}}^2 \leq \nonumber\\
&\qquad \lim_{\varepsilon' \to 0}\sum_{\alpha = 1}^M  \bigg(\int_0^t \mathcal{E}^{v_\varepsilon, v_{\varepsilon'}}_{\alpha, \ket{\Psi_0}}(\tau) d\tau + \int_0^t \mathcal{D}_{\alpha, \ket{\Psi_0}}^{v_\varepsilon, v_{\varepsilon'}}(\tau) d\tau\bigg).
\end{align*}
Let us first bound $\lim_{\varepsilon'\to 0}\int_0^t \mathcal{D}_{\alpha, \ket{\Psi_0}}^{v_\varepsilon, v_{\varepsilon'}}(\tau)d\tau$ --- note from lemma~\ref{lemma:pert_theory_bound} that $\mathcal{D}_{\alpha, \ket{\Psi_0}}^{v_\varepsilon, v_{\varepsilon'}}(\tau) = 4\norm{L_\alpha} \Vert{a_{\alpha, \uptau_\tau(v_{\alpha, \varepsilon} - v_{\alpha, \varepsilon'})}^-\ket{\Psi_0}}\Vert$, where $\uptau_\tau : L^2(\mathbb{R}) \to L^2(\mathbb{R})$ is the time-translation unitary group. Now, from the Cauchy-Schwarz inequality it follows that
\begin{align}\label{eq:lemma_reg_init_state_error_1}
&\norm{a_{\alpha, \uptau_\tau(v_{\alpha, \varepsilon} - v_{\alpha, \varepsilon'})}^-\ket{\Psi_0}}^2 =\sum_{n = 0}^\infty n\times \nonumber\\
&\quad \int_{\mathbb{R}^{n - 1}}\abs{\int_{\mathbb{R}} \big(\hat{v}^*_{\alpha, \varepsilon'}(\omega) - \hat{v}_{\alpha, \varepsilon}(\omega)\big) e^{i\omega_1 \tau} \phi_{\alpha, n}((\omega_1, \omega)) d\omega_1}^2d\omega, \nonumber\\
&\quad \leq \mathcal{N}_{1, k + 2} \int_{\mathbb{R}} \frac{\abs{\hat{v}_{\alpha, \varepsilon}(\omega) - \hat{v}_{\alpha, \varepsilon'}(\omega)}^2}{(1 + \omega^2)^{k + 2}} d\omega.
\end{align}
Furthermore, we note from Taylor's theorem that
\begin{align}\label{eq:lemma_reg_init_state_error_2}
&\abs{\hat{v}_\varepsilon(\omega) - \hat{v}_{\varepsilon'}(\omega)}^2,  \nonumber\\
&\qquad =\hat{\mu}(\omega) \abs{\hat{\rho}(\omega \varepsilon) - \hat{\rho}(\omega \varepsilon')}^2, \nonumber \\
&\qquad \leq \hat{\mu}(\omega) \omega^2 (\varepsilon - \varepsilon')^2 \sup_{\nu \in \mathbb{R}} \abs{\partial_\nu \hat{\rho}(\nu)}^2, \nonumber \\
&\qquad \leq \hat{\mu}(\omega) \omega^2 (\varepsilon - \varepsilon')^2,
\end{align}
where we have used that since $\partial_\nu \hat{\rho}(\nu)$ is the fourier transform of $x \rho(x)$, 
\[
\sup_{\nu \in \mathbb{R}} \abs{\partial_\nu \hat{\rho}(\nu)} \leq \int_{[-1, 1]} \abs{x} \abs{\rho(x)} dx \leq 1.
\]
It now follows from Eqs.~\ref{eq:lemma_reg_init_state_error_1} and \ref{eq:lemma_reg_init_state_error_2}
\begin{align*}
&\lim_{\varepsilon' \to 0} \int_0^t \mathcal{D}_{\alpha, \ket{\Psi_0}}^{v_\varepsilon, v_{\varepsilon'}}(\tau) d\tau \leq \nonumber\\
&\qquad 4\norm{L_\alpha} t \sqrt{\mathcal{N}_{1, k + 2}} \bigg(\int_{\mathbb{R}} \frac{\omega^2 \hat{\mu}(\omega)}{(1 + \omega^2)^{k + 2}} d\omega\bigg)^{1/2} \abs{\varepsilon },
\end{align*}
or equivalently, 
\begin{align}\label{eq:lemma_reg_d_bound_final}
&\lim_{\varepsilon' \to 0} \sum_{\alpha = 1}^M \int_0^t \mathcal{D}_{\alpha, \ket{\Psi_0}}^{v_\varepsilon, v_{\varepsilon'}}(\tau) d\tau \leq \nonumber\\
&\qquad \qquad O\bigg(\varepsilon\ \text{poly}\bigg(t, M, \sup_{\alpha}\norm{L_\alpha}, \mathcal{N}_{1, k + 2}\bigg)\bigg).
\end{align}
Next, we consider $\lim_{\varepsilon'\to 0} \int_0^t \mathcal{E}_{\alpha, \ket{\Psi_0}}^{v_\varepsilon, v_{\varepsilon'}}(\tau) d\tau$ --- for this, we use lemma \ref{lemma:gfunc_error_bound}(b). To use this lemma, we need to provide a bound on $c_{\mu_\alpha, \ket{\Psi_0}}$ which appears in the lemma statement and is defined in lemma \ref{lemma:inc_state_trunc}. Note from its definition that $c_{\mu_\alpha, \ket{\Psi_0}}$ bounds $\norm{a_{\alpha, \uptau_\tau v_\varepsilon}^- \ket{\Psi_0}}$ and
\begin{align*}
&\norm{a^-_{\alpha, \uptau_\tau v_{\alpha, \varepsilon}} \ket{\Psi_0}}^2 \leq \sum_{n = 0}^\infty n \times \nonumber\\
&\qquad \int_{\mathbb{R}^{n - 1}} \abs{\int_{\mathbb{R}} \hat{v}^*_{\varepsilon}(\omega) e^{i\omega \tau} \phi_{\alpha, n}((\omega_1, \omega)) d\omega_1}^2 d\omega, \nonumber\\
&\leq \mathcal{N}_{1, k + 1} \int_{\mathbb{R}}\frac{\abs{\hat{v}_{\alpha, \varepsilon}(\omega)}^2}{(1 + \omega^2)^{k + 1}} d\omega.
\end{align*}
Also, since $\sup_{\nu \in \mathbb{R}} \abs{\hat{\rho}(\nu)} \leq \int_{[-1, 1]} \abs{\rho(x)}dx = 1$, $\abs{\hat{v}_{\alpha, \varepsilon}(\omega)}^2 \leq \hat{\mu}(\omega)$, we obtain that
\[
\norm{a^-_{\alpha, \uptau_\tau v_{\alpha, \varepsilon}} \ket{\Psi_0}} \leq \sqrt{\mathcal{N}_{1, k + 1}} \bigg(\int_{\mathbb{R}} \frac{\hat{\mu}(\omega)}{(1 + \omega^2)^{k + 1}}d\omega\bigg)^{1/2}.
\]
Since assumption \ref{assump:radon_measure} holds, $\Delta^0_{\mu_\alpha, [0, t]}(\varepsilon), \Delta^1_{\mu_\alpha, [0, t]}(\varepsilon) \leq O(\varepsilon^p \text{poly}(t))$ for some $p > 0$. It then follows from lemma \ref{lemma:gfunc_error_bound}(b) that
\begin{align}\label{eq:lemma_reg_e_bound_final}
&\lim_{\varepsilon' \to 0} \sum_{\alpha = 1}^M \int_0^t \mathcal{E}_{\alpha, \ket{\Psi_0}}^{v_\varepsilon, v_{\varepsilon'}}(\tau) d\tau\leq \nonumber \\
& O\bigg(\varepsilon^p \text{poly}\bigg(t, M, \text{sup}_\alpha \norm{L_\alpha}, \sup_{\alpha, s \in [0, \tau]} \norm{[H_S(\tau), L_\alpha]}, \mathcal{N}_{1, k + 1}\bigg)\bigg)
\end{align}
Combining Eqs.~\ref{eq:lemma_reg_d_bound_final} and \ref{eq:lemma_reg_e_bound_final}, we obtain the lemma statement. $\hfill \square$
%In this section, we develop a certifiable Markovian dilation of a non-Markovian model. We use the well-known star-to-chain transformation for mapping the non-Markovian problem to a Hamiltonian simulation problem, and provide error bounds on this dilation.
%Next, we analyze the error incurred on approximating a non-Markovian model with its chain approximation. There are two sources of error in this approximation --- the first is in introducing a frequency cutoff into the model, and the next is in approximating environment in the resulting model by its chain representation. We analyze both of these errors separately --- for this analysis, we restrict ourselves to coupling functions whose Fourier transforms fall off sufficiently fast with frequency. Then, we consider models specified by a distributional coupling functions where an additional regularization step (as described in the previous section) is needed to map them to coupling functions in this class.

Next, we consider frequency truncation of the coupling functions --- after regularization, the coupling functions decay at high frequencies, and consequently a frequency cut-off can be introduced. To make this precise, we note that for a coupling function specified by $(\mu, \varphi)$ then the regularized coupling function $\hat{v}_\varepsilon$ falls off at least as $\omega^{-1}$ --- this is so because the fourier transform of a mollifier $\hat{\rho}$, $\hat{\rho}(\omega)$ falls off faster than any polynomial in $\omega$. Since $\hat{\mu}(\omega)$ grows atmost polynomially with $\omega$, the $\omega^{-1}$ decay of the regularized coupling function follows. The next lemma considers non-Markovian models with coupling functions fall off as $\omega^{-1}$, and bounds the error incurred on introducing a frequency cut-off.
\begin{lemma}\label{lemma:frequency_truncation_schwartz} Consider a non-Markovian model described by coupling functions 
\[
\bigg\{v_\alpha \in L^2(\mathbb{R}) \bigg| \norm{(\cdot) {\hat{v}_\alpha(\cdot)}}_{L^\infty} = \sup_{\omega \in \mathbb{R}} \abs{\omega \hat{v}_{\alpha}(\omega)} < \infty\bigg\}_{\alpha \in \{1, 2 \dots M\}},
\]
jump operators $\{L_\alpha\}_{\alpha \in \{1, 2 \dots M\}}$ and system Hamiltonian $H_S(t)$. For $\omega_c > 0$, consider a non-Markovian model described by coupling functions $v_{\omega_c} = \{v_{\omega_c} \in L^2(\mathbb{R}) | \hat{v}_{\alpha, \omega_c} := \hat{v}_\alpha \mathcal{I}_{[-\omega_c, \omega_c]}\}_{\alpha \in \mathbb{Z}_{\geq 0}}$ but with the same jump operators and system Hamiltonian. Let $\ket{\Psi(t)}$ and $\ket{\Psi_{\omega_c}(t)}$ be the state at time $t$ for both of these models starting with initial state $\ket{\Psi_0} \in \mathcal{H}_S \otimes \textnormal{F}_\infty^M(L^2(\mathbb{R})) $, then ,
\begin{align*}
&\norm{\ket{\Psi(t)} - \ket{\Psi_{\omega_c}(t)}}^2 \nonumber \\
&\leq  \frac{2}{\omega_c^{1/2}}\sum_{\alpha = 1}^M { \norm{L_\alpha} \norm{(\cdot) \hat{v}_\alpha(\cdot)}_{L^\infty}}\bigg(\norm{L_\alpha}\norm{v_\alpha}_{L^2} t^2 + 2\mu_{\ket{\Psi_0}}^{(1)} t \bigg).
\end{align*}
\end{lemma}
\noindent\emph{Proof}:It is useful to note that since $\abs{ \hat{v}_\alpha(\omega)} \leq \norm{(\cdot)\hat{v}_\alpha(\cdot)}_{L^\infty} / \omega$, we obtain that
\begin{align}\label{eq:norm_cutoff}
\norm{v_\alpha - v_{\alpha, \omega_c}}_{L^2}^2 = \int_{|\omega| \geq \omega_c} |\hat{v}_{\alpha}(\omega)|^2 d\omega \leq \frac{\norm{(\cdot)\hat{v}_\alpha(\cdot)}_{L^\infty}^2}{\omega_c}.
\end{align}
The proof of this lemma follows from lemma \ref{lemma:pert_theory_bound}. Consider the two error terms defined in lemma \ref{lemma:pert_theory_bound}. First, note that 
\begin{align*}
&\mathcal{D}_\alpha^{v, v_{\omega_c}}(\tau) = 4\norm{L_\alpha} \norm{a_{\alpha, \uptau_\tau(v_\alpha - v_{\alpha, \omega_c})}\ket{\Psi_0}}, \nonumber\\
&\qquad \leq 4 \mu^{(1)}_{\ket{\Psi_0}}\norm{L_\alpha} \norm{v_\alpha - v_{\alpha, \omega_c}}_{L^2}, \nonumber\\
&\qquad \leq {4\mu_{\ket{\Psi_0}}^{(1)}}\frac{\norm{L_\alpha}\norm{(\cdot)\hat{v}_\alpha(\cdot)}_{L^\infty}}{\sqrt{\omega_c}}.
\end{align*}
Furthermore, since $v_\alpha, v_{\alpha, \omega_c} \in L^2(\mathbb{R})$, we obtain that $\forall \tau, s \in [0, t]$
\begin{align*}
&\abs{\langle \uptau_\tau(v_\alpha - v_{\alpha, \omega_c}), \uptau_s v_\alpha\rangle}\nonumber\\
&\qquad = \abs{\int_{\mathbb{R}} (\hat{v}_\alpha^*(\omega) - \hat{v}_{\alpha, \omega_c}^*(\omega))\hat{v}_\alpha(\omega)e^{-i\omega (s - \tau)} d\omega}\nonumber\\
&\qquad \leq \norm{v_\alpha}_{L^2} \norm{v_\alpha - v_{\alpha, \omega_c}}_{L^2} \nonumber\\
&\qquad \leq \frac{\norm{v_\alpha}_{L_2} \norm{(\cdot) \hat{v}_\alpha(\cdot)}_{L^\infty}}{ \omega_c^{1/2}},
\end{align*}
and
\begin{align*}
&\abs{\langle  \uptau_s v_{\alpha, \omega_c}, \uptau_\tau(v_\alpha - v_{\alpha, \omega_c})\rangle}\nonumber\\
&\qquad= \abs{\int_{\mathbb{R}} (\hat{v}_\alpha(\omega) - \hat{v}_{\alpha, \omega_c}(\omega))\hat{v}^*_{\alpha, \omega_c}(\omega)e^{i\omega (s - \tau)} d\omega} \nonumber\\
&\qquad \leq \norm{v_{\alpha, \omega_c}}_{L^2} \norm{v_\alpha - v_{\alpha, \omega_c}}_{L^2} \nonumber\\
&\qquad \leq \frac{\norm{v_\alpha}_{L_2} \norm{(\cdot) \hat{v}_\alpha(\cdot)}_{L^\infty}}{\sqrt{\omega_c}}.
\end{align*}
Therefore, we obtain that $\forall \tau \in [0, t]$,
\[
\mathcal{E}_\alpha^{v, v_{\omega_c}}(\tau) \leq 4 t \frac{\norm{L_\alpha}^2\norm{v_\alpha}_{L_2} \norm{(\cdot) \hat{v}_\alpha(\cdot)}_{L^\infty}}{\omega_c^{1/2}}
\]
Substituting these estimates into lemma \ref{lemma:pert_theory_bound}, we then obtain the lemma statement.\hfill \(\square\)

Finally, we turn to the proof of the full proposition. The proof of this proposition simply puts together the error bounds provided in lemmas \ref{lemma:star_to_chain}, \ref{lemma:reg_error} and \ref{lemma:frequency_truncation_schwartz}.

\begin{repproposition}{theorem:final_dilation}
Consider a non-Markovian model specified by a system Hamiltonian $H_S(t)$, jump operators $\{L_\alpha\}_{\alpha \in \{1, 2 \dots M\}}$ and coupling functions $\{(\mu_\alpha, \varphi_\alpha)\}_{\alpha \in \{1, 2 \dots M\}}$ where $\mu_\alpha$ satisfy assumption \ref{assump:radon_measure} with $\hat{\mu}_\alpha(\omega) < O(\omega^{2k}$) for some $k > 0$. For $\ket{\Psi_0} := \ket{\sigma} \otimes \ket{\Phi_0} \in \mathcal{H}_S\otimes \textnormal{Fock}[L^2(\mathbb{R})]^{\otimes M}$, where $\ket{\sigma} \in \mathcal{H}_S$ and $\ket{\Phi_0}$ is an initial environment state that satisfies assumption \ref{assump:initial_state}, then $\exists$ a Markovian dilation of the  non-Markovian model with 
\begin{align*}
&N_m, B \leq O\bigg(\textnormal{poly}\bigg(\frac{1}{\epsilon}, t, M, \sup_{\alpha}\norm{L_\alpha},\nonumber \\
&\qquad \qquad \qquad \qquad   \sup_{\alpha, s \in [0,t]}\norm{[H_S(s), L_\alpha]},\nonumber\\
 &\qquad \qquad \qquad \qquad \mathcal{N}_{1, k + 1}, \mathcal{N}_{1, k + 2}, \mathcal{N}_{1, 0} \bigg)\bigg)
\end{align*}
whose system-environment state at time $t$ is within $\epsilon$ norm distance of the exact state.
\end{repproposition}
\noindent\emph{Proof}: Throughout this proof, for notational convenience, we will hide several polynomial factors with respect to $t, M, \sup_\alpha \norm{L_\alpha}, \sup_{\alpha, s\in[0, t]} \norm{[H_S(s), L_\alpha]}, \mathcal{N}_{1, 0}, \mathcal{N}_{1, k},$ $\mathcal{N}_{1, k + 1}$ in the big-$O$ notation i.e.~we will use $\tilde{O}$ to denote
\begin{align*}
&\tilde{O}(f) = O\bigg(f \ \textnormal{poly}\bigg(t, M, \sup_{\alpha}\norm{L_\alpha},\nonumber \\
&\qquad \qquad \qquad \qquad   \sup_{\alpha, s \in [0,t]}\norm{[H_S(s), L_\alpha]},\nonumber\\
 &\qquad \qquad \qquad \qquad \mathcal{N}_{1, k + 1}, \mathcal{N}_{1, k + 2},\mathcal{N}_{1, 0}\bigg)\bigg)
\end{align*}
It would also be convenient to note in the following proof that $\mu^{(1)}_{\ket{\Psi_0}} \leq \mathcal{N}_{1, 0}$. 

Consider first the regularization step --- we will denote the mollifier used $\rho$, regularization parameter used $\varepsilon$ and the regularized coupling functions $v_\varepsilon = \{v_{\alpha, \varepsilon} \in L^2(\mathbb{R}) \}_{\alpha \in \{1, 2 \dots M\}}$  here, from lemma \ref{lemma:reg_error}, it follows that a choice of the regularization parameter $\varepsilon = \tilde{O}(\epsilon)$ ensures that the regularization error is $O(\epsilon)$. Next, we consider the error incurred on introducing the frequency-cutoff (lemma \ref{lemma:frequency_truncation_schwartz}) in the regularized model. We note that since $\hat{\rho}(\omega)$, the fourier transform of the mollifier $\rho$, falls of with $\omega$ faster than any polynomial in $\omega$ and since $\hat{\mu}_\alpha(\omega) \leq O(\omega^{2k})$, then $\norm{(\cdot) \hat{v}_\alpha(\cdot)}_{L^\infty} \leq \sup_{\omega \in \mathbb{R}} \abs{\omega} \abs{\hat{\rho}(\omega \varepsilon)} \sqrt{\hat{\mu}_\alpha(\omega)} \leq O(\varepsilon^{-(k + 1)})$. Furthermore, $\norm{v_{\alpha, \varepsilon}}_{L^2} \leq  O(\varepsilon^{-(k + 1)})$ Now, from lemma \ref{lemma:frequency_truncation_schwartz}, it then follows that for a frequency cutoff $\omega_c$, the cutoff error scales as $\tilde{O}(\varepsilon^{-(k + 1)} / \omega_c^{1/4})$. Then, choosing $\omega_c = \tilde{O}(\text{poly}(\epsilon^{-1}, \varepsilon^{-1})) = \tilde{O}(\text{poly}(\varepsilon^{-1}))$ ensures that the frequency cut-off error is $O(\epsilon)$. Finally, from lemma \ref{lemma:star_to_chain}, it follows that $N_m = \tilde{O}(\text{poly}(\epsilon^{-1}, \omega_c)) = \tilde{O}(\epsilon)$ modes are needed in a chain dilation of the resulting model to reduce the error of the chain-dilation step to $O(\epsilon)$. This completes the proof. \hfill \(\square\)

\subsection{$k-$local Non-Markovian open system dynamics is in BQP}
\label{subsec:alg}
\noindent We next consider the $k-$local Non-Markovian open system problem. 
\begin{repproblem}{prob:k_local_non_mkv}[$k-$local non-Markovian dynamics] Consider a system of $n$ qudits $(\mathcal{H}_S = \big(\mathbb{C}^d\big)^{\otimes n})$ interacting with $M = \textnormal{poly}(n)$ baths with
\begin{enumerate}
\item[(a)] System Hamiltonian $H_S(t)$ is $k-$local i.e. $H_S(t) = \sum_{i = 1}^{N} H_i(t)$, where $N = \textnormal{poly}(n)$, and for $i \in \{1, 2 \dots N\}$, $H_i(t)$ is an operator acting on atmost $k$ qudits and $\norm{H_i(t)} \leq 1$.
\item[(b)] Jump operators $\{L_\alpha \}_{\alpha \in \{1, 2 \dots M\}}$ such that for $\alpha \in \{1, 2\dots M\}$, $L_\alpha$ acts on at-most $k$ qudits and $\norm{L_\alpha} \leq 1$.
\item[(c)] Coupling functions $\{(\mu_\alpha, \varphi_\alpha) \}_{\alpha \in \{1, 2 \dots M\}}$ such that for $\alpha \in \{1, 2 \dots M\}$, $\mu_\alpha$ satisfies the polynomial growth conditions (assumption \ref{assump:radon_measure}).
\item[(d)] Initial state $\ket{\Psi} = \ket{0}^{\otimes n} \otimes \ket{\Phi}$, where $\ket{\Phi}$ satisfies assumption \ref{assump:initial_state}.  Furthermore, the initial state is computable in the sense that for $v_1, v_2 \dots v_m \in L^2(\mathbb{R})$ and $P \in\mathbb{Z}_{>0}$, all the amplitudes
\begin{align*}
 \bra{\textnormal{vac}}\prod_{i = 1}^m \bigg(\int_{\mathbb{R}}v_i(\omega)a_\omega d\omega\bigg)^{n_i} \ket{\phi_\alpha} 
\end{align*}
with $n_1 + n_2 \dots n_m \leq P$ can be computed in $\textnormal{poly}(m, P)$ time on a classical or quantum computer.
\end{enumerate}
Denoting by $\rho_S(t)$ the reduced state of the system at time $t$ for this non-Markovian model, then for $\varepsilon > 0$ and $t = \textnormal{poly}(n)$, prepare a quantum state $\hat{\rho}$ such that $\norm{\hat{\rho} - \rho_S(t)}_\textnormal{tr} \leq \varepsilon$.
\end{repproblem}
To prove that this problem can be efficiently solved on a quantum computer, we proceed in three steps. First, we compute a Markovian dilation of the non-Markovian system with $N_m$ modes  --- from theorem \ref{theorem:final_dilation} it follows that $N_m = \text{poly}(n, \varepsilon^{-1})$ modes are needed to ensure that the error between the dynamics of the non-Markovian and its Markovian dilation is $< \varepsilon$. Next, we simulate the Markovian dilation on a quantum computer --- we thus consider a different problem, defined below.
\begin{problem}[$k-$local non-Markovian chain model]
\label{prob:chain_model}
Consider a system of $n$ qudits $(\mathcal{H}_S = \big(\mathbb{C}^d\big)^{\otimes n})$ interacting with $M = \textnormal{poly}(n)$ baths with
\begin{enumerate}
\item[(a)] System Hamiltonian $H_S(t)$ is $k-$local i.e. $H_S(t) = \sum_{i = 1}^{N} H_i(t)$, where $N = \textnormal{poly}(n)$, and for $i \in \{1, 2 \dots N\}$, $H_i(t)$ is an operator acting on atmost $k$ qudits and $\norm{H_i(t)} \leq 1$.
\item[(b)] Jump operators $\{L_\alpha \}_{\alpha \in \{1, 2 \dots M\}}$ such that for $\alpha \in \{1, 2\dots M\}$, $L_\alpha$ acts on at-most $k$ qudits and $\norm{L_\alpha} \leq 1$.
\item[(c)] Square-integrable coupling functions $\{v_\alpha \in L^2(\mathbb{R}) | \norm{v_\alpha}_{L^2} = \textnormal{poly}(n), \ \textnormal{supp}(\hat{v}_\alpha) \subseteq [-\omega_c, \omega_c]\}_{\alpha \in \{1, 2 \dots M\}}$ where $\omega_c = \textnormal{poly}(n)$.
\item[(d)] Single-particle environment dynamics specified by $\{\upnu_{\alpha, t} : L^2(\mathbb{R}) \to L^2(\mathbb{R}) \}_{\alpha \in \{1, 2 \dots M\}}$ where $\upnu_{\alpha, t}$ is the chain unitary group with $N_m = \textnormal{poly}(n)$ modes generated by $v_\alpha$.
\item[(e)] Initial state $\ket{\Psi} = \ket{0}^{\otimes n} \otimes \ket{\Phi}$, where $\ket{\Phi} = \ket{\phi_1} \otimes \ket{\phi_2}\otimes \dots \ket{\phi_M}$ satisfies assumption \ref{assump:initial_state}.  Furthermore, the initial state is computable in the sense that for $v_1, v_2 \dots v_m \in L^2(\mathbb{R})$ and $P \in\mathbb{Z}_{>0}$, all the amplitudes
\begin{align*}
 \bra{\textnormal{vac}}\prod_{i = 1}^m \bigg(\int_{\mathbb{R}}v_i(\omega)a_\omega d\omega\bigg)^{n_i} \ket{\phi_\alpha} 
\end{align*}
with $n_1 + n_2 \dots n_m \leq P$ can be computed in $\textnormal{poly}(m, P)$ time on a classical or quantum computer.
\end{enumerate}
Denoting by $\rho_S(t)$ the reduced state of the system at time $t$ for this non-Markovian model, then $t = \textnormal{poly}(n)$, prepare a quantum state $\hat{\rho}$ such that $\norm{\hat{\rho} - \rho_S(t)}_\textnormal{tr} \leq 1 /\textnormal{poly}(n)$.
\end{problem}
Since the Markovian dilation is still an infinite dimensional system (with an effective Hilbert space $\mathcal{H}_S\otimes \textnormal{Fock}[\mathbb{C}^{N_m}]^{\otimes M}$), this requires a truncation of Hilbert space to a finite-dimensional one, followed by simulating the finite-dimensional quantum dynamics. To show that the approximating finite-dimensional quantum system can be efficiently simulated, we use the Hamiltonian simulatability lemma from Ref.~\cite{aharonov2003adiabatic}, which we restate below
\begin{lemma}[Hamiltonian simulatability, Ref.~\cite{aharonov2003adiabatic}]
\label{lemma:hamil_simul}
Given a Hamiltonian $H(t)$ over a system of $n$ qudits such that for every $t \geq 0$ 
\begin{enumerate}
\item[(a)] $\forall$ computational basis element $\ket{a}$, the the set of computational basis elements $\ket{b}$ such that $\bra{a} H(t) \ket{b} \neq 0$ together with the elements $\bra{a} H(t)\ket{b}$ can be computed in $\textnormal{poly}(n)$ time, and
\item[(b)] $\int_0^t \norm{H(t')}dt' = \textnormal{poly}(n)$,
\end{enumerate}
then $\exists$ a quantum circuit over $n$ qudits of depth $\textnormal{poly}(n)$ which implements a unitary $\hat{U}$ such that $\norm{\hat{U} - U(t, 0)} \leq 1 / \textnormal{poly}(n)$, where $U(\cdot, \cdot)$ is the propagator corresponding to $H(t)$.
\end{lemma}
The next lemma, which shows the quantum simulability of problem \ref{prob:chain_model}, follows from the application of lemma~\ref{lemma:hamil_simul} together with a bound on the error incurred in truncating the infinite-dimensional Hilbert space of the environment. A detailed proof of this is in appendix \ref{app:hspace_trunc}

\begin{lemma} \label{lemma:chain_model_bqp}Problem \ref{prob:chain_model} can be solved on a quantum computer in run time $\textnormal{poly}(n)$.
\end{lemma}

\begin{reptheorem}{theorem:bqp_non_mkv}[$k-$local Non-Markovian dynamics $\in$ BQP] 
Problem \ref{prob:k_local_non_mkv} can be solved in $\textnormal{poly}(n)$ time on a quantum computer.
\end{reptheorem}
\noindent\emph{Proof}: An application of lemma \ref{theorem:final_dilation} approximates problem \ref{prob:k_local_non_mkv} to an instance of problem \ref{prob:chain_model}, and then the theorem statement follows from lemma \ref{lemma:chain_model_bqp}. \hfill\(\square\)

\twocolumngrid

\section{Conclusion}

In conclusion, our work identifies the class of tempered radon measures as memory kernels for which a unitary group generating non-Markovian system dynamics can be constructed. We therefore generalize the unitary group for Markovian dynamics (i.e.~with a delta function memory kernel) described in the theory of quantum stochastic calculus. We then consider the $k-$local many-body non-Markovian dynamics, and show that it can be efficiently simulated on quantum computers, thus establishing that this generalization is consistent with the Extended Church-Turing thesis.

Our work points to several important open questions about non-Markovian dynamics. The first is to understand if the growth conditions on the radon measure describing the memory kernel (assumption \ref{assump:radon_measure}) are necessary --- while there are radon measures that violate these conditions, it is possible that these growth conditions hold for any \emph{tempered} radon measure. Alternatively, violating these growth condition could lead to unphysical situations, such as ``infinitely long" memory times in the non-Markovian system. Formalizing these ideas would allow us to further sharpen the mathematical definition of a physically reasonable non-Markovian model. 

Second, can the unitary group describing non-Markovian dynamics constructed in this paper be further characterized? An important characterization would be to understand if this unitary group is strongly continuous (i.e.~is it associated with a self-adjoint Hamiltonian)? Similar questions have been previously answered for the unitary group for Markovian dynamics provided by a quantum stochastic differential equation \cite{chebotarev1997quantum, gregoratti2000hamiltonian, gregoratti2001hamiltonian}.

Finally, it would also be of interest to develop quantum algorithms for non-Markovian dynamics with better dependence on the problem size as well as the incurred approximation error by exploiting further structure in the non-Markovian model (e.g.~spatial locality, or availability of the Hamiltonian/jump operators as linear combination of unitaries) and using similar techniques that have been used in Hamiltonian or Lindbladian simulation problems \cite{berry2014exponential, berry2015hamiltonian, cleve2016efficient, haah2021quantum}.
\begin{acknowledgements}
I thank J.~I.~Cirac and D.~Malz for useful discussions. I acknowledge Max Planck Harvard research center for support from the quantum optics (MPHQ) postdoctoral fellowship.
\end{acknowledgements}

\twocolumngrid
\bibliography{references.bib}
\onecolumngrid
\appendix
\section{Proof of proposition \ref{thm:schr_sq_int}}
\label{app:sq_int}
In this section, we sketch a detailed proof of proposition 1, which deals with the well definition of non-Markovian dynamics with square integrable coupling functions. For convenience throughout this section, we will define $\ell$ to be the constant
\begin{align}
\ell = \sum_{\alpha = 1}^M \norm{v_\alpha}_{L^2} \norm{L_\alpha}.
\end{align}
\begin{replemma}{lemma:domain}
For all $t \in \mathbb{R}$,
\begin{enumerate}
    \item[(a)] $H(t):  \mathcal{H}_S\otimes \textnormal{F}_\infty^M[L^2(\mathbb{R})] \to \mathcal{H}$ is essentially self adjoint.
    \item[(b)] $H(t)$ is closable, and if $\overline{H(t)}:\textnormal{dom}[\overline{H(t)}] \to \mathcal{H}$ is its closure then $\mathcal{H}_S\otimes \textnormal{F}_1^M[L^2(\mathbb{R})] \subseteq \textnormal{dom}[\overline{H(t)}]$ and $\forall \ket{\Psi} \in \textnormal{dom}[\overline{H(t)}]$,
    $\overline{H(t)} \ket{\Psi} = \sum_{n = 0}^\infty H(t) \big(\Pi_n \ket{\Psi}\big).$
\end{enumerate}
\end{replemma}
\noindent\emph{Proof}:\\
\noindent\emph{(a)} is shown using Nelson's analytic vector theorem (Theorem X.39 of Ref.~\cite{reed1975ii}), and showing that all the vectors in $\mathcal{H}_S\otimes \textnormal{F}_0[L^2(\mathbb{R})]^{\otimes M}$ are analytic vectors of $H_S(t)$. We note that it follows easily from the definition of $H(t)$ that for every $n \in \mathbb{Z}_{\geq 0}$, $H(t)(\textnormal{id}\otimes\Pi_{\leq n})$ is a bounded operator and 
\[
\norm{H(t) \big(\textnormal{id}\otimes \Pi_{\leq n}\big)} \leq \norm{H_S(t)} + 2 \ell \sqrt{n + 1} .
\]
Recall that given an operator $O:\textnormal{dom}[O] \to \mathcal{H}$, a vector $x$ is an analytic vector of $O$ if $\sum_{n = 0}^\infty  t^n\norm{O^n x} / n! < \infty \ \forall \ t\in \mathbb{R}$. Let $\ket{\Psi} \in \mathcal{H}_S\otimes \textnormal{F}_0[L^2(\mathbb{R})]^{\otimes M}$ and let $N_0 \in \mathbb{Z}_{\geq 1}$ be the number of particles in the environment (i.e.~$\Pi_{>N_0}\ket{\Psi} = 0$). It then immediately follows that for any $k \in \mathbb{Z}_{\geq 0}$, $H^k(t) \ket{\Psi_0}$ has at most $N_0 + k$ particles, and thus
\[
\norm{H^k(t) \ket{\Psi_0}} \leq \big(\norm{H_S(t)} + 2\sqrt{N_0 + k + 1} \ell\big)^k \norm{\ket{\Psi_0}} \leq 2^k \big(\norm{H_S(t)}^k + (N_0 + k + 1)^{k / 2}\ell^k\big) \norm{\ket{\Psi_0}} ,
\]
and thus for $t \geq 0$
\begin{align*}
&\sum_{k = 0}^\infty \frac{t^k}{k!} \norm{H_S^k(t)\ket{\Psi_0}} \leq e^{2t \norm{H_S(t)}} \norm{\ket{\Psi_0}} + \sum_{k = 0}^\infty \frac{(2\ell t)^k}{k!} (N_0 + k + 1)^{k/2} \norm{\ket{\Psi_0}},\nonumber\\
&\qquad\leq e^{2t \norm{H_S(t)}} \norm{\ket{\Psi_0}} + \norm{\ket{\Psi_0}} + \sum_{k = 1}^\infty \frac{(2e\ell t)^k}{k^{k/2}} \bigg(1 + \frac{N_0 + k}{k}\bigg)^{k/2} \norm{\ket{\Psi_0}}.
\end{align*}
Wherein we have used the Stirling's approximation in the last estimate. The summation can now be seen to converge for any $t$ and hence $\ket{\Psi_0}$ is an analytic vector of $H(t)$. 
\\

\noindent\emph{(b)} We first consider a sequence $\{ \ket{\Psi_i} \in \mathcal{H}_S \otimes \textnormal{F}_\infty^M(L^2(\mathbb{R}))\}_{i \in \mathbb{N}}$ such that $\lim_{i \to \infty} \ket{\Psi_i} = 0$ and the sequence $\{H(t) \ket{\Psi_i} \}_{i \in \mathbb{N}}$ converges. We first show that under these conditions, $\lim_{i \to \infty} H(t) \ket{\Psi_i} = 0$ --- to see this, assume the contrary i.e.~$\lim_{i \to \infty} H(t) \ket{\Psi_i} = \ket{\Phi} \neq 0$. Since $\ket{\Phi} \neq 0$, $\exists N > 0,\big(\textnormal{id}\otimes\Pi_{\leq N}\big)\ket{\Phi} \neq 0$. Note that
\[
\norm{\big(\textnormal{id}\otimes \Pi_{\leq N} \big)H(t)} \leq  \norm{H_S(t)} + 2\ell \sqrt{N + 1},
\]
and therefore $\Pi_{\leq N} \ket{\Phi} = \lim_{i \to \infty}  \Pi_{\leq N} H(t)\ket{\Psi_i} = \Pi_{\leq N} H(t) \lim_{i \to \infty} \ket{\Psi_i} = 0$, where we have used that $\Pi_{\leq N}, \Pi_{\leq N} H(t)$ are bounded operators to swap the order of limits. Thus, we contradict our original assumption of $\ket{\Phi}\neq 0$ and hence $\ket{\Phi} = 0$. This shows that the operator $H(t)$ is closable.

Next, we consider $\ket{\Psi} \in \mathcal{H}_S\otimes \textnormal{F}_1^M[L^2(\mathbb{R})]$ --- we consider the sequence $\{\ket{\Psi_n} := \big( \textnormal{id}\otimes \Pi_{\leq n}\big) \ket{\Psi}\}_{n \in \mathbb{N}}$ which converges to $\ket{\Psi}$. Furthermore, we note that the sequence $\{H(t) \ket{\Psi_n}\}_{n \in \mathbb{N}}$ also converges, and converges to $\sum_{m = 0}^\infty H(t) \Pi_m \ket{\Psi}$ since
\[
\norm{H(t) \ket{\Psi_n} - \sum_{m = 0}^\infty H(t) \Pi_m \ket{\Psi}} \leq \norm{H_S(t)} \bigg({\sum_{m = n + 1}^\infty} \norm{\Pi_m \ket{\Psi}}^2\bigg)^{1/2} + 2\ell \bigg(\sum_{m = n + 1}^\infty (m + 1) \norm{\Pi_m \ket{\Psi}}^2 \bigg)^{1/2},
\]
and since $\norm{\ket{\Psi}} < \infty$ and $\mu^{(1)}_{\ket{\Psi}} < \infty$, $\sum_{m = n}^\infty \norm{\Pi_m \ket{\Psi}}^2, \sum_{m = n}^\infty m\norm{\Pi_m \ket{\Psi}}^2 \to 0$ as $n \to \infty$. Consequently, we obtain that $\ket{\Psi} \in \textnormal{dom}[\overline{H(t)}]$ and $\overline{H(t)} \ket{\Psi} = \sum_{n = 0}^\infty H(t) \Pi_{n} \ket{\Psi}$. \hfill\(\square\) \\

\noindent We are now poised to first investigate the existence of solution of the Schr\"odinger's equation for the time-dependent Hamiltonian defined in Definition \ref{def:hamil}. We restrict ourselves to initial states with only a finite number of particles in the environment i.e.~$\ket{\Psi(0)} \in \mathcal{H}_S\otimes \text{F}_\infty^M[L^2(\mathbb{R})]$, and use the density of $\ket{\Psi(0)} \in \mathcal{H}_S\otimes \text{F}_\infty^M[L^2(\mathbb{R})]$ to extend it to the entire system-environment Hilbert space. While $H(t)$, for every $t$, admits a self adjoint extension, since the equation under consideration is non-autonomous, this by itself does not imply that the solution of this equation exists. Instead, we analyze this equation by first truncating the number of particles in the environment, and analyzing the convergence of the obtained solution with the truncation.

\begin{definition}\label{def:hamil_p_particle}
For $p \in \mathbb{Z}_{\geq 0}$ and $t \in \mathbb{R}$, define $H^p(t):\mathcal{H} \to \mathcal{H}$ via $H^p(t) = \Pi_{\leq p} H(t) \Pi_{\leq p}$.
\end{definition}
Several properties of $H^p(t)$ follows trivially from its definition.
\begin{lemma} $H^p(t)$ has the following properties
\begin{enumerate}
    \item[(a)] $H^p(t)$ is a bounded operator for all $t \in \mathbb{R}$.
    \item[(b)] $H^p(t)$ is norm continuous with respect to $t$.
\end{enumerate}
\end{lemma}
\noindent\emph{Proof}: \\
\noindent (a) This follows straightforwardly by noting that $\forall t \in \mathbb{R}, \alpha \in \{1, 2 \dots M\}$, $\norm{a_{\alpha, \tau_t v_\alpha}^-\Pi_{\leq p}} \leq \sqrt{p}\norm{v_\alpha}_{L^2}$ and $\norm{a_{\alpha, \tau_t v_\alpha}^+ \Pi_{\leq p}} \leq \sqrt{p + 1}\norm{v_\alpha}_{L^2}$.\\

\noindent(b) For any $\delta > 0$, note that
\[
\norm{H^p(t + \delta) - H^p(t)} \leq \norm{H_S(t + \delta) - H_S(t)} + \sum_{\alpha = 1}^M \sqrt{p}\norm{L_\alpha} \norm{(\uptau_{t + \delta} - \uptau_{t})v_\alpha}_{L^2}.
\]
Since $H_S(t)$ is norm continuous, $\norm{H_S(t+ \delta) - H_S(t)} \to 0$ as $\delta \to 0$, and since $\uptau_t$ is strongly continuous in $t$, $\norm{(\uptau_{t + \delta} - \uptau_t)v_\alpha}_{L^2} \to 0$ as $\delta \to 0$, thus showing from the above estimate that $H^p(t)$ is norm continuous. \hfill\(\square\)
\begin{lemma}\label{lemma:particle_num_bound}
For any $p \in \mathbb{N}$ and $\tau, s \in \mathbb{R}$, there exists a unitary operator $U^p(\tau, s):\mathcal{H}\to \mathcal{H}$ which is norm continuous and differentiable with respect to both $s$ and $\tau$ and which satisfies
\[
i \frac{d}{d\tau}U^p(\tau, s) = H^p(\tau) U^p(\tau, s) \text{ with } U^p(s, s) = \textnormal{id}.
\]
Furthermore, let $\ket{\Phi} \in \mathcal{H}_S \otimes \textnormal{F}_\mathcal{S}[L^2(\mathbb{R})]$, and for $t > 0$, consider $M^{(k)}_{\ket{\Phi}}(t) $ defined by
\[
M^{(0)}_{\ket{\Phi}}(t) = \norm{\ket{\Phi}}^2, M^{(k)}_{\ket{\Phi}}(t) = 2 \mu^{(k)}_{\ket{\Phi}} +  2^{2k - 3} \ell^2 t^2 \bigg(\norm{\ket{\Phi}}^2 + M^{(k - 1)}_{\ket{\Phi}}(t)\bigg)^{2} \ \text{for } k \geq 1,
\]
then $\forall \tau, s \in [0, t]$,  $\mu^{(k)}_{U^p(\tau, s)\ket{\Phi}}  \leq M^{(k)}(t) \ \forall p \in \mathbb{Z}_{\geq 0}$.
\end{lemma}
\noindent\emph{Proof}: Since $H^p(\tau)$ is both norm continuous in $\tau$ and bounded, the existence, norm continuity and differentiability of $U^p(\tau, s)$ follows follows directly from Dyson expansion (see theorem X.69 of Ref.~\cite{reed1975ii}). For part (b), we use the Sch\"odinger equation. Note that
\[
U^p(\tau, s)\ket{\Phi} = \Pi_{> p} \ket{\Phi} + U^p(\tau, s)\Pi_{\leq p} \ket{\Phi},
\]
and furthermore, $\mu^{(k)}_{\Pi_{\leq p}\ket{\Phi}}  \leq \mu^{(k)}_{\ket{\Phi}} \ \forall k \in \mathbb{Z}_{\geq 1}$. For convenience of notation, we set $\ket{\Psi^p(\tau, s)} = U^p(\tau, s) \Pi_{\leq p}\ket{\Phi}$. From the Schroedinger's equation, it follows that
\begin{align*}
\frac{d}{d\tau}\mu^{(k)}_{\ket{\Psi^p(\tau, s)}}   = \sum_{\alpha = 1}^M \sum_{n = 0}^{p - 1} \big((n + 1)^k - n^k\big) \textnormal{Im}\bra{\Psi^p(\tau, s)} \Pi_{n + 1}a^+_{\alpha, \uptau_\tau v_\alpha}L_\alpha \Pi_{n}\ket{\Psi^p(\tau, s)}.
\end{align*}
and therefore
\begin{align*}
&\bigg|\frac{d}{d\tau}\mu^{(k)}_{\ket{\Psi^p(\tau, s)}} \bigg|\leq 2 \ell  \sum_{n=0}^{p-1}\sqrt{n+1}\big((n + 1)^l - n^l\big)\norm{\Pi_{n + 1}\ket{\Psi^p(\tau, s)}}\norm{\Pi_n\ket{\Psi^p(\tau, s)}},
\end{align*}
Since $(n + 1)^l - n^l = (n + 1)^{l - 1} + n (n + 1)^{l - 2} + n^2 (n + 1)^{l - 3} \dots n^{l - 1}$ for $l \in \mathbb{Z}_{\geq 1}$, we obtain that
\begin{align*}
\bigg|\frac{d}{d\tau}\mu^{(k)}_{\ket{\Psi^p(\tau, s)}} \bigg|& \leq 2 \ell \sum_{n=0}^{p-1}\sum_{s = 1}^k (n + 1)^{k-s + 1/2}n^{s-1} \norm{\Pi_{n + 1}\ket{\Psi^p(\tau, s)}}\norm{\Pi_n\ket{\Psi^p(\tau, s)}}
\end{align*}
An application of the Cauchy-Schwarz inequality yields that $\forall s \in \{1, 2 \dots k\}$
\begin{align*}
&\sum_{n = 0}^{p - 1}(n + 1)^{k - s + 1/2}n^{s - 1}\norm{\Pi_{n + 1}\ket{\Psi^p(\tau, s)}}\norm{\Pi_n\ket{\Psi^p(\tau, s)}} \nonumber\\
&\qquad \qquad \leq \bigg(\sum_{n = 0}^{p- 1}(n + 1)^k \norm{\Pi_{n + 1}\ket{\Psi^p(\tau, s)}}^2 \bigg)^{1/2} \bigg(\sum_{n = 0}^{p - 1}n^{2s-2}(n + 1)^{k + 1 - 2s}\norm{\Pi_n \ket{\psi^p(\tau, s)}}^2 \bigg)^{1/2} \nonumber \\
&\qquad\qquad\leq \big(\mu^{(k)}_{\ket{\Psi^p(t, s)}}\big)^{1/2} \bigg(\sum_{n = 0}^{p - 1} (n + 1)^{k - 1}\norm{\Pi_n \ket{\Psi^p(\tau, s)}}^2\bigg)^{1/2}
\end{align*}
Noting that 
\[
\sum_{n = 0}^{p - 1}(n + 1)^{k - 1}\norm{\Pi_n \ket{\Psi^p(\tau, s)}}^2 \leq 2^{k - 2} \sum_{n = 0}^{p - 1} (n^{k - 1} + 1) \norm{\Pi_n \ket{\Psi^p(\tau, s)}}^2 = 2^{k - 2} \bigg( \mu^{(0)}_{\ket{\Psi^p(\tau, s)}} + \mu^{(k - 1)}_{\ket{\Psi^p(\tau, s)}}\bigg).
\]
Noting that $\mu^{(0)}_{\ket{\Psi^p(\tau, s)}} = \mu^{(0)}_{\Pi_{\leq p}\ket{\Phi}} \leq \mu^{(0)}_{\ket{\Phi}}$, we thus obtain
\[
\abs{\frac{d}{dt}\mu^{(k)}_{\ket{\Psi^p(\tau, s)}}} \leq 2^{k - 1}\ell \big(\mu^{(k)}_{\ket{\Psi^p(\tau, s)}}\big)^{1/2} \bigg( \mu^{(0)}_{\ket{\Phi}} + \mu^{(k - 1)}_{\ket{\Psi^p(\tau, s)}}\bigg),
\]
Integrating which we obtain that for $\tau \geq s$
\[
\big(\mu^{(k)}_{\ket{\Psi^p(\tau, s)}}\big)^{1/2} - \big(\mu^{(k)}_{\Pi_{\leq p}\ket{\Phi}}\big)^{1/2} \leq 2^{k -2} \ell\bigg( (\tau - s)\mu^{(0)}_{\ket{\Phi}}  + \int_s^{\tau}  \mu^{(k - 1)}_{\ket{\Psi^p(\tau', s)}} d\tau'\bigg) \leq 2^{k -2} \ell\bigg( t\mu^{(0)}_{\ket{\Phi}}  + \int_s^{\tau}  \mu^{(k - 1)}_{\ket{\Psi^p(\tau', s)}} d\tau'\bigg),
\]
and for $\tau < s$
\[
\big(\mu^{(k)}_{\ket{\Psi^p(\tau, s)}}\big)^{1/2} - \big(\mu^{(k)}_{\Pi_{\leq p}\ket{\Phi}}\big)^{1/2} \leq 2^{k -2} \ell\bigg( (s - \tau)\mu^{(0)}_{\ket{\Phi}}  + \int_\tau^{s}  \mu^{(k - 1)}_{\ket{\Psi^p(\tau', s)}} d\tau'\bigg) \leq 2^{k -2} \ell\bigg( t\mu^{(0)}_{\ket{\Phi}}  + \int_\tau^{s}  \mu^{(k - 1)}_{\ket{\Psi^p(\tau', s)}} d\tau'\bigg),
\]
Since $\forall k \in \mathbb{Z}_{\geq 0}$, $\mu^{(k)}_{U^p(\tau, s)\ket{\Phi}} = \mu^{(k)}_{\Pi_{> p}\ket{\Phi}} + \mu^{(k)}_{\ket{\Psi^p(\tau, s)}} $These equations can be recursively solved to obtain the functions $M_{\ket{\Phi_0}}^{(k)}(t)$. Note that $\mu^{(0)}_{\ket{\Psi^p(\tau, s)}} = \mu^{(0)}_{\Pi_{\leq p}\ket{\Phi}} \leq \bra{\Phi}\Phi\rangle$, and hence $M^{(0)}_{\ket{\Phi}}(t)$ can be set to $\norm{\ket{\Phi}}^2$. Assuming that $\mu^{(k - 1)}_{U^p(\tau, s)\ket{\Phi}} \leq M^{(k - 1)}_{\ket{\Phi}}(t) \ \forall \tau, s \in [0, t]$, we then obtain that
\[
\mu^{(k)}_{\ket{\Psi^p(\tau, s)}} \leq  \bigg(\big(\mu^{(k)}_{\Pi_{\leq p}\ket{\Phi}}\big)^{1/2} + 2^{k-2}\ell t\bigg(\bra{\Phi}\Phi\rangle + M^{(k - 1)}_{\ket{\Phi}}(t)\bigg)\bigg)^2 \leq 2\mu^{(k)}_{\Pi_{\leq p}\ket{\Phi}} + 2^{2k - 3}\ell^2 t^2 \bigg(\bra{\Phi}\Phi\rangle + M^{(k - 1)}_{\ket{\Phi}}(t)\bigg)^2,
\]
and since $\mu^{(k)}_{U^p(\tau, s)\ket{\Phi}} \leq \mu^{(k)}_{\ket{\Phi}} + \mu^{(k)}_{\ket{\Psi^p(\tau, s)}}$, we can choose $M^{(k)}_{\ket{\Phi}} = 2\mu^{(k)}_{\ket{\Phi}} + 2^{2k-3}\ell^2 t^2\big(\bra{\Phi}\Phi\rangle + M^{(k - 1)}_{\ket{\Phi}}(t)\big)^2$, which proves the lemma statement. \hfill\(\square\)

It is important to note that the bounds on $\mu_l^p(t)$ are uniform in $p$ --- we will exploit this in the following proofs to show the existence and differentiability of $\ket{\Psi^p(t)}$.
\begin{lemma}
\label{lemma:useful}
Let $\ket{\Psi_0} \in \mathcal{H}_S\otimes \textnormal{F}_\mathcal{S}^\infty[L^2(\mathbb{R})]$, then
\begin{enumerate}
    \item[(a)] $\forall t, s  \geq 0$, $\lim_{p\to\infty}U^p(t, s) \ket{\Psi_0}$ exists and $\in \mathcal{H}_{S}\otimes \textnormal{F}_\mathcal{S}^\infty[L^2(\mathbb{R})]$.
    \item[(b)] $\forall t, s  \geq 0$, $\lim_{p\to\infty} H^p(t)U^p(t, s) \ket{\Psi_0} = \overline{H(t)}\lim_{p \to \infty}U^p(t, s)\ket{\Psi_0}$ and $\overline{H(t)} \lim_{p \to \infty} U^p(t, s) \ket{\Psi_0}$ is strongly continuous in $t$.
    \item[(c)] $\forall t, s \geq 0$, $\lim_{p \to \infty} U^p(t, s) H^p(s) \ket{\Psi_0} = \lim_{p \to \infty} U^p(t, s) \overline{H(s)} \ket{\Psi_0}$ and $ \lim_{p \to \infty} U^p(t, s) \overline{H(s)} \ket{\Psi_0}$ is strongly continuous in $s$.
    \item[(d)] $\exists g, h \in \textnormal{C}^0(\mathbb{R})$ such that $\norm{H^p(t) U^p(t, s)\ket{\Psi_0}} \leq g(t)$ and $\norm{H^p(t) \ket{\Psi_0}} \leq h(t) \ \forall \ t\geq 0$ and $p \in \mathbb{Z}_{\geq 0}$.
\end{enumerate}
\end{lemma}
\noindent\emph{Proof}:\\
\noindent\emph{(a)} To prove the existence of limit, we appeal to the completeness of $\mathcal{H}$ and show that the sequence $\{U^p(t, s) \ket{\Psi_0}\}_{p \in \mathbb{Z}_{\geq 0}}$ is Cauchy. Consider $p, q \in \mathbb{N}$ with $p > q$ --- we note that
\[
\norm{U^p(t, s) \ket{\Psi_0} - U^q(t, s) \ket{\Psi_0}} \leq \norm{U^p(t, s) \Pi_{\leq q} \ket{\Psi_0} - U^q(t, s) \Pi_{\leq q} \ket{\Psi_0}} + 2 \norm{\Pi_{> q} \ket{\Psi_0}}.
\]
Furthermore, since both $U^p(t, s)$ and $U^q(t, s)$ are norm (and thus strongly) differentiable with respect to $t$ and $s$, we obtain that
\begin{align}\label{eq:pert_theory_bd}
&\norm{U^p(t, s) \Pi_{\leq q} \ket{\Psi_0} - U^q(t, s) \Pi_{\leq q} \ket{\Psi_0}} = \norm{\int_s^t \frac{d}{d\tau}\big(U^p(s, \tau) U^q(\tau, s)\big) \Pi_{\leq q} \ket{\Psi_0} d\tau} \nonumber\\
&\qquad \qquad \leq \int_s^t \norm{\big(H^p(\tau) - H^q(\tau)\big) U^q(\tau, s) \Pi_{\leq q} \ket{\Psi_0}}d\tau .
\end{align}
Furthermore, since $p > q$, we obtain that
\begin{align}\label{eq:pert_theory_state}
\ket{\Phi_{p, q}(\tau, s)} := \big(H^p(\tau) - H^q(\tau)\big)U^q(\tau, s)\Pi_{\leq q}\ket{\Psi_0}  =  \big(H^p(s) - H^q(s)\big)\Pi_{q}U^q(\tau, s)\Pi_{\leq q}\ket{\Psi_0},
\end{align}
and thus
\begin{align}\label{eq:norm_estimate}
\norm{\ket{\Phi_{p, q}(s)}} \leq \ell \sqrt{q + 1}  \norm{\Pi_{q} U^q(\tau, s)\Pi_{\leq q}\ket{\Psi_0}}.
\end{align}
Using the bound from lemma \ref{lemma:particle_num_bound}, we obtain that
$\norm{\Pi_q U^q(\tau, s) \Pi_{\leq q}\ket{\Psi_0}}  \leq \sqrt{M_{\Pi_{\leq q}\ket{\Psi(0)}}^{(2)}(\max(s, t))} / q \leq \sqrt{M_{\ket{\Psi_0}}^{(2)}(\max(s, t))} / q$ and thus
\[
\int_{s}^t \norm{\ket{\Phi_{p, q}(\tau, s)}} d\tau  \leq  \frac{\ell \abs{t - s} \sqrt{q + 1}}{q} \sqrt{M^{(2)}_{\Pi_{\leq q}\ket{\Psi_0}}(\max(s, t))} \leq \frac{\ell \abs{t - s} \sqrt{q + 1}}{q} \sqrt{M^{(2)}_{\ket{\Psi_0}}(\max(s, t))},
\]
and hence
\[
\norm{U^p(t, s) \ket{\Psi_0} - U^q(t, s) \ket{\Psi_0}}  \leq \frac{\ell \abs{t - s} \sqrt{q + 1}}{q} \sqrt{M^{(2)}_{\ket{\Psi_0}}(\max(s, t))} + 2\norm{\Pi_{\leq q}\ket{\Psi_0}}.
\]
Thus, $\norm{U^p(t, s) \ket{\Psi_0} - U^q(t, s) \ket{\Psi_0}} \to 0$ as $p, q \to \infty$, thus implying that the sequence $\{U^p(t, s)\ket{\Psi_0}\}_{p \in \mathbb{N}}$ is Cauchy, and hence converges. Furthermore, from lemma \ref{lemma:particle_num_bound}, the moments $\mu^{(k)}_{U^p(t, s) \ket{\Psi_0}}$ are bounded uniformly in $p$ for all $k \in \mathbb{Z}_{\geq 0}$, and hence from the dominated convergence theorem it follows that all the particle number moments of $\lim_{p \to \infty}U^p(t, s) \ket{\Psi_0}$ are also bounded --- this shows that $\lim_{p \to \infty} U^p(t, s) \ket{\Psi_0} \in \mathcal{H}_S \otimes \textnormal{F}_{\mathcal{S}}^M[L^2(\mathbb{R})]$. \\  \ \\

\noindent\emph{(b)} For $\ket{\Psi_0} \in \mathcal{H}_S \otimes \textnormal{F}_\mathcal{S}^M[L^2(\mathbb{R})]$, since $\norm{\lim_{p\to \infty} U^p(t, s) \Pi_{> p} \ket{\Psi_0}} = \lim_{p \to \infty} \norm{\Pi_{> p}\ket{\Psi_0}} = 0$, we obtain that
\[
\overline{H(t)} \lim_{p\to \infty} U^p(t, s) \ket{\Psi_0} = \overline{H(t)} \lim_{p \to \infty} U^p(t, s) \Pi_{\leq p} \ket{\Psi_0}.
\]
We already established in part (a) that the sequence $\{U^p(t, s) \ket{\Psi_0}\}_{p \in \mathbb{Z}_{\geq 0}}$, and hence the sequence $\{U^p(t, s) \Pi_{\leq p}\ket{\Psi_0}\}_{p \in \mathbb{Z}_{\geq 0}}$, converges. We now show that the sequence $\{H(t) U^p(t, s) \Pi_{\leq p}\ket{\Psi_0} \}_{p \in \mathbb{Z}_{\geq 0}}$ also converges. To see this, we note that for $p, q \in \mathbb{Z}_{\geq 0}$ with $q \leq p$, 
\begin{align}\label{eq:hamiltonian_diff_lemma_useful}
\norm{H(t) U^p(t, s) \Pi_{\leq p} \ket{\Psi_0} - H(t) U^q(t, s) \Pi_{\leq q} \ket{\Psi_0} } \leq  \int_s^t \norm{H(t) U^p(t, \tau)\ket{\Phi_{p, q}(\tau, s)} } d\tau+ \norm{H(t)U^p(t, s)\ket{\Gamma_{p, q}(t, s)}},
\end{align}
where $\ket{\Phi_{p, q}(\tau, s)}$ is defined in Eq.~\ref{eq:pert_theory_state} and $\ket{\Gamma_{p, q}(t, s)} = \big(\Pi_{\leq p} - \Pi_{\leq q}\big) \ket{\Psi_0}$. Now, from the definition of $H(t)$, it follows that
\[
\norm{H(t) U^p(t, s) \ket{\Gamma_{p, q}(t, s)}} \leq \norm{H_S(t)} \norm{ \ket{\Gamma_{p, q}(t, s)}} + 2 \ell \bigg( \norm{\ket{\Gamma_{p, q}(t, s)}}^2 + \mu^{(1)}_{U^p(t, s) \ket{\Gamma_{p, q}(t, s)}}\bigg)^{1/2}.
\]
Furthermore, using lemma \ref{lemma:particle_num_bound}, we obtain that
\[
\mu^{(1)}_{U^p(t, s) \ket{\Gamma_{p, q}(t, s)}} \leq 2\bigg(\mu^{(1)}_{\Gamma_{p, q}(t, s)} + \ell^2 \textnormal{max}^2(s, t) \norm{\ket{\Gamma_{p, q}(t, s)}}^4\bigg).
\]
Since $\ket{\Psi_0} \in \mathcal{H}_S\otimes \textnormal{F}_\mathcal{S}^M[L^2(\mathbb{R})]$, we obtain that
\begin{align*}
&\norm{\ket{\Gamma_{p, q}(t, s)}} \leq  \norm{\Pi_{> q} \ket{\Psi_0}} + \norm{\Pi_{> p}\ket{\Psi_0}} \to 0 \textnormal{ as } p, q \to \infty \ \text{and}, \\
&\mu^{(1)}_{\ket{\Gamma_{p, q}(t, s)}} \leq \mu^{(1)}_{\ket{\Pi_{>q} \ket{\Psi_0}}} + \mu^{(1)}_{\ket{\Pi_{>p} \ket{\Psi_0}}} \to 0 \textnormal{ as } p, q \to \infty,
\end{align*}
and therefore it follows from the previous estimates that $\norm{H(t) U^p(t, s) \ket{\Gamma_{p, q}(t, s)}} \to 0$ as $p, q \to \infty$. Consider now the second term in Eq.~\ref{eq:hamiltonian_diff_lemma_useful} --- since $U^p(t, \tau) \ket{\Phi_{p, q}(\tau, s)} \in \mathcal{H}_S \otimes \textnormal{F}_\infty^M[L^2(\mathbb{R})]$, we obtain that
\[
\int_s^t \norm{H(\tau)U^p(t, \tau) \ket{\Phi_{p, q}(\tau, s)}}d\tau \leq \int_s^t \norm{H_S(\tau)} \norm{\ket{\Phi_{p, q}(\tau, s)}} d\tau + 2\ell \int_s^t \bigg(\norm{\ket{\Phi_{p, q}(\tau, s)}}^2 + \mu_{U^p(t, \tau)\ket{\Phi_{p, q}(\tau, s)}}^{(1)}\bigg)^{1/2} d\tau.
\]
Using Eq.~\ref{eq:norm_estimate} and the bound from lemma \ref{lemma:particle_num_bound}, we obtain that
\[
\int_s^t \norm{H_S(\tau)} \norm{\ket{\Phi_{p, q}(\tau, s)}} d\tau \leq \frac{\ell \sqrt{q + 1}}{q} \sqrt{M^{(2)}_{\ket{\Psi_0}}(t_\textnormal{max})} \int_s^t \norm{H_S(\tau)}d\tau.
\]
It follows from lemma \ref{lemma:particle_num_bound} that for $\tau \in [\textnormal{min}(t, s), \textnormal{max}(t, s)]$,
\[
{\mu^{(1)}_{U^p(t, \tau)\ket{\Phi_{p, q}(\tau, s)}}} \leq 2 \bigg(\mu^{(1)}_{\ket{\Phi_{p, q}(\tau, s)}} + \ell^2 \textnormal{max}^2(s, t)\norm{\ket{\Phi_{p, q}(\tau, s)}}^4\bigg),
\]
and using Eq.~\ref{eq:norm_estimate} it follows that for $\tau \in [\min(t, s), \max(t, s)]$,
\[
\mu^{(1)}_{\ket{\Phi_{p, q}(\tau, s)}} \leq \frac{(q + 1)^2}{q^3} \ell^2 M^{(3)}_{\Pi_{\leq q}\ket{\Psi_0}}(\max(s, t)) \leq  \frac{(q + 1)^2}{q^3} \ell^2 M^{(3)}_{\ket{\Psi_0}}(\max(s, t)).
\]
From these estimates, it thus follows that 
\[
\int_s^t \norm{H(\tau) U^p(t, \tau) \ket{\Phi_{p, q}(\tau, s)}} d\tau \to 0 \textnormal{ as } p, q \to \infty.
\]
Therefore, from Eq.~\ref{eq:hamiltonian_diff_lemma_useful}, it follows that the sequence $\{H(t)U^p(t, s) \Pi_{\leq p}\ket{\Psi_0}\}_{p \in \mathbb{N}}$ converges --- since $H(t)$ is a closable operator, it then follows that
\[
\lim_{p \to \infty} H(t) U^p(t, s) \Pi_{\leq p}\ket{\Psi_0} = \overline{H(t)} \lim_{p \to \infty} U^p(t, s) \Pi_{\leq p}\ket{\Psi_0} = \overline{H(t)} \lim_{p \to \infty} U^p(t, s) \ket{\Psi_0}.
\]
Finally, we show that $\lim_{p\to \infty} H^p(t) U^p(t, s) \ket{\Psi_0} = \lim_{p\to \infty} H(t) U^p(t, s) \Pi_{\leq p} \ket{\Psi_0}$. We begin by noting that
\[
H^p(t) U^p(t, s) \ket{\Psi_0} = H^p(t) U^p(t, s) \Pi_{\leq p} \ket{\Psi_0} + \Pi_{> p} \ket{\Psi_0} \implies \lim_{p \to \infty} H^p(t) U^p(t, s) \ket{\Psi_0} = \lim_{p \to \infty} H^p(t) U^p(t, s) \Pi_{\leq p} \ket{\Psi_0} .
\]
Furthermore,
\begin{align*}
&\norm{(H(t) - H^p(t)) U^p(t, s) \Pi_{\leq p} \ket{\Psi_0}} ^2 \leq  \nonumber \\
&\qquad \qquad (p +1)\ell^2 \norm{\Pi_p  U^p(t, s) \ket{\Psi_0}}^2 \leq \frac{p + 1}{p^2}\ell^2 M^{(2)}_{\Pi_{\leq p}\ket{\Psi_0}}(\max(t, s)) \leq \frac{p + 1}{p^2}\ell^2 M^{(2)}_{\ket{\Psi_0}}(\max(t, s)).
\end{align*}
and thus $\lim_{p \to \infty} H^p(t) U^p(t, s) \Pi_{\leq p}\ket{\Psi_0} = \lim_{p \to \infty} H(t) U^p(t, s) \Pi_{\leq p}\ket{\Psi_0}$. Hence, we obtain that $\overline{H(t)} \lim_{p \to \infty} U^p(t, s) \ket{\Psi_0} = \lim_{p \to \infty} H^p(t) U^p(t, s) \ket{\Psi_0}$.

Now, we investigate the continuity of $\overline{H(t)} \lim_{p\to \infty} U^p(t, s) \ket{\Psi_0} = \lim_{p \to \infty} H^p(t) U^p(t, s) \Pi_{\leq p}\ket{\Psi_0}$ with respect to $t$. For $p \in \mathbb{Z}_{\geq 0}$ and $\delta > 0$, define $\Delta^p(\delta)$ va
\[
\Delta^p(\delta) = \norm{\bigg(H^p(t + \delta) U^p(t + \delta, s) -  H^p(t) U^p(t, s) \bigg)\Pi_{\leq p}\ket{\Psi_0}}.
\]
We need to show that $\lim_{\delta \to 0} \lim_{p\to \infty}\Delta^p(\delta) = 0$. To see this, we note that
\[
\Delta^p(\delta) \leq \norm{\bigg(H^p(t + \delta) - H^p(t)\bigg) U^p(t, s)\Pi_{\leq p} \ket{\Psi_0}} + \norm{H^p(t + \delta)\bigg(U^p(t + \delta, s) - U^p(t, s)\bigg) \Pi_{\leq p} \ket{\Psi_0}}.
\]
Now
\begin{align*}
 &\norm{\bigg(H^p(t + \delta) - H^p(t)\bigg) U^p(t, s) \ket{\Psi_0}}  \nonumber \\
 &\qquad \qquad \leq \norm{H_S(t + \delta) - H_S(t)} \norm{\ket{\Psi_0}} +\bigg(\norm{\ket{\Psi_0}}^2 + \mu^{(1)}_{U^p(t, s) \Pi_{\leq p}\ket{\Psi_0}}\bigg)^{1/2} \sum_{\alpha = 1}^M  \norm{L_\alpha}  \norm{\bigg(\uptau_{\alpha, t + \delta} - \uptau_{\alpha, t}\bigg) v_\alpha}_{L^2}, \nonumber \\
 &\qquad \qquad \leq \norm{H_S(t + \delta) - H_S(t)} \norm{\ket{\Psi_0}} +\bigg(\norm{\ket{\Psi_0}}^2 + M^{(1)}_{ \ket{\Psi_0}}\bigg)^{1/2} \sum_{\alpha = 1}^M  \norm{L_\alpha}  \norm{\bigg(\uptau_{\alpha, t + \delta} - \uptau_{\alpha, t}\bigg) v_\alpha}_{L^2},
\end{align*}
and consequently by the strong continuity of $\uptau_{\alpha, t}$,
\[
\lim_{\delta \to 0} \lim_{p \to \infty}\norm{\bigg(H^p(t + \delta) - H^p(t)\bigg) U^p(t, s) \ket{\Psi_0}}   = 0.
\]
Furthermore,
\[
 \norm{H^p(t + \delta)\bigg(U^p(t + \delta, s) - U^p(t, s)\bigg) \Pi_{\leq p} \ket{\Psi_0}} \leq \int_{t}^{t+ \delta} \norm{H^p(\tau + \delta) H^p(\tau) U^p(\tau, s) \Pi_{\leq p}\ket{\Psi_0}}d\tau.
\]
It follows from lemma \ref{lemma:particle_num_bound} that $\norm{H^p(\tau + \delta) H^p(\tau) U^p(\tau, s) \ket{\Psi_0}}$ is bounded above by a constant independent of $p$ and continuous in $\tau$. Thus, we obtain that
\[
\lim_{\delta \to 0 }\lim_{p \to \infty}  \norm{H^p(t + \delta)\bigg(U^p(t + \delta, s) - U^p(t, s)\bigg) \Pi_{\leq p} \ket{\Psi_0}} = 0.
\]
Thus, we obtain that $\lim_{\delta \to 0} \lim_{p \to \infty} \Delta^p(\delta) = 0$.
 \\ \ \\
 \emph{(c)} The proof of this part closely follows that of part (b), with only minor modifications which we outline here. We can show that $\lim_{p\to \infty} \norm{H^p(s) \ket{\Psi_0} - \overline{H(s)}\ket{\Psi_0}} = 0$, which would imply that $\lim_{p\to \infty} U^p(t, s) H^p(s) \ket{\Psi_0} = \lim_{p\to \infty} U^p(t, s) \overline{H(s)}\ket{\Psi_0}$, in two steps --- first, we establish that $\lim_{p \to \infty} H^p(s) \ket{\Psi_0} = \lim_{p \to \infty} H^p(s)\Pi_{\leq p} \ket{\Psi_0}$ using the fact that all the particle-number moments of $\ket{\Psi_0}$ are finite. Then, we can show that $\lim_{p\to \infty} H^p(s) \Pi_{\leq p} \ket{\Psi_0} = \lim_{p \to \infty} H(s) \Pi_{\leq p} \ket{\Psi_0}$ by analyzing the norm $\norm{(H(s) - H^p(s)) \Pi_{\leq p}\ket{\Psi_0}}$. By showing that the sequence $\{H(s) \Pi_{\leq p} \ket{\Psi_0} \}_{p \in \mathbb{N}}$ converges, and using the closability of $H(t)$, we then obtain that $\lim_{p \to \infty} H(s) \Pi_{\leq p} \ket{\Psi_0} = \overline{H(s)} \lim_{p \to \infty}  \Pi_{\leq p} \ket{\Psi_0}$. Finally, using lemma \ref{lemma:domain}b, you obtain that $\overline{H(s)}\lim_{p \to \infty} \Pi_{\leq p}\ket{\Psi_0} = \overline{H(s)} \ket{\Psi_0}$. To prove that $\lim_{p\to \infty}U^p(t, s) \overline{H(s)} \ket{\Psi_0} = \lim_{p\to \infty} U^p(t, s) H^p(s) \ket{\Psi_0}$ is strongly continuous, we can again analyze $\Delta^p(\delta)$, where
 \begin{align*}
 \Delta^p(\delta) &= \norm{\bigg(U^p(t, s+\delta) H^p(s+ \delta) - U^p(t, s) H^p(s)\bigg) \ket{\Psi_0}}, \nonumber\\
  &\leq \norm{\bigg(U^p(t, s + \delta) - U^p(t, s)\bigg) H^p(s + \delta) \ket{\Psi_0}} +  \norm{U^p(t, s)\bigg(H^p(t, s + \delta) -H^p(s)\bigg) \ket{\Psi_0}}.
 \end{align*}
Using lemma \ref{lemma:particle_num_bound}, we can show that $\lim_{p\to \infty} \lim_{\delta \to 0}\Delta^p(\delta) = 0$.
 \\ \ \\
\emph{(d)} This follows straightforwardly from lemma \ref{lemma:particle_num_bound}, and noting that 
\[
\norm{H^p(t) U^p(t, s) \ket{\Psi_0)}} \leq \norm{H_S(t)} \norm{\ket{\Psi_0}} + 2\ell \bigg(\norm{\ket{\Psi_0}}^2 + M^{(1)}_{\ket{\Psi_0}}(\max(t, s)) \bigg)^{1/2}.
\]
which yields the upper bound $g(t)$. Similarly,
\[
\norm{H^p(t) \ket{\Psi_0}} \leq \norm{H_S(t)} \norm{\ket{\Psi_0}} + 2\ell \bigg(\norm{\ket{\Psi_0}}^2 +\mu^{(1)}_{\ket{\Psi_0}}\bigg)^{1/2},
\]
which yields the upper bound $h(t)$.
\hfill \(\square\)

\begin{repproposition}{thm:schr_sq_int}
Given a non-Markovian model with square integrable coupling functions and for $t, s \in \mathbb{R}$, there exists a unique isometry $U(t, s) : \mathcal{H}_S \otimes \textnormal{F}_\infty^M[L^2(\mathbb{R})] \to \mathcal{H}_S \otimes \textnormal{F}_\infty^M[L^2(\mathbb{R})] \subseteq \mathcal{H}$ which is strongly continuous and differentiable in both $t, s$ and satisfies
\begin{align}\label{eq:prop_schr_eq}
\frac{d}{dt} U(t, s) = -i \overline{H(t)} U(t, s),\ \frac{d}{ds} U(t, s) = i U(t, s) \overline{H(s)},
\end{align}
with $U(s, s) = \textnormal{id} \ \forall s \in \mathbb{R}$.
\end{repproposition}
\noindent\emph{Proof}: We first construct the unitary group $U(t, s)$ --- given a state $\ket{\Psi_0} \in \mathcal{H}_S \otimes \textnormal{F}_\mathcal{S}^M[L^2(\mathbb{R})]$, we let
\[
U(t, s) \ket{\Psi_0} = \lim_{p \to \infty} U^p(t, s) \ket{\Psi_0}.
\]
It follows from lemma \ref{lemma:useful} that $U(t, s)$ is well defined, that $U(t, s) \ket{\Psi_0} \in \mathcal{H}_S \otimes \textnormal{F}_\mathcal{S}^M[L^2(\mathbb{R})]$ and that $U(t, s)$ is an isometry. Now, we note that since $U^p(t, s)$ is the propagator corresponding to $H^p(t)$, 
\[
U(t, s) \ket{\Psi_0} = \lim_{p\to \infty} U^p(t, s) \ket{\Psi_0} = \ket{\Psi_0}- i \lim_{p\to \infty} \int_s^t H^p(\tau) U^p(\tau, s) \ket{\Psi_0}d\tau.
\]
From lemma \ref{lemma:useful}(d), it follows that $\norm{H^p(\tau) U^p(\tau, s) \ket{\Psi_0}}$ is bounded above by a continuous (and thus integrable) function of $\tau$, and hence from the dominated convergence theorem, we obtain that
\[
U(t, s)\ket{\Psi_0} = \ket{\Psi_0} - i \int_s^t \lim_{p \to \infty}  H^p(\tau) U^p(\tau, s)\ket{\Psi_0}d\tau.
\]
Finally, using lemma \ref{lemma:useful}(b), we obtain that $\lim_{p\to \infty} H^p(\tau) U^p(\tau, s) \ket{\Psi_0} = \overline{H(\tau)}U(\tau, s) \ket{\Psi_0}$, and since $ \overline{H(\tau)}U(\tau, s) \ket{\Psi_0}$ is strongly continuous in $\tau$, $U(t, s) \ket{\Psi_0}$ is strongly differentiable in $t$. Thus,
\[
U(t, s) \ket{\Psi_0} =  \ket{\Psi_0} - i \int_s^t \overline{H(\tau)} U(\tau, s)\ket{\Psi_0}d\tau \implies \frac{d}{dt}U(t, s) \ket{\Psi_0} = -i \overline{H(t)} U(t, s)\ket{\Psi_0}.
\]
which shows that $d U(t, s)/dt = -i\overline{H(t)} U(t, s)$ (where derivatives are understood as strong derivatives) with $U(s, s) = \textnormal{id}$. A similar argument can be made using lemmas \ref{lemma:useful}(c) and \ref{lemma:useful}(d) to show that $d U(t,s) \ket{\Psi_0} / ds = iU(t, s) \overline{H(s)}$. 

Since $\forall t \in \mathbb{R}$, $H(t)$ is essentially self adjoint, $\overline{H(t)}$ is self adjoint --- from this, it immediately follows that if the solution to Eq.~\ref{eq:prop_schr_eq} exists, the it must be unique. To see this, we simply note that the self-adjointness of $\overline{H(t)}$ implies that $\norm{U(t, s)\ket{\Psi_0}} = \norm{U(s, s)\ket{\Psi_0}}$, and hence $\ket{\Psi_0} = 0 \implies \ket{\Psi_0} = 0 \ \forall t \geq 0$. Now if there were two distinct solutions $U_1(t, s), U_2(t, s)$ to Eq.~\ref{eq:prop_schr_eq}, then $\big(U_1(t, s) - U_2(t, s)\big)\ket{\Psi_0}$ would be a non-zero vector, which leads to a contradiction since by essential self adjointness of $H(t)$, $\norm{\big(U_1(t, s) - U_2(t, s)\big)\ket{\Psi_0}} = 0$. \hfill\(\square\)

\section{Proof of lemma \ref{lemma:main_error_bound_mu}}\label{app:distributional_mem_ker}
\begin{replemma}{lemma:main_error_bound_mu}
Consider $\mu \in \mathcal{M}(\mathbb{R})$ with the Lesbesgue decomposition $\mu = \mu_c + \mu_d$ with $\phi_c \in \textnormal{C}^0(\mathbb{R})$ given by $\phi_c(x) = \mu_c((-\infty, x]) \ \forall \ x \in \mathbb{R}$, and $\mu_d \cong \sum_{i \in I} \alpha_i \delta(x - y_i)$ for some $\{\alpha_i \in \mathbb{C}\}_{i \in I}, \{y_i \in \mathbb{R}\}_{i \in I}$ and finite and countably infinite index set $I$. Given a compact interval $[a, b] \subseteq \mathbb{R}$ and $f \in \textnormal{C}^1(\mathbb{R})$, define $\langle\mu_{[a, b]}^*, f\rangle $ by
\begin{align*}
&\langle \mu^*_{[a, b]}, f\rangle =\langle \mu_{c, [a, b]}^*, f\rangle +  \langle \mu_{d, [a, b]}^*, f\rangle \ \text{where} \\ 
&\langle\mu_{c, [a, b]}^*, f\rangle =\nonumber\\
&\qquad  f(b) \phi_c(b) - f(a) \phi_c(a) - \int_a^b \phi_c(x) f'(x) dx \ \text{ and}\\
&\langle\mu_{d, [a, b]}^*, f\rangle = \frac{1}{2} \sum_{i \in I | y_i \in \{a, b\}} \alpha_i f(y_i) + \sum_{i \in I | y_i \in (a, b)} \alpha_i f(y_i).
\end{align*}
Then, for every compact intervals $[a, b] \subseteq \mathbb{R}$, $\exists \Delta^0_{\mu; [a, b]}(\varepsilon), \Delta^1_{\mu; [a,b]}(\varepsilon) > 0 $ where $ \Delta^0_{\mu; [a, b]}(\varepsilon), \Delta^1_{\mu; [a,b]}(\varepsilon) \to 0$ as $\varepsilon \to 0$ such that $\forall \varepsilon \in (0, (b- a) / 2)$ and for any even (symmetric about $0$) positive function $\alpha \in \textnormal{C}_c^\infty(\mathbb{R})$ with $\textnormal{supp}(\alpha) \subseteq [-\varepsilon,\varepsilon]$ and $\int_{[-\varepsilon, \varepsilon]}\alpha(x) dx=1$,
\begin{align*}
&\abs{ \langle \mu^*_{ [a, b]}, f\rangle  - \langle \mu, \alpha \star (f\cdot \mathcal{I}_{[a, b]}) \rangle} \leq \nonumber\\
&\qquad \Delta^0_{\mu; [a, b]}(\varepsilon) \sup_{x \in [a, b]} \abs{f(x)} + \Delta^1_{\mu; [a, b]}(\varepsilon)\sup_{x \in [a, b]} \abs{f'(x)}.
\end{align*}
The functions $\Delta^0_{\mu; [a, b]}, \Delta^1_{\mu; [a, b]}$ will be called the error functions corresponding to $\mu$.
\end{replemma}
\noindent\emph{Proof}: Let $\rho \in \textnormal{C}_c^\infty(\mathbb{R})$ be a symmetric and positive-valued function with $\textnormal{supp}(\rho) \subseteq [-\varepsilon, \varepsilon]$ for some $\varepsilon > 0$. Let $f \in \textnormal{C}^1(\mathbb{R})$. For $[a, b] \subseteq \mathbb{R}$, define ${f}^\rho_{[a, b]} = \big(f \cdot \mathcal{I}_{[a, b]}\big)\star \rho$. 
\\ \emph{Analysis of the continuous part.} Now, we consider $\langle \mu_c, f^\rho_{[a, b]}\rangle$ --- since $f^\rho_{[a, b]} \in \textnormal{C}^1_c(\mathbb{R})$, we note that
\[
\langle \mu_c, f^\rho_{[a, b]}\rangle = - \int_{\mathbb{R}} \phi_c(x) f^{\rho'}_{[a, b]}(x) dx.
\]
We note that $\forall x \in \mathbb{R}$,
\[
 f^{\rho'}_{[a, b]}(x) = \int_a^b \rho'(x - y) f(y) dy = f(a) \rho(x - a) - f(b) \rho(x - b) + \int_a^b f'(y)\rho(x - y) dy.
\]
Therefore,
\[
\langle \mu_c, f^\rho_{[a, b]}\rangle = \int_{\mathbb{R}} f(b)\phi_c(y)\rho(y - b) dy - \int_{\mathbb{R}} f(a)\phi_c(y) \rho(y - a) dy - \int_{\substack{y \in \mathbb{R} \\ x \in [a, b]}} \phi_c(y) f'(x) \rho(y - x) dx dy.
\]
Since $\varepsilon < (b - a) / 2 \implies \textnormal{supp}(\rho) \subseteq [-(b - a)/2, (b - a)/2]$, this can be rewritten with integrals being only over compact intervals,
\[
\langle \mu_c, f^\rho_{[a, b]}\rangle = \int_{\frac{3a - b}{2}}^{\frac{3b - a}{2}} f(b)\phi_c(y)\rho(y - b) dy -  \int_{\frac{3a - b}{2}}^{\frac{3b - a}{2}}  f(a)\phi_c(y) \rho(y - a) dy - \int_{y = \frac{3a - b}{2}}^{\frac{3b - a}{2}} \int_{x = a}^b \phi_c(y) f'(x) \rho(y - x) dx dy.
\]
Now, since $\phi_c$ is continuous, it is uniformly continuous over the compact interval $[(3a - b) / 2, (3b - a)/2]$. Thus, $\exists \delta_{\mu_c; a, b}(\varepsilon)$ where $\delta_{\mu_c; [a, b]}(\varepsilon) \to 0$ as $\varepsilon\to 0$ such that $\forall y, y' \in [(3a - b) / 2, (3b - a)/2]$ with $\abs{y - y'}< \varepsilon$, $\abs{\phi_c(y) - \phi_c(y')} < \delta_{\mu_c; [a, b]}(\varepsilon)$. Using this, we obtain that $\forall x \in [a, b]$
\[
\abs{\phi_c(x) - \int_{y \in[(3a - b) / 2, (3b - a)/2]} \phi_c(y) \rho(x - y) dy} ={\int_{y \in [x-\varepsilon, x + \varepsilon]} \abs{\big(\phi_c(x) - \phi_c(y)\big)}\rho(x - y) dy} \leq \delta_{\mu_c; [a, b]}.
\]
It then follows that
\begin{align*}
&\abs{\langle \mu_c^*, f\rangle - \langle \mu_c, f^\rho_{[a, b]}\rangle} \leq \sum_{x \in \{a, b\}} \abs{f(x)}\abs{\phi_c(x) - \int_{[(3a - b) / 2, (3b - a)/2]} \phi_c(y)\rho(y - x)dy} + \nonumber\\
&\qquad \qquad \qquad \int_{x\in[a, b]}\abs{\phi_c(y) - \int_{y \in [(3a - b) / 2, (3b - a)/2]}\phi_c(y)\rho(x - y) dy} \abs{f'(x)} dx,
\end{align*}
and consequently,
\begin{subequations}\label{eq:mu_c_error_estimate}
\begin{align}
\abs{\langle \mu_c^*, f\rangle - \langle \mu_c, f^\rho_{[a, b]}\rangle} \leq \delta_{\mu_c; [a, b]}(\varepsilon)\bigg(2\sup_{x \in [a, b]}|f(x)| + \int_a^b \abs{f'(x)}dx\bigg) \leq \Delta^0_{\mu_c; [a, b]}(\varepsilon)\sup_{x \in [a, b]}|f(x)| + \Delta^1_{\mu_c; [a, b]}\sup_{x \in [a, b]} \abs{f'(x)},
\end{align}
where 
\begin{align}
\Delta^0_{\mu_c; [a, b]}(\varepsilon) = 2\delta_{\mu_c; [a, b]}(\varepsilon) \text{ and }\Delta^1_{\mu_c; [a, b]}(\varepsilon) = (b - a)\delta_{\mu_c; [a,b]}(\varepsilon),
\end{align}
\end{subequations}
both of which $\to 0$ as $\varepsilon \to 0$.
\\
\emph{Analysis of the atomic part.} We now consider $\langle \mu_d, f^\rho_{[a, b]}\rangle$ --- since $f^\rho_{[a, b]} \in \textnormal{C}_c^0(\mathbb{R})$, we obtain that
\[
\langle \mu_d, f^\rho_{[a, b]}\rangle = \sum_{i \in I} \alpha_i f^\rho_{[a, b]}(y_i).
\]
Therefore,
\[
\abs{\langle \mu_d^*, f\rangle - \langle \mu_d, f^\rho_{[a, b]}\rangle } \leq \sum_{i \in I | y_i \notin [a, b]} \abs{\alpha_i} \abs{f^\rho_{[a, b]}(y_i)} + \sum_{i \in I | y_i \in(a, b)} \abs{\alpha_i} \abs{f(y_i) - f^\rho_{[a, b]}(y_i)} + \sum_{i \in I | y_i \in \{a, b\}} \abs{\alpha_i} \abs{\frac{f(y_i)}{2} - f^\rho_{[a, b]}(y_i)}.
\]
Since $\text{supp}(f^\rho_{[a, b]}) \subseteq [a - \varepsilon, a + \varepsilon]$, and $\norm{f^\rho_{[a, b]}}_{L^\infty} \leq \sup_{x \in [a, b]}|f(x)|$, we obtain that
\[
\sum_{i \in I | y_i \notin [a, b]} \abs{\alpha_i} \abs{f^\rho_{[a, b]}(y_i)} \leq \bigg(\sup_{x\in [a, b]} |f(x)|\bigg) \bigg(\sum_{i \in I | y_i \in [a - \varepsilon, a)} |\alpha_i| + \sum_{i \in I | y_i \in (b , b + \varepsilon]} \abs{\alpha_i} \bigg).
\]
Furthermore, we note that for $x \in [a + \varepsilon, b - \varepsilon]$, we obtain that
\[
\abs{f(x) -  f^\rho_{[a, b]}(x)} \leq {\int_{[-\varepsilon,  \varepsilon]} \abs{f(x) - f(x - y)}\rho( y) dy} \leq \sup_{y \in [a, b]}\abs{f'(y)} \int_{[-\varepsilon, \varepsilon]}|y|\rho(y) dy \leq \varepsilon \sup_{y \in [a, b]}\abs{f'(y)},
\]
and thus we obtain that
\begin{align*}
&\sum_{i \in I | y_i \in (a, b)}\abs{\alpha_i} \abs{f(y_i) - f_{[a, b]}^\rho(y_i)} \leq 2\sup_{y \in [a, b]} \abs{f(y)} \sum_{\substack{i \in I | y_i \in (a, a + \varepsilon] \\ \text{or } y_i \in (b - \varepsilon, b] }} \abs{\alpha_i} + \varepsilon \sup_{y \in [a, b]} \abs{f'(y)} \sum_{i \in I | y_i \in (a + \varepsilon, b - \varepsilon)} \abs{\alpha_i}.
\end{align*}
Similarly, we can note that
\begin{align*}
&\abs{\frac{1}{2}f(a) -  f^\rho_{[a, b]}(a)} \leq \int_{[0, \varepsilon]} \abs{f(a) - f(a+ y)} \rho(y) dy \leq \sup_{y \in [a, b]}\abs{f'(y)} \int_{[0, \varepsilon]} y\rho(y) \leq \frac{\varepsilon}{2} \sup_{y \in [a, b]}\abs{f'(y)} \ \text{and}, \\
&\abs{\frac{1}{2}f(b) -  f^\rho_{[a, b]}(b)} \leq \int_{[0, \varepsilon]} \abs{f(b) - f(b - y)} \rho(y) dy \leq \sup_{y \in [a, b]}\abs{f'(y)} \int_{[0, \varepsilon]} y\rho(y) \leq \frac{\varepsilon}{2} \sup_{y \in [a, b]}\abs{f'(y)},
\end{align*}
and thus we obtain that
\begin{align*}
\sum_{i \in I | y_i \in \{a, b\}} \abs{\alpha_i} \abs{\frac{f(y_i)}{2} - f^\rho_{[a, b]}(y_i)} \leq \frac{\varepsilon}{2} \sup_{y \in [a, b]}\abs{f'(y)} \sum_{i \in I | y_i \in \{a, b\}} \abs{\alpha_i}.
\end{align*}
Therefore, we obtain that
\begin{subequations}\label{eq:mu_d_error_estimate}
\begin{align}
\abs{\langle \mu_d^*, f\rangle - \langle \mu_d, f^\rho_{[a, b]}\rangle } \leq \Delta_{\mu_d; [a,b]}^0(\varepsilon) \sup_{y \in [a, b]} \abs{f(y)} + \Delta_{\mu_d; [a,b]}^1(\varepsilon) \sup_{y \in [a, b]} \abs{f'(y)},
\end{align}
where
\begin{align}
\Delta_{\mu_d; [a,b]}^0(\varepsilon) = \sum_{\substack{i \in I | y_i \in [a-\varepsilon, a) \text{ or} \\ y_i \in (b, b+\varepsilon]}} \abs{\alpha_i} + 2\sum_{\substack{i \in I | y_i \in (a, a + \varepsilon] \\ \text{or } y_i \in (b - \varepsilon, b] }} \abs{\alpha_i} \text{ and } \Delta^1_{\mu_d; [a, b]}(\varepsilon) = \varepsilon\bigg(\sum_{i \in I | y_i \in (a+ \varepsilon, b - \varepsilon)}\abs{\alpha_i} + \frac{1}{2}\sum_{i \in I | y_i \in \{a, b\}} \abs{\alpha_i}\bigg).
\end{align}
\end{subequations}
We can note that, by construction, $\Delta^0_{\mu_d; [a, b]}(\varepsilon), \Delta^1_{\mu_d; [a, b]}(\varepsilon) \to 0$ as $\varepsilon \to 0$. Using Eqs.~\ref{eq:mu_c_error_estimate} and \ref{eq:mu_d_error_estimate}, we obtain the lemma statement. \hfill \(\square\)

\section{Proof of lemma \ref{lemma:inc_state_trunc}} \label{app:lemma_inc_state_trunc}
\begin{replemma}{lemma:inc_state_trunc}
Let $(\mu,\varphi)$ be a distributional coupling function. Given two mollifiers $\rho, \sigma \in \textnormal{C}_c^\infty(\mathbb{R})$ and $\varepsilon, \delta > 0$, let $v_\varepsilon$ and $v_\delta \in L^2(\mathbb{R})$ be the $\varepsilon, \rho-$ and $\delta, \sigma-$regularization of $(\mu, \varphi)$ respectively. Let $\mathcal{H}_S$ be Hilbert space, then, 
\begin{enumerate}
\item[(a)]$\forall \ket{\Phi} \in \mathcal{H}_S \otimes \textnormal{F}_{\infty, \mathcal{S}}^M$, $\exists c_{\mu, \ket{\Phi}} > 0$, $\forall \tau \geq 0$, $\forall \alpha \in \{1, 2 \dots M\}, \varepsilon > 0$ such that $\norm{a_{\alpha, \uptau_\tau v_\varepsilon}^- \ket{\Phi}} \leq c_{\mu, \ket{\Phi}}$.
\item[(b)] $\forall \ket{\Phi} \in \mathcal{H}_S \otimes \textnormal{F}_{\infty, \mathcal{S}}^M$, $\exists c_{\mu,\rho, \ket{\Phi}}, d_{\mu, \sigma, \ket{\Phi}} > 0$, $\forall \tau \geq 0, \forall \alpha \in \{1, 2 \dots M\}$, $\varepsilon, \delta > 0$ such that $\norm{a_{\alpha, \uptau_\tau(v_\varepsilon - v_\delta)}^- \ket{\Phi}} \leq c_{\mu,\rho, \ket{\Phi}}\varepsilon + c_{\mu, \sigma, \ket{\Phi}} \delta$.
\end{enumerate}
\end{replemma}

\noindent\emph{Proof}: Any state $\ket{\Psi} \in \mathcal{H}_S \otimes \textnormal{F}_{\infty, \mathcal{S}}^M$ can be expressed as
\[
\ket{\Psi} = \sum_{j = 1}^N \ket{\sigma_j} \otimes \ket{u_j}^{\otimes n_j},
\]
for some $N \in \mathbb{Z}_{\geq 1}$, and
\[
\{\ket{\sigma_j} \in \mathcal{H}_S\}_{j \in \{1, 2 \dots N\}}, \bigg\{\ket{u_j} = \bigoplus_{\alpha = 1}^M \ket{u_{\alpha, j}}, u_{\alpha, j} \in \mathcal{S}(\mathbb{R}) \ \forall \ \alpha \in \{1, 2 \dots M\}\bigg \}_{j \in \{1, 2 \dots N\}}\ \text{and } \{n_j \in \mathbb{Z}_{\geq 0}\}_{j \in\{1,2 \dots N\}}.
\]
\noindent\emph{(a)} We obtain that
\begin{align*}
&\norm{a^{-}_{\alpha, \uptau_\tau v_\varepsilon} \ket{\Phi}} \leq\sum_{j = 1}^N \sqrt{n_j} \norm{\sigma_j}\norm{u_j}^{n_j - 1} \bigg| \int_{\mathbb{R}} \sqrt{\hat{\mu}(\omega)} \hat{\rho}(\omega \varepsilon) u_{\alpha, j}(\omega) d\omega\bigg|, \nonumber\\
&\qquad\leq \sum_{j = 1}^N \sqrt{n_j} \norm{\sigma_j}\norm{u_j}^{n_j - 1} \int_{\mathbb{R}} \bigg|\sqrt{\hat{\mu}(\omega)} \hat{\rho}(\omega \varepsilon) u_{\alpha, j}(\omega)\bigg| d\omega.
\end{align*}
Note that by assumption, $|\sqrt{\hat{\mu}(\omega)}|$ is a continuous function of at-most polynomial growth. Since $\forall j \in \{1, 2\dots N\}, \alpha \in \{1, 2 \dots M\}$, $u_{\alpha, j} \in \mathcal{S}(\mathbb{R})$, 
\[
\norm{\sqrt{\hat{\mu}(\omega)} u_{\alpha, j}(\omega) ( 1 + \omega^2)}_{L^\infty} < \infty.
\]
Furthermore, note that since $\rho$ is a mollifier,
\[
\norm{\hat{\rho}}_{L^\infty} \leq \frac{1}{\sqrt{2\pi}} \int_{\mathbb{R}} |\rho(s)| ds = \frac{1}{\sqrt{2\pi}}.
\]
Therefore,
\begin{align*}
&\norm{a^{-}_{\alpha, \uptau_\tau v_\varepsilon} \ket{\Phi}}  \leq \sum_{j = 1}^N  \sqrt{\frac{n_j}{2\pi} } \norm{\sigma_j}\norm{u_j}^{n_j - 1}\norm{\sqrt{\hat{\mu}(\omega)} u_{\alpha, j}(\omega) ( 1 + \omega^2)}_{L^\infty}  \int_{\mathbb{R}} \frac{d\omega}{1 + \omega^2},  \nonumber\\
&\qquad \leq \sup_{\alpha \in \{1, 2 \dots M\}} \bigg(\sum_{j =1}^N \sqrt{\frac{\pi n_j}{2}}\norm{\sigma_j}\norm{u_j}^{n_j - 1}\norm{\sqrt{\hat{\mu}(\omega)} u_{\alpha, j}(\omega) ( 1 + \omega^2)}_{L^\infty} \bigg),
\end{align*}
where the bound can be identified as the constant $c_{\mu, \ket{\Psi}}$ in the lemma statement. \\ 

\noindent\emph{(b)} We obtain that
\begin{align*}
&\norm{a^{-}_{\alpha, \uptau_\tau(v_\varepsilon - v_{\delta})} \ket{\Phi}} \leq\sum_{j = 1}^N \sqrt{n_j} \norm{\sigma_j}\norm{u_j}^{n_j - 1} \bigg| \int_{\mathbb{R}} u_{\alpha, j}(\omega) \sqrt{\hat{\mu}(\omega)} \big({\hat{\rho}(\omega\varepsilon) }- {\hat{\sigma}(\omega \delta)}\big)d\omega \bigg| \nonumber\\
&\qquad\leq \sum_{j = 1}^N \sqrt{n_j} \norm{\sigma_j}\norm{u_j}^{n_j - 1}  \int_{\mathbb{R}} \bigg|\sqrt{\hat{\mu}(\omega)} u_{\alpha, j}(\omega)\big(\hat{\rho}(\omega \varepsilon) - \hat{\sigma}(\omega \delta)\big) \bigg| d\omega.
\end{align*}
Again, since by assumption $|\sqrt{\hat{\mu}(\omega)}|$ is a function of at-most polynomial growth, and $\forall \alpha \in \{1, 2 \dots M\}, j \in \{1, 2 \dots N\}$, $u_{\alpha, j} \in \mathcal{S}(\mathbb{R})$ and therefore,
\[
\norm{\sqrt{\hat{\mu}(\omega)} u_{\alpha, j}(\omega)(1 + \omega^2)^2}_{L^\infty} < \infty.
\]
Furthermore, since $\rho, \sigma \in \textnormal{C}_c^\infty(\mathbb{R}) \subseteq \mathcal{S}(\mathbb{R})$, $\hat{\rho}, \hat{\sigma} \in \mathcal{S}(\mathbb{R})$. In particular, $\norm{\hat{\rho}'}_{L^\infty} , \norm{\hat{\sigma}'}_{L^\infty} < \infty$. Furthermore,
\[
\hat{\rho}(0) - \hat{\sigma}(0) = \frac{1}{\sqrt{2\pi}} \int_{\mathbb{R}} \rho(s)ds - \frac{1}{\sqrt{2\pi}}\int_{\mathbb{R}}\sigma(s) ds = 0,
\]
since the mollifiers are, by definition, normalized to have unit area. Thus, using the Taylor's remainder theorem, we obtain that
\[
\big|\hat{\rho}(\omega \varepsilon) - \hat{\sigma}(\omega \delta)\big |  \leq  \norm{\hat{\rho}'}_{L^\infty}\big|\omega \big| \varepsilon + \norm{\hat{\sigma}'}_{L^\infty}\big|\omega \big| \delta \ \forall \ \omega \in \mathbb{R}.
\]
We thus obtain that
\[
\norm{a^-_{\alpha, \uptau_\tau(v_\varepsilon - v_\delta)} \ket{\Phi}} \leq \max_{\alpha\in\{1, 2 \dots M\}}\sum_{j = 1}^N \sqrt{n_j} \norm{\sigma_j}\norm{u_j}^{n_j - 1} \norm{\sqrt{\hat{\mu}(\omega)} u_{\alpha, j}(\omega)(1 + \omega^2)^2}_{L^\infty} \int_{\mathbb{R}}\frac{\norm{\hat{\rho}'}_{L^\infty}\big|\omega \big| \varepsilon + \norm{\hat{\sigma}'}_{L^\infty}\big|\omega \big| \delta}{(1 + \omega^2)^2} d\omega.
\]
Noting that $\int_{\mathbb{R}} |\omega| / (1 + \omega^2)^2 d\omega <\infty$, we obtain the lemma statement with the constants
\[
c_{\mu, \rho, \ket{\Phi}} = \max_{\alpha \in \{1, 2\dots M\}} \sum_{j = 1}^N\sqrt{n_j} \norm{\sigma_j}\norm{u_j}^{n_j - 1} \norm{\sqrt{\hat{\mu}(\omega)} u_{\alpha, j}(\omega)(1 + \omega^2)^2}_{L^\infty} \norm{\hat{\rho}'}_{L^\infty} \int_{\mathbb{R}}\frac{|\omega|}{(1 + \omega^2)^2}d\omega.  \ \ \ \ \ \ \ \ \ \ \square
\]
%%%%%%%%%%%%%%%%%%%%%%%%%%%%Appendix: Chain Approximation %%%%%%%%%%%%%%%%%%%%%%%%%%%%%%%
\section{Chain approximation}\label{app:chain_appx}
We first analyze the unitary group describing the single-particle dynamics of an environment that has been approximated by a 1D chain of a discrete set of modes generated from the star-to-chain transformation. This unitary group should approximate the time-translation unitary group --- following Ref.~\cite{trivedi2021convergence}, we provide a bound on the distance between two unitary groups as a function of the number of modes in the chain.
\begin{definition}[Chain unitary group on $L^2(\mathbb{R})$]\label{def:chain_unitary} 
Given $v \in L^2(\mathbb{R})$ with $\textnormal{supp}(\hat{v}) \in [-\omega_c, \omega_c]$ for some $\omega_c > 0$, a chain unitary group with $N_m$ modes generated by $v$ is the strongly continuous single parameter unitary group $\upnu_t : L^2(\mathbb{R}) \to L^2(\mathbb{R})$ defined by
\[
\upnu_t f = \sum_{\beta = 1}^{N_m} c_\beta(t) \varphi_\beta + \bigg(f - \sum_{\beta = 1}^N \langle \varphi_\beta, f\rangle \varphi_\beta \bigg),
\]
where 
\begin{enumerate}
\item[(a)] $\{\varphi_\beta \in L^2(\mathbb{R}) \}_{\beta \in \{1, 2 \dots N_m\}}$, called the mode functions, are a set of orthonormal functions (i.e.~$\langle \varphi_\alpha, \varphi_\beta \rangle = \delta_{\alpha, \beta}$ that are given by
\[
\hat{\varphi}_\alpha(\omega) = \frac{p_\alpha(\omega) \hat{v}(\omega)}{\big[\int_{-\omega_c}^{\omega_c} p_\alpha^2(\omega) |\hat{v}(\omega)|^2 d\omega \big]^{1/2}} \ \forall \ \alpha \in \{1, 2 \dots N_m\},
\]
where $p_\alpha$ is a degree $\alpha - 1$ polynomial generated by the following recursion starting from $p_1(\omega) = 1, B_1 = 0$,
\begin{align}\label{eq:lanczos_recurrence}
a_{\alpha} = \frac{\int_{-\omega_c}^{\omega_c} \omega p^2_\alpha(\omega) |\hat{v}(\omega)|^2d\omega}{\int_{-\omega_c}^{\omega_c}p_\alpha^2(\omega)|\hat{v}(\omega)|^2 d\omega}, p_{\alpha + 1}(\omega) = (\omega - A_\alpha) p_\alpha(\omega) - B_{\alpha - 1} p_{\alpha - 1}(\omega), B_\alpha = \frac{\int_{-\omega_c}^{\omega_c}p^2_{\alpha - 1}(\omega) |\hat{v}(\omega)|^2 d\omega}{\int_{-\omega_c}^{\omega_c}p^2_\alpha(\omega)|\hat{v}(\omega)|^2 d\omega}.
\end{align}
\item[(b)] The coefficients $\{c_\beta(t) \in \mathbb{C}\}_{\beta \in \{1, 2 \dots M\}}$ are given by the dynamical law: $c_\beta(0) = \langle \varphi_\beta, f\rangle$ for $\beta \in \{1, 2 \dots M\}$, together with
\begin{align}
i\frac{d}{dt} \begin{bmatrix}
c_1(t) \\
c_2(t) \\
c_3(t) \\
\vdots \\
c_M(t)
\end{bmatrix} = 
\begin{bmatrix}
\omega_1 & t_1 & 0 & 0 &\dots & 0 \\
t_1 & \omega_2 & t_2 & 0 &\dots & 0 \\
0 & t_2 & \omega_3 & t_3 & \dots & 0 \\
\vdots & \vdots & \vdots & \vdots & \ddots & \vdots \\
0 & 0 & 0 & 0 & \dots & \omega_M
\end{bmatrix}
\begin{bmatrix}
c_1(t) \\
c_2(t) \\
c_3(t) \\
\vdots \\
c_M(t)
\end{bmatrix}
\end{align}
where $\omega_\alpha = A_\alpha \ \forall \alpha \in \{1, 2 \dots M\}$ and $t_\alpha = \sqrt{B_{\alpha + 1}} \ \forall \ \alpha \in \{1, 2 \dots N_m\}$, 
\end{enumerate}
\end{definition}
For completeness, we provide a simple and well-known upper bound on the coefficients $\{\omega_\alpha\}_{\alpha \in \{1,2 \dots N_m\}}$ and $\{t_\alpha\}_{\alpha\in \{1, 2 \dots N_m - 1\}}$ which will be useful in the following sections.
\begin{lemma}[Upper bound on $\omega_\alpha, t_\alpha$ (Ref.~\cite{chin2010exact})]
\label{lemma:upper_bound_chain_coeffs}
Given a chain unitary group with $N_m$ modes generated by $v \in L^2(\mathbb{R})$ with $\textnormal{supp}(\hat{v}) \subseteq [-\omega_c, \omega_c]$ for $\omega_c \geq 0$, then
\[
\abs{\omega_\alpha} \leq \omega_c \ \text{and } t_\alpha \leq \omega_c \ \forall \ \alpha \in \{1, 2 \dots N_m\},
\]
where $\{\omega_\alpha \}_{\alpha \in \{1, 2 \dots N_m\}}$ and $\{t_\alpha \}_{\alpha \in \{1, 2 \dots N_m\}}$ are the parameters of the chain unitary group defined in definition \ref{def:chain_unitary}.
\end{lemma}

\begin{lemma}[Ref.~\cite{trivedi2021convergence}]
\label{lemma:truncation_chain_group}
Given $v \in L^2(\mathbb{R}) \cap L^\infty(\mathbb{R})$ with $\textnormal{supp}(f) \in [-\omega_c, \omega_c]$, let $\upnu_t : L^2(\mathbb{R}) \to L^2(\mathbb{R})$ be the chain unitary group with $N_m$ modes generated by $v$ (definition \ref{def:chain_unitary}), then $\forall t \geq 0$,
\[
\frac{1}{2}\norm{\uptau_t v - \upnu_t v}_{L^2}^2 \leq \norm{v}_{L^2}^2  N_m^2 e^{N_m} \bigg(\frac{2\omega_c t}{N_m}\bigg)^{N_m},
\]
where $\uptau_t : L^2(\mathbb{R}) \to L^2(\mathbb{R})$ is the translation group $[(\uptau_t f)(x) = f(x + t) \ \forall \ f \in L^2(\mathbb{R})]$.
\end{lemma}
\noindent\emph{Proof}: We define the polynomial $\pi_\alpha$ of degree $\alpha - 1$ via
\[
\pi_\alpha = \frac{\norm{v}_{L^2}}{\norm{p_i \hat{v}}_{L^2}} p_\alpha  \ \text{for } \alpha \in \{1, 2 \dots N_m\}.
\]
We will denote by $A \in \mathbb{R}^{N_m \times N_m}$ the matrix
\[
A = \begin{bmatrix}
\omega_1 & t_1 & 0 & 0 &\dots & 0 \\
t_1 & \omega_2 & t_2 & 0 &\dots & 0 \\
0 & t_2 & \omega_3 & t_3 & \dots & 0 \\
\vdots & \vdots & \vdots & \vdots & \ddots & \vdots \\
0 & 0 & 0 & 0 & \dots & \omega_{N_m}
\end{bmatrix},
\]
We will denote by $\lambda_i \in \mathbb{R}$ and $u^i\in \mathbb{R}^{N_m}, i \in \{1, 2 \dots N_m\}$ the eigenvalues and eigenvectors of the matrix $A$. We note that if $(\lambda \in \mathbb{R}, u \in \mathbb{R}^{N_m})$ is an eigenvalue, eigenvector pair of $A$, then
\begin{align*}
&\big(\omega_1 - \lambda\big) u_1 + t_1 u_{2} = 0, \\
&\big(\omega_i - \lambda\big) u_i + t_{i - 1} u_{i - 1} + t_i u_{i + 1} = 0 \ \text{for }i\in \{2, 3 \dots N_m - 1\}, \\
&\big(\omega_{N_m} - \lambda \big)u_{N_m} + t_{N_m - 1} u_{N_m - 1} = 0.
\end{align*}
Solving these recursions, we obtain that
\[
u_i = u_1 \pi_i(\lambda) \ \text{and } \pi_{N_m + 1}(\Omega) = 0 \ (\text{or } p_{N_m + 1}(\Omega) = 0).
\]
Therefore, the eigenvalues $\lambda_1, \lambda_2 \dots \lambda_{N_m}$ are the roots of the polynomial $p_{N_m + 1}$, and the eigenvectors are given by
\[
u^i_j = \frac{\pi_j(\lambda_i)}{N_i} \ \text{where } N_i = \bigg(\sum_{j = 1}^{N_m}\pi_j^2(\lambda_i) \bigg)^{1/2}
\]
It can be noted that the matrix $A$ is hermitian, and consequently, its eigenvectors for an orthonormal basis for $\mathbb{R}^{N_m}$, which implies that
\begin{align}\label{eq:orthonormality}
\sum_{j = 1}^{N_m} \pi_j(\lambda_i) \pi_j(\lambda_{i'}) = N_i^2 \delta_{i, i'} \text{ and } \sum_{i = 1}^{N_m} \pi_j(\lambda_i) \pi_{j'}(\lambda_i) = N_i^2 \delta_{j, j'}.
\end{align}
We next compute $\upnu_t v$ --- noting that $v \propto \varphi_1$, we obtain that
\[
\upnu_t v = \sum_{\beta = 1}^{N_m} c_\beta(t) \varphi_\beta, \ \text{where } \begin{bmatrix}
c_1(t) \\
c_2(t) \\
\vdots \\
c_{N_m}(t)
\end{bmatrix} = \norm{v}e^{-iAt} 
\begin{bmatrix}
1\\
0 \\
\vdots \\
0
\end{bmatrix},
\]
which can be rewritten as
\[
e^{-iAt} \begin{bmatrix}
1 \\
0 \\
\vdots \\
0
\end{bmatrix} = \sum_{i = 1}^{N_m} \frac{\pi_1(\lambda_i)}{N_i}e^{-i\lambda_i t} u^i \implies c_j(t) =  \norm{v}_{L^2}\sum_{i = 1}^{N_m} \frac{\pi_1(\lambda_i) \pi_j(\lambda_i)}{N_i^2}e^{-i\lambda_i t}.
\]
We now consider
\begin{align*}
&\frac{1}{2}\norm{\uptau_t v - \upnu_t v}_{L^2}^2 = \norm{v}_{L^2}^2 - \sum_{j = 1}^{N_m} \text{Re}\bigg(c_j^*(t) \int_{\mathbb{R}} \hat{\varphi}_j^*(\omega) \hat{v}(\omega) e^{-i\omega t} d\omega \bigg) \nonumber\\
&\qquad=  \norm{v}_{L^2}^2 - \sum_{i, j = 1}^{N_m} \frac{\pi_1(\lambda_i)\pi_j(\lambda_i)}{N_i^2} \int_{-\omega_c}^{\omega_c} \pi_j(\omega) |\hat{v}(\omega)|^2 \cos((\omega - \lambda_i)t) d\omega.
\end{align*}
We next use the Gauss quadrature theorem --- we note that the polynomials $p_1, p_2 \dots p_{N_m}$ are the polynomials that would be used to approximate the integral of $f(\omega) |\hat{v}(\omega)|^2$ in the interval $[-\omega_c, \omega_c]$ with Gaussian quadrature with $N_m$ interpolating points. In particular, for every $N_m \in \mathbb{Z}_{> 1}$, $\exists w \in [0, \infty)^{N_m}$ with $\norm{w}_{1} = 1$ such that for all polynomials $q$ of degree $\leq 2 N_m - 1$, 
\[
\frac{1}{ \norm{v}_{L_2}^2}\int_{-\omega_c}^{\omega_c} q(\omega) |\hat{v}(\omega)|^2d\omega = \sum_{i = 1}^{N_m} w_i q(\lambda_i).
\]
Note that from the Taylor's remainder theorem, it follows that, 
\[
\forall \omega \in [-\omega_c, \omega_c], \ \cos(\omega t) = q_{N_m}(\omega) + r_{N_m}(\omega),
\]
where $q_{N_m}$ is a polynomial of degree $N_m$ with $q_{N_m}(0) = 1$, and 
\begin{align}\label{eq:estimate_inf_norm_remainder}
\sup_{\omega \in [-2\omega_c, 2\omega_c]} |r_{N_m}(\omega)| \leq (2\omega_c t)^{N_m + 1} / (N_m + 1)!.
\end{align}
We thus obtain that
\begin{align*}
&\frac{1}{2}\norm{\uptau_t v - \upnu_t v}_{L^2}^2  =\nonumber\\
 &\norm{v}_{L^2}^2 - \sum_{i, j = 1}^{N_m} \frac{\pi_1(\lambda_i)\pi_j(\lambda_i)}{N_i^2} \int_{-\omega_c}^{\omega_c} \pi_j(\omega) |\hat{v}(\omega)|^2 q_{N_m}(\omega - \lambda_i) d\omega - \sum_{i, j = 1}^{N_m}\frac{\pi_1(\lambda_i)\pi_j(\lambda_i)}{N_i^2} \int_{-\omega_c}^{\omega_c} \pi_j(\omega) |\hat{v}(\omega)|^2 r_{N_m}(\omega - \lambda_i) d\omega.
\end{align*}
Now, from the Gauss quadrature theorem, it follows that since for $j \in \{1, 2 \dots N_m\}$ degree of $\pi_j(\omega) q_{N_m}(\omega - \lambda_j) \leq 2N_m - 1$
\begin{align*}
\int_{-\omega_c}^{\omega_c} \pi_j(\omega) | \hat{v}(\omega)|^2 q_{N_m}(\omega - \lambda_i) d\omega = \norm{v}_{L_2}^2\sum_{k = 1}^{N_m} w_k \pi_j(\lambda_k) q_{N_m}(\lambda_k - \lambda_i),
\end{align*}
and therefore
\begin{align*}
 \sum_{i, j = 1}^{N_m} \frac{\pi_1(\lambda_i)\pi_j(\lambda_i)}{N_i^2} \int_{-\omega_c}^{\omega_c} \pi_j(\omega) |\hat{v}(\omega)|^2 q_{N_m}(\omega - \lambda_i) d\omega = \norm{v}_{L^2}^2 \sum_{i, j, k = 1}^{N_m}w_k \frac{\pi_j(\lambda_i) \pi_j(\lambda_k)}{N_i^2} q_{N_m}(\lambda_k - \lambda_i).
\end{align*}
where we have used that $\pi_1(\omega) = 1 \ \forall \omega\in\mathbb{R}$. Furthermore, using Eq.~\ref{eq:orthonormality}, we obtain that
\[
\sum_{i, j = 1}^{N_m} \frac{\pi_1(\lambda_i)\pi_j(\lambda_i)}{N_i^2} \int_{-\omega_c}^{\omega_c} \pi_j(\omega) |\hat{v}(\omega)|^2 q_{N_m}(\omega - \lambda_i) d\omega = \norm{v}_{L^2}^2 \norm{w}_{1} = \norm{v}_{L_2}^2,
\]
and therefore
\[
\frac{1}{2}\norm{\uptau_t v - \upnu_t v}_{L^2}^2 = -\sum_{i, j = 1}^{N_m} \frac{\pi_j(\lambda_i)}{N_i^2} \int_{-\omega_c}^{\omega_c} \pi_j(\omega) |\hat{v}(\omega)|^2 r_{N_m}(\omega - \lambda_i) d\omega.
\]
Since for $i \in \{1, 2 \dots N_m\}, \lambda_i \in [-\omega_c, \omega_c]$ and therefore
\[
\frac{1}{2}\norm{\uptau_t v - \upnu_t v}_{L^2}^2 \leq \sum_{i, j = 1}^{N_m}\bigg|\frac{\pi_j(\lambda_i)}{N_i^2} \bigg| \sup_{\omega\in[-2\omega_c, 2\omega_c]} |r_{N_m}(\omega)| \int_{-\omega_c}^{\omega_c} |\pi_j(\omega)| |\hat{v}(\omega)|^2 d\omega
\]
Note that $\forall i, j \in \{1, 2 \dots N_m\}, |\pi_j(\lambda_i) | \leq N_i$ and $N_i \geq 1$. Using this and the estimate in Eq.~\ref{eq:estimate_inf_norm_remainder}, we obtain that
\[
\frac{1}{2}\norm{\uptau_t v - \upnu_t v}_{L^2}^2 \leq \frac{(2\omega_c t)^{N_m + 1}}{(N_m + 1)!} \sum_{i, j = 1}^{N_m} \int_{-\omega_c}^{\omega_c} |\pi_j(\omega)| |\hat{v}(\omega)|^2 d\omega \leq  \frac{(2\omega_c t)^{N_m + 1}}{(N_m + 1)!} \sum_{i, j = 1}^{N_m} \norm{\hat{v} \pi_j}_{L^2} \norm{v}_{L^2} = \frac{(2\omega_c t)^{N_m + 1}}{(N_m + 1)!} N_m^2 \norm{v}_{L^2}^2.
\]
Finally, using Stirling's approximation to estimate $(N_m + 1)! \geq  (N_m + 1)^{N_m + 1} e^{-N_m} \geq N_m^{N_m + 1} e^{-N_m}$, we obtain that
\[
\frac{1}{2}\norm{\uptau_t v - \upnu_t v}_{L^2}^2 \leq \norm{v}_{L^2}^2  N_m^2  \bigg(\frac{2e\omega_c t}{N_m}\bigg)^{N_m},
\]
which proves the lemma statement. \hfill \(\square\)

\begin{replemma}{lemma:star_to_chain}[Star-to-chain transformation]
Consider a non-Markovian model with system Hamiltonian $H_S(t)$, square-integrable coupling functions $\{v_\alpha \in \text{L}^2(\mathbb{R})\}_{\alpha \in \{1, 2 \dots M\}}$ and jump operators $\{L_\alpha\}_{\alpha \in \{1, 2 \dots M\}}$. Furthermore, assume that $\exists\ \omega_c > 0$ such that for $\abs{\omega} \geq \omega_c, v_\alpha(\omega) = 0$. Then there exists a chain dilation of this non-Markovian model with $N_m$ modes and bandwidth $\leq \omega_c$ such that
\begin{align*}
&\norm{\ket{\Psi(t)} - \ket{\hat{\Psi}(t)}} \leq 4\ell t \bigg(1 + \ell^2 t^2  + \mu^{(1)}_{\ket{\Psi_0}}\bigg)^{1/2} N_m \bigg(\frac{2e\omega_c t}{N_m}\bigg)^{N_m/2},
\end{align*}
where $\ket{\Psi(t)}$ and $\ket{\hat{\Psi}(t)}$ are the system-environment states obtained from the model and its chain dilation, $\ket{\Psi(0)} = \ket{\hat{\Psi(0)}} = \ket{\Psi_0}$, $\ell = \sum_{\alpha = 1}^M \norm{v_\alpha}_{L^2} \norm{L_\alpha}$ and $\mu^{(1)}_{\ket{\Psi_0}}$ is the initial expectation value of the particle number operator of the environment.
\end{replemma}
\noindent\emph{Proof}: Let $U(t, s)$ and $\hat{U}(t, s)$ be the propagators corresponding to the exact model and its chain dilation respectively --- we note that both $U(t, s)\ket{\Psi}$ and $\hat{U}(t, s)\ket{\Psi}$ are strongly differentiable with respect to $t$ and $s$ if $\ket{\Psi} \in \mathcal{H}_S \otimes \textnormal{F}_1^M[L^2(\mathbb{R})]$. Consider now,
\[
\norm{\ket{\Psi(t)} - \ket{\hat{\Psi}(t)}} = \norm{\ket{\Psi_0} - \hat{U}(0, t) U(t, 0) \ket{\Psi_0}}.
\]
Now, 
\begin{align*}
&\frac{d}{dt}\bigg(\hat{U}(0, t) U(t, 0) \ket{\Psi_0}\bigg) = i\sum_{\alpha = 1}^M \hat{U}(0, t) \big(L_\alpha a_{\alpha, \upnu_{\alpha, t} v_\alpha - \uptau_{\alpha, t} v_\alpha}^+ +  L_\alpha^\dagger a_{\alpha, \upnu_{\alpha, t} v_\alpha - \uptau_{\alpha, t} v_\alpha}^- \big) \ket{\Psi(t)},
\end{align*}
where $\uptau_{\alpha, t} : L^2(\mathbb{R}) \to L^2(\mathbb{R})$ and $\upnu_{\alpha, t} : L^2(\mathbb{R}) \to L^2(\mathbb{R})$ are the time-translation and chain-unitary groups on the $\alpha^\text{th}$ bath respectively. We can thus obtain the estimate,
\[
\norm{\frac{d}{dt} \bigg(\hat{U}(0, t) U(t, 0) \ket{\Psi_0}\bigg)} \leq \sum_{\alpha = 1}^M \norm{L_\alpha} \bigg(\norm{a_{\alpha, \upnu_{\alpha, t} v_\alpha - \uptau_{\alpha, t} v_\alpha}^+ \ket{\Psi(t)}} + \norm{a_{\alpha, \upnu_{\alpha, t} v_\alpha - \uptau_{\alpha, t} v_\alpha}^- \ket{\Psi(t)}}\bigg).
\]
Moreover,
\begin{align*}
&\norm{a_{\alpha, \upnu_{\alpha, t} v_\alpha - \uptau_{\alpha, t} v_\alpha}^+ \ket{\Psi(t)}}^2 \leq \norm{\upnu_{\alpha, t} v_\alpha - \uptau_{\alpha, t} v_\alpha}_{L^2}^2 \sum_{n = 0}^\infty (n + 1)\norm{\Pi_n \ket{\Psi_\uptau(t)}}^2 = \norm{\upnu_{\alpha, t} v_\alpha - \uptau_{\alpha, t} v_\alpha}_{L^2}^2 \big(1 + \mu_{\ket{\Psi(t)}}^{(1)}\big) \ \text{and}\\
&\norm{a_{\alpha, \upnu_{\alpha, t} v_\alpha - \uptau_{\alpha, t} v_\alpha}^- \ket{\Psi(t)}}^2 \leq \norm{\upnu_{\alpha, t} v_\alpha - \uptau_{\alpha, t} v_\alpha}_{L^2}^2 \sum_{n = 0}^\infty n\norm{\Pi_n \ket{\Psi(t)}}^2 = \norm{\upnu_{\alpha, t} v_\alpha - \uptau_{\alpha, t} v_\alpha}_{L^2}^2 \mu_{\ket{\Psi(t)}}^{(1)}
\end{align*}
Finally, using lemma \ref{lemma:particle_num_bound}, we obtain that
\[
\mu^{(1)}_{\ket{\Psi(t)}} \leq 2{\mu^{(1)}_{\ket{\Psi_0}}} +2t^2 \bigg( \sum_{\alpha = 1}^M \norm{L_\alpha} \norm{v_\alpha}_{L^2} \bigg)^2.
\]
From these estimates, we thus obtain
\[
\norm{\ket{\Psi(t)} - \ket{\hat{\Psi}(t)}} \leq 2t \bigg(1 + 2{\mu_{\ket{\Psi_0}}^{(1)}} + 2t^2 \bigg( \sum_{\alpha = 1}^M \norm{L_\alpha} \norm{v_\alpha}_{L^2} \bigg)^2\bigg)^{1/2} \sum_{\alpha = 1}^M \norm{L_\alpha} \sup_{s \in [0,t]}\norm{\upnu_{\alpha, s} v_\alpha - \uptau_{\alpha, s} v_\alpha}_{L^2}
\]
The lemma now follows using this estimate together with lemma~\ref{lemma:truncation_chain_group}. $\hfill \square$

\section{Proof of lemma \ref{lemma:chain_model_bqp} }\label{app:hspace_trunc}
\begin{replemma}{lemma:chain_model_bqp}Problem \ref{prob:chain_model} can be solved on a quantum computer in run time $\textnormal{poly}(n)$.
\end{replemma}
\noindent\emph{Proof}: For notational simplicity, we will denote by $a_{\alpha, j}$ for $\alpha \in \{1, 2 \dots M\}$ and $j \in \{1, 2 \dots N_m\}$ the annihilation operator corresponding to the $j^\text{th}$ chain mode of the $\alpha^\text{th}$ bath, which have the commutation relations $[a_{\alpha, j}, a_{\alpha', j'}^\dagger] = \delta_{\alpha, \alpha'} \delta_{j, j'}$. Problem \ref{prob:chain_model} is then equivalent to the simulation of the Hamiltonian defined on the Hilbert space $\mathcal{H}_S \otimes \textnormal{Fock}[\mathbb{C}^{N_m}]^{\otimes M}$
\begin{align}\label{eq:chain_hamiltonian}
H(t) = H_S(t) + \sum_{\alpha = 1}^M \norm{v_\alpha}_{L^2} \big( L_\alpha a_{\alpha, 1}^\dagger + L_\alpha^\dagger a_{\alpha, 1}\big) + \sum_{\alpha = 1}^M \sum_{j = 1}^{N_m} \omega_{\alpha, j}a_{\alpha, j}^\dagger a_{\alpha, j} + \sum_{\alpha = 1}^M \sum_{j = 1}^{N_m - 1} t_{\alpha, j}\bigg(a_{\alpha, j} a_{\alpha, j + 1}^\dagger + a_{\alpha, j + 1} a_{\alpha, j}^\dagger\bigg),
\end{align}
where $\{\omega_{\alpha, j}\}_{\alpha \in \{1, 2 \dots M\}, \{1, 2 \dots N_m\}}, \{t_{\alpha, j}\}_{\alpha \in \{1, 2 \dots M\}, \{1, 2 \dots N_m\}}$ are the chain parameters corresponding to the unitary group $\upnu_{\alpha, t}$. We first truncate this model into a finite-dimensional model --- to do so, we first derive a bound on the expectation number of particles in the $M$ baths coupling to the system. Denoting by $\mu_\alpha^{(1)} = \langle \sum_{j = 1}^{N_m} a_{\alpha, j}^\dagger a_{\alpha, j}\rangle$ and $\mu_\alpha^{(2)} = \langle \sum_{j = 1}^{N_m} (a_{\alpha, j}^\dagger a_{\alpha, j})^2\rangle$, we obtain from Heisenberg's equations of motion that
\begin{align*}
&\frac{d}{dt} \mu_\alpha^{(1)} = -i\norm{v_\alpha}_{L^2}\bigg( \langle L_\alpha a_{\alpha, 1}^\dagger \rangle - \textnormal{c.c.}\bigg) \leq 2\norm{v_\alpha}_{L^2} \norm{L_\alpha} \sqrt{\langle a_{\alpha,1}^\dagger a_{\alpha, 1}\rangle} \leq 2\norm{v_\alpha}_{L^2} \norm{L_\alpha} \sqrt{\mu_\alpha^{(1)}}, \nonumber\\
&\frac{d}{dt}\mu_\alpha^{(2)} = -i\norm{v_\alpha}_{L^2}\bigg(2\langle L_\alpha \big(a_{\alpha, 1}^\dagger\big)^2 a_{\alpha, 1}\rangle + \langle L_\alpha a_{\alpha, 1}^\dagger \rangle - \textnormal{c.c.}\bigg) \leq 2\norm{v_\alpha}_{L^2}\norm{L_\alpha} \sqrt{\mu_\alpha^{(1)}}\bigg( 2\sqrt{\mu_\alpha^{(2)}} + 1\bigg)
\end{align*}
integrating which yields
\[
\mu_\alpha^{(1)}(t) \leq \bigg(\sqrt{\mu_\alpha^{(1)}(0)} + \norm{v_\alpha}_{L^2} \norm{L_\alpha} t\bigg)^2,
\]
and
\[
\sqrt{\mu_\alpha^{(2)}(t)} - \frac{1}{2}\log\big(1 + 2\sqrt{\mu_\alpha^{(2)}(t)}\big) \leq\sqrt{\mu_\alpha^{(2)}(0)} - \frac{1}{2}\log\big(1 + 2\sqrt{\mu_\alpha^{(2)}(0)}\big) + 2\norm{v_\alpha}_{L^2}\norm{L_\alpha}\bigg(\sqrt{\mu_\alpha^{(1)}(0)} t + \frac{1}{2}\norm{v_\alpha}_{L^2} \norm{L_\alpha} t^2\bigg),
\]
or equivalently
\[
\sqrt{\mu_\alpha^{(2)}(t)} - \bigg(\frac{\mu_\alpha^{(2)}(t)}{4}\bigg)^{1/4}  \leq \sqrt{\mu_\alpha^{(2)}(0)}+ 2\norm{v_\alpha}_{L^2}\norm{L_\alpha}\bigg(\sqrt{\mu_\alpha^{(1)}(0)} t + \frac{1}{2}\norm{v_\alpha}_{L^2} \norm{L_\alpha} t^2\bigg),
\]
where we have used that for $x \geq 0$, $0\leq \log(1 + x)\leq \sqrt{x}$. We thus obtain that if $t = \textnormal{poly}(n)$, $\mu_\alpha^{(2)}(t) = \textnormal{poly}(n)$. With this bound, we consider truncating the Hamiltonian --- given $p \in \mathbb{Z}_{>1}$, we consider the projector $\mathcal{P}_p = \textnormal{id}\otimes \Pi_{\leq p}^{\otimes M}$, where $\Pi_{\leq p}$ is a projector onto the space with less than or equal to $p$ particles defined on $\textnormal{Fock}[\mathbb{C}^{N_m}]$. Also, we define $\mathcal{Q}_p = \textnormal{id} - \mathcal{P}_p$. Denoting by $\ket{\Psi(t)}$ the state corresponding to the Hamiltonian under consideration (Eq.~\ref{eq:chain_hamiltonian}) at time $t$ and by $U_{\mathcal{P}}(t, 0)$ the propagator corresponding to the Hamiltonian $\mathcal{P} H(t) \mathcal{P}$, then
\[
\norm{ \ket{\Psi(t)} - U_\mathcal{P}(t, 0) \mathcal{P}\ket{\Psi_0}} \leq \norm{\mathcal{Q}_p\ket{\Psi(t)}} + \int_0^t \norm{\mathcal{P}_p H(s) \mathcal{Q}_p \ket{\Psi(s)}}ds.
\]
Both the terms in the above estimate can be easily bounded from above in terms of $p$ --- note that
\begin{align}\label{eq:projector_q_estimate}
\norm{\mathcal{Q}_p\ket{\Psi(t)}}^2 \leq \sum_{\alpha = 1}^M \bra{\Psi(t)} \textnormal{id}\otimes\big(\textnormal{id}^{\otimes (\alpha - 1)} \otimes \Pi_{> p} \otimes \textnormal{id}^{M- \alpha}\big)\ket{\Psi(t)} \leq \frac{1}{p}\sum_{\alpha = 1}^M \mu_\alpha^{(1)}(t) = \frac{1}{p}\textnormal{poly}(n),
\end{align}
where we have used $M = \textnormal{poly}(n)$. Furthermore, noting that $\mathcal{P}_p, \mathcal{Q}_p$ commute with any system operators, and for $\alpha \in \{1, 2 \dots M\}, i, j \in \{1, 2 \dots N_m\}$, $\mathcal{P}_p a_{\alpha, i}^\dagger a_{\alpha, j} \mathcal{Q}_p = 0$ and $\mathcal{P}_p a_{\alpha, i}^\dagger \mathcal{Q}_{p} = 0$, we obtain that for $s \in (0, t)$, 
\[
\norm{\mathcal{P}_p H(s) \mathcal{Q}_p \ket{\Psi(s)}} \leq \sum_{\alpha = 1}^M \norm{v_\alpha}_{L^2} \norm{L_\alpha} \norm{\mathcal{P}_p a_{\alpha, 1} \mathcal{Q}_p \ket{\Psi(s)}} \leq  \sum_{\alpha = 1}^M \norm{v_\alpha}_{L^2} \norm{L_\alpha} \norm{a_{\alpha, 1} \mathcal{Q}_p \ket{\Psi(s)}}.
\]
For $\alpha \in \{1, 2 \dots M\}$, we obtain that
\[
\norm{ a_{\alpha, 1} \mathcal{Q}_p \ket{\Psi(s)}}^2 = \bra{{\Psi(s)}} \mathcal{Q}_p a_{\alpha, 1}^\dagger a_{\alpha, 1}\mathcal{Q}_p \ket{\Psi(s)} =  \bra{{\Psi(s)}} a_{\alpha, 1}^\dagger a_{\alpha, 1}\mathcal{Q}_p \ket{\Psi(s)} \leq \bra{\Psi(s)}\big(a_{\alpha, 1}^\dagger a_{\alpha, 1}\big)^2\ket{\Psi(s)} \bra{\Psi(s)} \mathcal{Q}_p \ket{\Psi(s)},
\]
and consequently using Eq.~\ref{eq:projector_q_estimate},
\[
\norm{ a_{\alpha, 1} \mathcal{Q}_p \ket{\Psi(s)}}^2 \leq \frac{\mu^{(2)}_\alpha(s)}{p}\sum_{\alpha' = 1}^M \mu^{(1)}_{\alpha'}(s).
\]
Therefore, $\int_0^t \norm{\mathcal{P}_p H(s) \mathcal{Q}_p(s) \ket{\Psi(s)}}ds \leq {\textnormal{poly}(n)}/{\sqrt{p}}$. Thus, we obtain the estimate
\[
\norm{\ket{\Psi(t)} - U_{\mathcal{P}}(t, 0) \mathcal{P}\ket{\Psi_0}} \leq \frac{\textnormal{poly}(n)}{\sqrt{p}}.
\]
Hence, to ensure that the error is within the truncation is below $1 / \textnormal{poly}(n)$, we need to choose $p = \textnormal{poly}(n)$.

Finally, we apply lemma \ref{lemma:hamil_simul} to prove the simulatability of the hamiltonan $\mathcal{P}_p H(t) \mathcal{P}_p$ --- we need to show that on a (suitably chosen) basis, for any choice of computational basis $\ket{b}$, $H(t)\ket{b}$ can be efficiently computed as a sparse vector. We consider the basis set of the form $\mathcal{B} = \mathcal{B}_S \times \mathcal{B}_1 \times \mathcal{B}_2 \dots \times \mathcal{B}_M$, where $\mathcal{B}_S$ is the computational basis for the $n$ qudit system, and for $\alpha \in \{1, 2 \dots M\}$, $\mathcal{B}_\alpha$ is the subset of Fock state basis for the $\alpha^\text{th}$ bath with number of particles less than $p$ i.e.
\[
\mathcal{B}_\alpha = \bigg\{\big(a_{\alpha, 1}^\dagger\big)^{n_1} \big(a_{\alpha, 2}^\dagger\big)^{n_2} \dots \big(a_{\alpha, N_m}^\dagger\big)^{n_{N_m}}\ket{\text{vac}}  \bigg| n_1, n_2 \dots n_{N_m} \in \mathbb{Z}_{\geq 0} \text{ with } n_1 + n_2 \dots n_{N_m}\leq p \bigg\}
\]
Consider now the Hamiltonian $\mathcal{P}_p H(t) \mathcal{P}_p$ --- it can be expressed as sum the following terms:
\begin{itemize}
\item $\mathcal{P}_p H_S(t) \mathcal{P}_p$ --- Since by assumption is expressible only as $\textnormal{poly}(n)$ operators that act on at-most $k$ qudits, it immediately follows that $\mathcal{P}_p H_S(t)\mathcal{P}_p \ket{b} = H_S(t)\ket{b}$ can be classically efficiently computed for $\ket{b} \in \mathcal{B}$.
\item $\mathcal{P}_p \omega_{\alpha, j} a_{\alpha, j}^\dagger a_{\alpha, j} \mathcal{P}_p$ for $\alpha \in \{1, 2 \dots M\}, j \in \{1, 2 \dots N_m\}$ --- this term is diagonal in the basis $\mathcal{B}$. Furthermore, since there are only $N_m M = \textnormal{poly}(n)$ such terms, $\mathcal{P}_p\sum_{\alpha = 1}^M \sum_{j = 1}^{N_m} \omega_{\alpha, j} a_{\alpha, j}^\dagger a_{\alpha, j} \mathcal{P}_p \ket{b}$ can be efficiently computed $\forall \ket{b} \in \mathcal{B}$.
\item $\mathcal{P}_p t_{\alpha, j} (a_{\alpha, j} a_{\alpha, j + 1}^\dagger + a_{\alpha, j + 1} a_{\alpha, j}^\dagger) \mathcal{P}_p$ for $\alpha \in \{1, 2 \dots M\}, j \in \{1, 2 \dots N_m - 1\}$ --- applying this term on $\ket{b} \in \mathcal{B}$ produces a vector with at-most two non-zero elements when represented on the same basis. Furthermore, since there are only $(N_m - 1)M = \text{poly}(n)$ such terms, $\mathcal{P}_p \sum_{\alpha = 1}^M \sum_{j = 1}^{N_m - 1}t_{\alpha, j} (a_{\alpha, j} a_{\alpha, j + 1}^\dagger + a_{\alpha, j + 1} a_{\alpha, j}^\dagger) \mathcal{P}_p\ket{b}$ can be efficiently computed $\forall \ket{b}\in \mathcal{B}$.
\item $\mathcal{P}_p \big(L_\alpha a_{\alpha, 1}^\dagger + L_\alpha^\dagger a_{\alpha, 1}\big) \mathcal{P}_p$ for $\alpha \in \{1, 2 \dots M\}$ --- since $L_\alpha$ only acts on at-most $k$ qudits, applying this term on $\ket{b} \in \mathcal{B}$ produces a vector with at-most $2d^k$ non-zero elements when represented on the same basis. Furthermore, since there are only $M =\text{poly}(n)$ such terms, $\mathcal{P}_p \sum_{\alpha = 1}^M \big(L_\alpha a_{\alpha, 1}^\dagger + L_\alpha^\dagger a_{\alpha, 1}\big) \mathcal{P}_p \ket{b}$ can be efficiently computed $\forall \ket{b} \in \mathcal{B}$.
\end{itemize}
It thus follows that $\mathcal{P}_p H(t) \mathcal{P}_p\ket{b}$ can be efficiently computed $\forall \ket{b} \in \mathcal{B}$. Finally, we note from lemma \ref{lemma:upper_bound_chain_coeffs} that $\abs{\omega_{\alpha, j} }, t_{\alpha, j} \leq \omega_c$ for all $\alpha \in \{1, 2 \dots M\}, j \in \{1,2 \dots N_m\}$
\[
\norm{\mathcal{P}_p H(t) \mathcal{P}_p} \leq \norm{H_S(t)} +2\sqrt{p + 1} \sum_{\alpha = 1}^M \norm{L_\alpha} \norm{v_\alpha}_{L^2} + p M N_m \omega_c + 2\omega_c (p + 1) M (N_m - 1),
\]
where we have used the estimates $\norm{\mathcal{P}_p a_{\alpha, j} \mathcal{P}_p}, \norm{\mathcal{P}_p a_{\alpha, j}^\dagger \mathcal{P}_p} \leq \sqrt{p + 1}$, $\norm{\mathcal{P}_p a_{\alpha, j}^\dagger a_{\alpha, j }\mathcal{P}_p} \leq p$. Noting that by assumption $\norm{L_\alpha} \leq 1, \norm{H_S(t)}, p, M, N_m, \omega_c, t = \text{poly}(n)$, we obtain that $\int_0^t \norm{\mathcal{P}_p H(s) \mathcal{P}_p} ds \leq O(\textnormal{poly}(n))$. Thus, from lemma \ref{lemma:hamil_simul}, we can show that there is a circuit with depth $\text{poly}(n)$ that approximates the propagator corresponding to $\mathcal{P}_p H(s) \mathcal{P}_p$ with evolution time $t$ within $1 / \text{poly}(n)$ spectral norm error.Furthermore, the initial state can be efficiently represented on the basis $\mathcal{B}$ since it is efficiently projectable (assumption \ref{assump:initial_state}b), and hence the reduced system state $\rho_S(t)$ can be efficiently simulated on this quantum circuit. \hfill \(\square \)

\end{document}